Al-Azhar University
Faculty of Engineering
Electrical Engineering Department

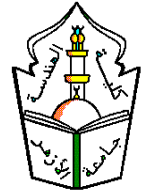

# Performance Evaluation of Wireless Multi-Carrier (MC) Communication Systems

A thesis
Submitted to Electrical Engineering Department
Faculty of Engineering – Al-Azhar University for the Degree of

**Master of Science**

In

Electrical Communications Engineering

By

**Mohammed Sobhy Ahmed El-Bakry**

B.Sc. of Electronics and Communications Engineering – Institute of Aviation Engineering and Technology (IAET)

Under the Supervision of

**Prof. Dr. Abd El-Hady Abd El-Azim Ammar**

Electrical Engineering Department
Faculty of Engineering – Al-Azhar University

**Dr. Hamed Abd El-Fattah El-Shenawy**

Electrical Communications Department
Higher Institute of Engineering – El-Shourok Academy

Cairo – Egypt
2014

Al-Azhar University
Faculty of Engineering
Electrical Engineering Department

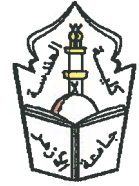

# Performance Evaluation of Wireless Multi-Carrier (MC) Communication Systems

A thesis
Submitted to Electrical Engineering Department
Faculty of Engineering – Al-Azhar University for the Degree of

**Master of Science**

In

Electrical Communications Engineering

By

**Mohammed Sobhy Ahmed El-Bakry**

B.Sc. of Electronics and Communications Engineering – Institute of Aviation Engineering and Technology (IAET)

**Approved by examining committee**

**Prof. Dr. El-Sayed Mostafa Saad**
Faculty of Engineering – Helwan University

**Prof. Dr. Hesham Mohamed El-Badaway**
National Telecommunication Institute (NTI)

**Prof. Dr. Abd El-Hady Abd El-Azim Ammar**
Faculty of Engineering – Al-Azhar University

Cairo-Egypt
2014

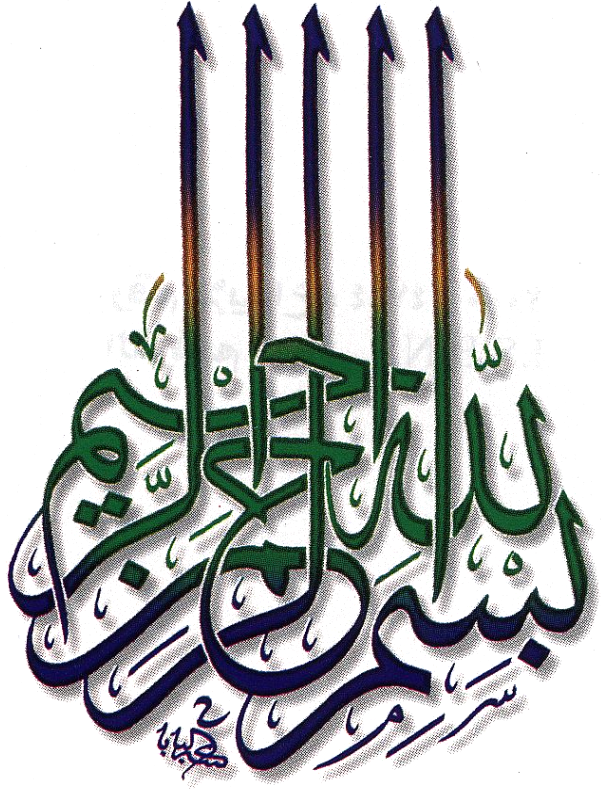

قُلْ إِنَّ صَلَاتِي وَنُسُكِي وَمَحْيَايَ وَمَمَاتِي لِلَّهِ رَبِّ الْعَالَمِينَ ﴿١٦٢﴾ لَا شَرِيكَ لَهُ ۖ وَبِذَٰلِكَ أُمِرْتُ وَأَنَا أَوَّلُ الْمُسْلِمِينَ ﴿١٦٣﴾

سورة الأنعام

صدق الله العظيم

# Acknowledgment

In the Name of Allah, the Most Merciful, the Most Compassionate all praise be to ALLAH, the Lord of the worlds; and prayers and peace be upon Mohamed His servant and messenger. First and foremost, I must acknowledge my limitless thanks to ALLAH, the Ever Magnificent; the Ever-Thankful, for His help and bless. I am totally sure that this work would have never become truth, without His guidance.

I would like to express my sincere appreciation to my supervisors Prof. Dr. Abd El-Hady Abd El-Azim Ammar, Professor of Communication Engineering, Al-Azhar University and Dr. Hamed Abd El-Fattah El-Shenawy, Doctor of Communication Engineering, Higher Institute of Engineering, El-Shourok Academy for their deep interest in the subject of this study, their guidance, invaluable discussions in numerous aspects, and their encouragement throughout.

I would also like to thank my committee members, Prof. Dr. El Sayed Mostafa Saad and Prof. Dr. Hesham Mohamed El-Badaway, for the assistance they provided at all levels of the research project. I also want to thank you for letting my defense be an enjoyable moment, and for your brilliant comments and suggestions, thanks to you.

I would like to appreciate the help offered by the Institute of Aviation Engineering and Technology staff for their support and for giving me the motivation to finish this study. Also, I would like to thank my all friends specially Dr. Ahmed Gomaa , Eng. Shimaa El-Sayed El-Tantawy, Eng. Ramez Mohammed El-Askary, Eng. Ihab Samy and Mrs. Mona Mohammed Ali for real support during my postgraduate study and thesis works.

I would like to acknowledge the support of all members of my family, my father, my mother, my father-in-law, my brother and my sisters for endowing me with the right tips to successful life in this world.

Last but not least, I would like to express my feelings for my wife Sherihan and my children Esraa and Anas for their emotional support.

*Mohammed Sobhy – September 10th 2014*

# Dedication

I dedicate my dissertation work to ALLAH, the major source of strength and success in my life. This work is dedicated to the sake of ALLAH, my Creator and my Master.

I dedicate my dissertation work to my uncle Mohammed and my aunt Amaal mercy ALLAH on them.

I dedicate my dissertation work to my teachers.

I dedicate my dissertation work to my loving parents, my brother Ahmed, my sisters Hanaa and Sayeda, all members of my family, my loving wife Sherihan and my children Esraa and Anas.

# Table of contents

# List of tables

# List of figures

# List of abbreviations

## A
| | |
|---|---|
| AWGN | Additive white gaussian noise |
| AM-AM | Amplitude-to- amplitude distortion |
| ADC | Analog-to-digital converter |
| ADSL | Asymmetric digital subscriber line |

## B
| | |
|---|---|
| BER | Bit error rate |
| BLUE | Best linear unbiased estimator |
| BPSK | Binary phase shift keying |

## C
| | |
|---|---|
| CIR | Channel impulse response |
| CP | Cyclic prefix |
| CFO | Carrier frequency offset |
| CPE | Common phase error |
| CIR | Carrier-to-interference power ratio |

## D
| | |
|---|---|
| DFT | Discrete Fourier transform |
| DAC | Digital to-analog converter |
| DAB | Digital audio broadcasting |
| DA | Data-aided |
| DC | Direct current |
| DFL | Decision feedback loop |
| DSP | Digital signal processing |
| DMT | Discrete multi-tone |

## F
| | |
|---|---|
| FFT | Fast Fourier transform |
| FFO | Fractional carrier frequency offset |
| FDM | Frequency division multiplexing |
| FEC | Forward error correction |
| FL | Frequency loop |
| FD | Frequency domain |

## G
| | |
|---|---|
| GI | Guard interval |

## H
| | |
|---|---|
| HPA | High power amplifier |

## I
| | |
|---|---|
| ISI | Inter-symbol interference |
| ICI | Inter-carrier interference |
| IFFT | Inverse fast Fourier transform |
| IDFT | Inverse discrete Fourier transform |
| IQ | In-phase and quadrature |
| IF | Intermediate frequency |
| IFO | Integer carrier frequency offset |
| IBO | Input back-off |

## L

| | |
|---|---|
| LOS | Line of sight |
| LO | Local oscillator |
| LOS | Line of sight |
| LF | Loop filter |
| LTE | Long-term evolution |

## M

| | |
|---|---|
| MC | Multi-carrier |
| MCM | Multi-carrier modulation |
| MLE | Maximum likelihood estimation |
| MSE | Mean square error |
| M-PSK | M-ary phase shift keying |
| M-QAM | M-ary quadrature amplitude modulation |
| MG | Mirror gain |

## N

| | |
|---|---|
| NDA | Non-data-aided |
| NLOS | Non-line-of-sight |
| NCO | Numerically controlled oscillator |

## O

| | |
|---|---|
| OFDM | Orthogonal frequency division multiplexing |
| OBO | Output back-off |

## P

| | |
|---|---|
| PAPR | Peak-to-average power ratio |
| PA | power amplifier |
| PN | Pseudonoise |
| PLL | Phase locked loop |
| PD | Phase detector |
| PED | phase error detector |
| PPM | Parts per million |
| PI filter | Proportional plus integral filter |

## Q

| | |
|---|---|
| QPSK | Quadrature phase shift keying |
| 16-QAM | 16-Quadrature amplitude modulation |
| 64-QAM | 64-Quadrature amplitude modulation |

# R

| | |
|---|---|
| RF | Radio frequency |
| RMS | Root mean square |
| RMSE | Root mean square error |

# S

| | |
|---|---|
| SCM | Single carrier modulation |
| STO | Symbol timing offset |
| SNR | Signal to noise ratio |
| SIR | Signal to interference ratio |
| S/P | Serial to parallel |
| SG | Signal gain |

# T

| | |
|---|---|
| TF-DFL | Time-frequency decision-feedback loop |
| TL | Time loop |

# V

| | |
|---|---|
| VCO | Voltage controlled oscillator |
| VDSL | Very high speed digital subscriber line |

# W

| | |
|---|---|
| WLAN | Wireless local area network |
| WSS | Wide sense stationary |
| WiMAX | Worldwide interoperability for microwave access |

# List of symbols

| | |
|---|---|
| $\bar{\tau}$ | Mean excess delay |
| $\tau_{RMS}$ | RMS delay spread |
| $\tau_{max}$ | Channel delay spread |
| $B_s$ | Signal bandwidth |
| $B_c$ | Coherence bandwidth |
| $B_D$ | Doppler spread |
| $T_C$ | Coherence time |
| $v$ | Relative velocity of the transmitter with respect to the receiver |
| $c$ | Speed of light |
| $f_c$ | Center carrier frequency |
| $f_d$ | Doppler frequency shift |
| $\theta$ | Angle between the direction of motion of the mobile and direction of arrival of the scattered waves |
| $a_p$ | The amplitude of $p^{th}$ path |
| $\tau_p$ | The propagation delay of $p^{th}$ path |
| $\theta_p$ | The phase shift of $p^{th}$ path |
| $L_h$ | The length of channel impulse response |
| $\delta(t)$ | Delta function |
| $h(\tau)$ | The channel impulse response |
| $\Delta f$ | Subcarrier spacing |
| $N_u$ | Number of subcarriers during useful duration |
| $N_{cp}$ | The cyclic prefix length |
| $N_{sym}$ | OFDM symbol Length after adding the cyclic prefix |
| $T_u$ | Useful OFDM symbol duration |
| $T_g$ | Guard interval |
| $T_{sym}$ | Complete OFDM symbol duration |
| $\Re$ | Data rate |
| $G$ | The fractional overhead |
| $\Phi_k(t)$ | $k^{th}$ subchannel waveform |
| $f_k$ | $k^{th}$ subcarrier frequency |
| $SNR_{loss}$ | SNR loss due to CP insertion |
| $Z_m$ | Transmitted complex data symbols at $m^{th}$ subcarrier |
| $T_s$ | Sampling time |
| $z_l(t)$ | The $l^{th}$ Baseband OFDM signals in the continuous-time domain |
| $z_{n,l}$ | The $l^{th}$ transmitted OFDM symbol at the $n^{th}$ sample in time domain |
| $z_{n,l}^c$ | The $l^{th}$ transmitted OFDM symbol with the CP |
| $Z_{m,l}$ | The $l^{th}$ transmitted OFDM symbol at the $k^{th}$ subcarrier in frequency domain |
| $z_l^c(t)$ | The anlage signal of the $l^{th}$ transmitted OFDM symbol |
| $r_l^c(t)$ | The received signal in the continuous-time domain |

| Symbol | Description |
|---|---|
| $\otimes$ | The linear convolution operation |
| $w_l(t)$ | The AWGN in in the continuous-time domain |
| $\sigma_n^2$ | The noise variance |
| $r_{n,l}^c$ | The $l^{th}$ received OFDM symbol at the $n^{th}$ samples with CP |
| $w_{n,l}$ | The $l^{th}$ AWGN at the $n^{th}$ samples in time domain |
| $r_{n,l}$ | The $l^{th}$ received OFDM symbol at the $n^{th}$ samples after CP removal |
| $R_{k,l}$ | The $l^{th}$ received OFDM symbol at the $k^{th}$ subcarrier in frequency domain |
| $H_{k,l}$ | The $l^{th}$ channel frequency response at the $k^{th}$ subcarrier in frequency domain |
| $W_{k,l}$ | The $l^{th}$ AWGN at the $k^{th}$ subcarrier in frequency domain |
| $\theta_r$ | Phase imbalance at the receiver |
| $\mu$ | Amplitude imbalance at the receiver |
| $\tilde{X}_k$ | The $k^{th}$ subcarrier after the FFT in the OFDM receiver distorted by IQ imbalance |
| $\xi$ | Signal Gain |
| $\sigma$ | Mirror Gain |
| $P_{in_{sat}}$ | The input saturation power of an amplifier |
| $\bar{P}_{in}$ | The average input power |
| $x_{n,l}$ | The $l^{th}$ received OFDM symbol at the $n^{th}$ samples with phase noise in time domain |
| $X_{k,l}$ | The $l^{th}$ received OFDM symbol at the $k^{th}$ subcarrier with phase noise in frequency domain |
| $\delta$ | The symbol timing offset |
| $\alpha(\delta)$ | An attenuation factor |
| $\beta_{k,l}$ | Accounts for ISI and ICI |
| $\tilde{Y}_{k,l}$ | The output of the DFT of the $k^{th}$ subcarrier of the $l^{th}$ symbol |
| $\delta f$ | The carrier frequency offset |
| $\varepsilon$ | The normalized frequency offset |
| $\varepsilon_i$ | The integer carrier frequency offset |
| $\varepsilon_f$ | The fractional carrier frequency offset |
| $y_{n,l}$ | The $l^{th}$ received OFDM symbol at the $n^{th}$ sample in time domain with CFO |
| $Y_{k,l}$ | The $l^{th}$ received OFDM symbol at the $k^{th}$ subcarrier in frequency domain with CFO |
| $\lvert k - \varepsilon_i \rvert_N$ | The value of $(k - \varepsilon_i)$ reduced to interval $[0, N_u - 1]$ |
| $\Lambda_0$ | The attenuation and the phase rotation factor due to the FFO |
| $\Lambda_{m-k}$ | The ICI coefficient between $m^{th}$ and $k^{th}$ subcarriers due to the FFO |
| $I_{k,l}$ | The ICI component |
| $\gamma(\varepsilon)$ | The SNR loss due to CFO |
| $\hat{\varepsilon}$ | The estimated normalized CFO |
| $arg(.)$ | The argument operation |
| $e_\varepsilon$ | The estimation error |

| Symbol | Description |
|---|---|
| $L_{av}$ | The number of samples used for averaging |
| $\mathcal{H}(z)$ | The phase transfer function of the first order PLL |
| $\mathcal{H}_e(z)$ | The phase error transfer function of the first order PLL |
| $\theta_i(z)$ | The Z-transforms of input phase sequence $\theta_i(n)$ |
| $\theta_e(z)$ | The Z-transforms of phase error sequence $\theta_e(n)$ |
| $\hat{\theta}(z)$ | The Z-transforms of output phase sequence $\hat{\theta}(n)$ |
| $\alpha'$ | The proportional gain of the loop filter of first order PLL |
| $y'_{n,l}$ | The $l^{th}$ received corrected OFDM symbol at the $n^{th}$ sample in time domain |
| $Y'_{k,l}$ | The $l^{th}$ received corrected OFDM symbol at the $k^{th}$ subcarrier in frequency domain |
| $\varepsilon_{k,l}$ | The error increment |
| $\hat{\varphi}_l$ | The time loop output sequence |
| $\hat{\psi}_{k,l}$ | The frequency loop output sequence |
| $a_{k,l}$ | The real and image parts of $Y'_{k,l}$ |
| $b_{k,l}$ | The image parts of $Y'_{k,l}$ |
| $e^I_{k,l}$ | The error between the real part of $Y'_{k,l}$ and the decision output |
| $I_{k,l}$ | The real part of the decision output |
| $Q_{k,l}$ | The image part of the decision output |
| $e^Q_{k,l}$ | The error between the image part of $Y'_{k,l}$ and the decision output |
| $\text{sgn}(.)$ | The signum function |
| $\bar{\varepsilon}_l$ | The average value of $\varepsilon_{k,l}$ |
| $\varphi$ | The phase difference between two identical parts |
| $N_0$ | The power spectral density of the AWGN |
| $var[\hat{\varepsilon}|\varepsilon,\{Y_k\}]$ | The variance of estimation error for the moose estimator |
| $u(t)$ | The first half of the first training symbol of the Schmidl and Cox estimator |
| $\hat{\varphi}$ | The estimated phase difference between two identical parts |
| $P_{S\&C}(d)$ | The correlation between the two halves of parts of the Schmidl and Cox estimator |
| $v$ | The integer number |
| $F_{1,k}$ | The received first training symbol in frequency domain |
| $F_{2,k}$ | The received second training symbol in frequency domain |
| $v_k$ | The differentially modulated even frequencies of second training symbol |
| $X_E$ | The set of indices for even frequency components |
| $W_E$ | The number of even frequencies with the PN sequence |
| $g$ | The spanning the range of possible frequency offsets |
| $\hat{g}$ | The estimated value of integer number $v$ |
| $B(g)$ | The metric function |
| $\hat{\varepsilon}_f$ | The estimated FFO |
| $\hat{\varepsilon}_i$ | The estimated IFO |
| $Q$ | The number of identical parts of Morelli and Mengali estimator |

| Symbol | Description |
|---|---|
| $\Psi(q)$ | The correlation between identical parts of the training of Morelli and Mengali estimator |
| $M$ | The length in sampling interval of each part of the training symbol |
| $P$ | Design parameter |
| $D(n)$ | The real-valued random envelope |
| $\alpha(q)$ | The noise term |
| $w_{M\&M}(q)$ | The weighting function |
| $\varphi(q)$ | The estimation of phase differences of the correlation |
| $D$ | An integer number |
| $p(j)$ | The location of $j^{th}$ pilot tone |
| $\mathcal{Z}_{l,p(j)}$ | The pilot tone at location $p(j)$ in FD at the $l^{th}$ OFDM symbol |
| $L_F$ | The known pilot symbol pairs in $l^{th}$ and $(l+D)^{th}$ OFDM symbols |
| $\varepsilon_{trial}$ | The trial frequency value |
| $S_f$ | The sets of the left pilot positions in one OFDM symbol, |
| $S_r$ | The sets of the right and all pilot positions in one OFDM symbol |
| $S$ | The sets of the all pilot positions in one OFDM symbol |
| $\mathcal{R}_{k,l}$ | The $l^{th}$ received clustered pilot tones at the $k^{th}$ subcarrier in frequency domain |
| $\mathcal{P}_s$ | The signal power of $\mathcal{R}_{k,l}$ |
| $\mathcal{P}_{ICI}$ | The ICI power of $\mathcal{R}_{k,l}$ |
| $SIR_k$ | The signal to interference ratio |
| $Re(.)$ | The real part |
| $Im(.)$ | The imaginary part |
| $\sigma_s^2$ | The variance of transmitted data |
| $\hat{\varphi}_k$ | The estimate of the phase pilot |
| $var(\hat{\varphi}_k)$ | The variance of the estimate of the phase pilot |
| $E\{Y_{k,l-1}^* Y_{k,l}\}$ | The expectation of the conjugate product of the received pilot tone |
| $var[Y_{k,l-1}^* Y_{k,l}]$ | The variance of the conjugate product of the received pilot tone |
| $H(z)$ | The phase transfer function of the type-2 control loop |
| $H_e(z)$ | The phase error transfer function of the type-2 control loop |
| $k_p$ | The proportional gain of the loop filter of the type-2 control loop |
| $k_I$ | The integral gain of the loop filter of the type-2 control loop |
| $\theta_n$ | The loop bandwidth |
| $\zeta$ | The damping factor |
| $F(s)$ | The transfer function of the PI filter |
| $\omega_n$ | The natural frequency |
| $C$ | The capacitor symbol |
| $R$ | The resistance symbol |
| $K_d$ | The PD gain |
| $K_o$ | The VCO gain |
| $\theta_i(s)$ | The input phase signal of the analog PLL |
| $\hat{\theta}(s)$ | The VCO's phase signal of the analog PLL |
| $\theta_e(s)$ | The phase error signal of the analog PLL |
| $Q'$ | The indication voltage signal |
| $B_T$ | The tracking range |
| $f_{max}$ | The maximum frequency of tracing range |
| $f_{min}$ | The minimum frequency of tracing range |
| $f_i$ | The input frequency |
| $f_o$ | The output frequency |
| $B_{ac}$ | The acquisition range |
| $f_2$ | The highest frequency of acquisition range |
| $f_1$ | The lowest frequency of acquisition range |
| $B_L$ | The noise bandwidth |
| $T_p$ | The acquisition time |
| $\Delta\omega_0$ | The initial frequency offset between the input and the VCO signals |

# Abstract

With the increasing demand for more wireless multimedia applications, it is desired to design wireless communication systems with higher data rate. Furthermore, the frequency spectrum has become a limited and valuable resource, making it necessary to utilize the available spectrum efficiently and coexist with other wireless systems.

Orthogonal frequency division multiplexing (OFDM) is a multicarrier modulation technique that has become a viable method for wireless communication systems due to the high spectral efficiency, immunity to frequency selective multipath fading, and being flexible to integrate with other techniques. However, the high-peak-to-average power ratio and sensitivity to carrier frequency synchronization errors are the major drawbacks for OFDM systems.

The carrier frequency offset (CFO) leads to inter-carrier interference (ICI) and deteriorates OFDM system performance. Thus, an accurate CFO compensation technique is required in order to achieve acceptable performance.

The CFO is the mismatch between the received carrier frequency and the carrier frequency generated by the local oscillator in the receiver. CFO occurs due to mismatch between the local oscillator at transmitter and receiver, and due to Doppler shifts due to relative motion between transmitter and receiver. CFO is divided into integer CFO (IFO) and fractional CFO (FFO).

A survey on different frequency synchronization schemes is made. These schemes are categorized into data-aided (DA) and non-data aided (NDA) schemes. In DA schemes additional training symbols are inserted into the transmitted data such as preamble or pilot tones. These schemes yields better performance but they are not bandwidth efficient. In NDA schemes the transmitted data are used without any other additional information. These schemes are bandwidth efficient but its performance is less than the DA schemes.

Frequency synchronization process is performed via two stages: acquisition and tracking. Acquisition means coarse estimation for CFO. It has wide range but low accuracy and deals with IFO. Tracking means fine estimation for CFO. It has narrower range and finer accuracy and deals with FFO.

This thesis is concerned with DA scheme and CFO tracking for OFDM system. OFDM system model is developed first without CFO and then with CFO. The system performance is evaluated via simulation. The bit error rate (BER), constellation diagram, the phase output and phase error were taken as performance measures. The performance is evaluated for different types of modulation over different channel conditions.

In this thesis, a dual bandwidth for CFO tracking based on type-2 control loop to reduce the acquisition time of PLL is proposed. It is proved that the proposed scheme is significantly faster than [20], and improves the system BER. Also, further refinement to the dual bandwidth scheme is improved by using clustered pilot tones. The improved dual bandwidth scheme is better than the dual bandwidth scheme.

# List of publications

Mohammed. S. El-Bakry, Hamed. F. El-Shenawy, Abd El-Hady. A. Ammar "Improved Carrier Frequency Offset Correction for OFDM Systems Based on Type-2 Control Loop", 31$^{st}$ National Radio Science Conference (NRSC), pp.157-166, 28-30 April 2014.

Mohammed. S. El-Bakry, Hamed. F. El-Shenawy, Abd El-Hady. A. Ammar "Improved Carrier Frequency Offset Correction for OFDM Systems Based on Dual Bandwidth and Clustered Pilot Tones", submitted to Wireless Personal Communications (WPC) and under reviewing.

# Chapter 1. Introduction

## 1.1 Thesis motivation

With the growing demand for wireless multimedia applications, it is desirable to design wireless systems with higher data rates. The limited frequency bandwidth, which often can be seen as an obstacle to the development of telecommunication, is also the propulsion for the evaluation of the wireless technology. The wireless technology is required to operate at high data rates, at high carrier frequencies under the environment of high mobility and large spectral interference, while the data transmission still remains reliable and supports multiple users.

Orthogonal frequency division multiplexing (OFDM) technology is at the core of multicarrier systems that play a crucial role in achieving the above requirements. OFDM modulation is widely used in communication systems to meet the demand for ever increasing data rates. The major advantage of OFDM over single-carrier transmission is its ability to deal with severe channel conditions without complex equalization. An OFDM signal can be viewed as a number of narrowband signals combined together based on orthogonality principle rather than one wideband signal, thus the complexity of the receiver is significantly reduced.

Unfortunately, The OFDM system suffers from different impairments such as the high peak-to-average power ratio (PAPR). The sensitivity to synchronization errors, such as carrier frequency offset (CFO) and symbol timing offset (STO), phase noise and in-phase and quadrature (IQ) imbalances. The above impairments deteriorate performance of OFDM transmission.

## 1.2 Thesis objectives

The OFDM is highly spectral efficiency techniques used for communication over frequency selective fading channel. The OFDM modulation is widely used in increasing data rates but suffer from several impairments. This thesis studies some impairment of OFDM systems. This thesis evaluates the BER performance for OFDM system in different cases. This thesis focuses on the frequency synchronization impairments due to CFO.

The frequency synchronization is the important to maintain the orthogonality between subcarriers. Loss in frequency synchronization is caused by a number of issues. CFO caused by Doppler shift due to relative motion between transmitter and receiver or the mismatch of local oscillators (LOs) between the transmitter and the receiver.

The CFO is usually divided in two parts, the fractional frequency offset (FFO); the integer frequency offset (IFO). FFO introduces inter-carrier interference (ICI) between sub-carriers. IFO does not introduce ICI between sub-carriers, but does introduce a cyclic shift of data subcarriers. Finally both deteriorate the system performance and increase the bit error rate (BER). Frequency synchronization process is performed via two stages: acquisition and tracking.

This thesis presents a survey on CFO compensation schemes. This thesis focuses on the fine recovery of FFO to restore the frequency synchronization and improves the system performance. This thesis proposes dual bandwidth scheme for residual CFO tracking.

## 1.3 Thesis organization

This thesis is organized into six chapters as follows:

Chapter 1 presents the thesis motivation, the thesis objectives and the thesis organization.

Chapter 2 presents wireless communication channels and the wireless channel parameters. The basic principles of OFDM technique are discussed. Also, the concept of the guard interval (GI) and cyclic prefix (CP) necessary to avoid ISI and ICI respectively in dispersive channels is explained. The advantages and disadvantages of OFDM are also presented.

Chapter 3 introduces the OFDM system model without impairments. The OFDM challenges are discussed. The effect of IQ imbalance on the performance of OFDM system is presented. The reasons of the large PAPR of the OFDM systems and its effect of on the performance of power amplifier (PA) are explained. The effects symbol timing offset (STO) on performance of OFDM technique are presented. Finally, the carrier frequency offset (CFO) for OFDM technique is introduced. The reasons of the CFO and the OFDM system model with CFO are introduced. The effects of the IFO and FFO on the performance of OFDM systems are presented. Also, the BER performance degradation due to the CFO is presented.

Chapter 4 explains the frequency synchronization stages and The CFO compensation schemes. The non-data aided (NDA) and data aided (DA) schemes are discussed. The idea, advantages and disadvantages of each scheme are introduced. The difference between CFO estimation using conventional pilot and CFO estimation using clustered pilot tones is introduced. The definitions and the main parameters of the PLL are presented. The dual bandwidth scheme of the PLL is proposed. The improved dual bandwidth scheme is investigated.

Chapter 5 presents the numerical results for OFDM system in the WLAN parameters for different cases. The BER performance of OFDM system for different modulation techniques in both AWGN and Raleigh multipath fading channel is achieved. The BER performance of OFDM system in AWGN with and without CP is investigated. The BER performance of OFDM system for different CP length, different channel delay spread and different FFT size for different modulation techniques in Raleigh multipath fading is investigated. The BER performance of OFDM system for the proposed dual bandwidth scheme and the improved dual bandwidth scheme in both AWGN and Raleigh multipath fading for different modulation techniques is implemented. Also, the constellation diagram and time response for the output phase and phase error for different modulation techniques in Raleigh multipath fading are investigated.

Chapter 6 concludes and summarizes the results obtained in this thesis and suggests some directions for future research.

# Chapter 2. OFDM Basics

## 2.1 Introduction

This chapter presents an overview for wireless communications system. It starts with a general presentation for communication through multipath radio channel. The basic principles of OFDM technique are introduced also. The orthogonality condition, the benefits of usage of the guard interval, the cyclic prefix and its drawbacks are discussed. Finally, advantages and challenges of OFDM transmission are discussed.

## 2.2 Wireless communication channels

In a wireless communication system, there are multipath channels between the transmitter and the receiver. Whereas the transmitter sends the modulated signal, there are multiple indirect paths (i.e. non-line-of-sight (NLOS)) between the transmitter and the receiver beside the direct path (i.e. line of sight (LOS)) [1] as shown in figure 2.1.

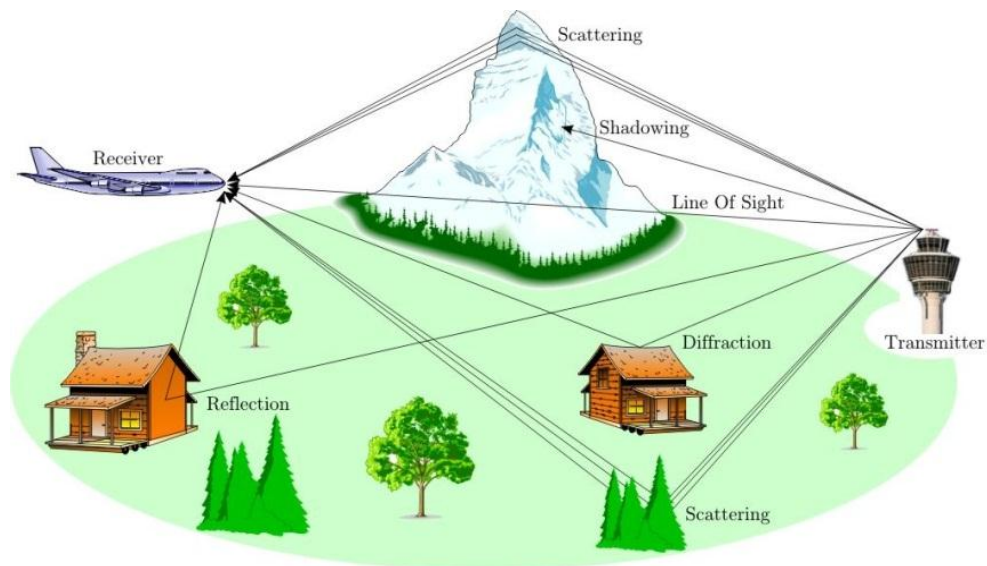

**Figure 2.1: Multipath signal propagation.**

These NLOS paths are due to three basic effects: reflection, refraction and diffraction. Thus, the transmitted signal through the wireless channel contains multiple replicas (echoes) of the transmitted signal. At the receiver, the multipath components experience different path loss, delay and angle of arrival, and interfere with each other constructively or destructively. Such an effect is known as small-scale fading or simply fading. The wireless channel is referred to as a multipath channel, and the multipath components of the wireless channel are characterized by their delay spread and Doppler spread.

## 2.2.1 Delay Spread

The multipath delay spread is used to characterize the time dispersion of a wireless channel, which is due to multiple replicas of the transmitted signal arriving at the receiver with different delays. In general, it is a measure of the time difference between the first significant multipath component (usually the LOS component) and the last multipath component. The delay spread is usually quantified by the mean excess delay ($\bar{\tau}$) and the root mean square (RMS) delay spread ($\tau_{RMS}$) [1]. Typical values of RMS delay spread are on the order of microseconds in outdoor mobile radio channels and on the order of nanoseconds in indoor radio channels.

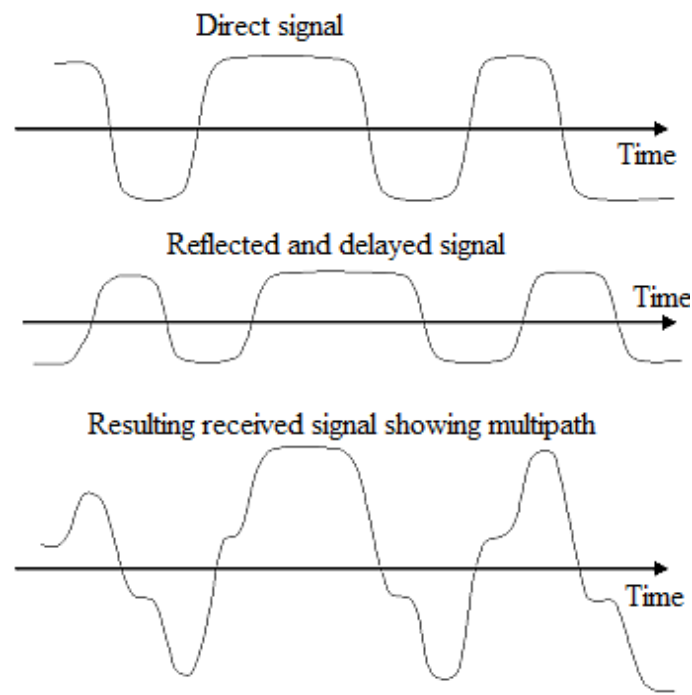

**Figure 2.2 : Multipath delay spread.**

The main effects of delay spread on the received signal are frequency selective fading and inter-symbol interference (ISI). Frequency selective fading is meaning that the channel does not affect all frequency components of the signal equally, which results in severely distortion of the received signal. On the other hand, ISI is the interference between consecutive symbols. Due to the reception of multiple copies of the signal with different delays, energy from one symbol can spread to the following symbol. If not dealt with, frequency selective fading and ISI can result in a significant degradation in system performance. Figure 2.2 shows the effect of ISI due to delay spread on the received signal.

The coherence bandwidth ($B_c$) is used to characterize the channel in frequency domain. The coherence bandwidth is a statistical measure of the range of frequencies over which the channel can be considered "flat" (i.e., all frequency components of transmitted signal undergo the same attenuation and phase shift in transmission through the channel) [1]. The coherence bandwidth and the RMS delay spread are approximately inversely proportional to each other.

The channel is said to be frequency selective fading if the bandwidth of the transmitted signal ($B_s$) is larger than coherence bandwidth i.e. $B_s > B_c$ as shown in figure 2.3 (a). In the case of frequency selective fading the transmitted signal is distorted by the channel and ISI is occurs. The channel is said to be frequency flat if the bandwidth of the transmitted signal ($B_s$) is smaller than coherence bandwidth i.e. $B_s < B_c$ as shown in figure 2.3 (b). In the case of frequency flat channel, the delay spread is small compared to the symbol duration and usually can be neglecting it [2].

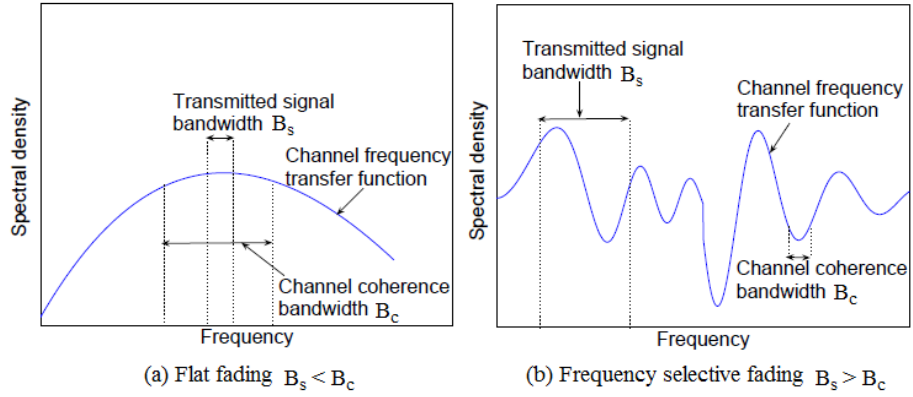

**Figure 2.3 : Flat fading versus frequency selective fading.**

The channel does not remain constant over time because of the changes in the multipath environment. The delay spread and coherence bandwidth do not give information on the time variations of the channel caused by either relative motion between the mobile and base station, or by movement of objects in the channel. So the Doppler spread and the coherence time can be studied.

### 2.2.2 Doppler Spread

The Doppler spread of the channel ($B_D$) is used as a measure of the spectral broadening due to time variations of the channel and is defined as the range of frequencies over which the Doppler spectrum is non-zero [1],[2]. The amount of spectral broadening depends on the Doppler frequency shift which is given by:

$$f_d = \frac{v \cdot f_c}{c} \cos\theta \qquad (2.1)$$

Where $f_d$ is the Doppler frequency shift, $c$ is the speed of light, $v$ is the relative velocity of the transmitter with respect to the receiver, $f_c$ is the center carrier frequency and $\theta$ is the angle between the direction of motion of the mobile and direction of arrival of the scattered waves.

In the time domain, the coherence time ($T_C$) is a statistical measure of the time duration over which the channel impulse response is invariant (static) [1], [2]. In other words, it is the time over which the channel is constant. Thus, it is defined as the time duration over which the power of two received signals has a strong correlation. The Doppler spread and coherence time are inversely proportional to one another [1], [2].

$$T_C \approx \frac{1}{B_D} \qquad (2.2)$$

The Doppler spread and coherence time are indicators of the time varying nature of the channel. It means that when two symbols that are passed through the channel with an interval greater than $T_C$, they will be affected differently by the channel.

The channel is said to be fast fading channel if the coherence time is smaller than the symbol duration ($T_u$) i.e. $T_c < T_u$. In the case of the fast fading channel the signal will distort significantly since the channel will change during the transmission of the symbol. In practice, fast fading only occurs for very low data rate transmission. The channel is said to be slow fading if the coherence time is larger than the symbol duration i.e. $T_c > T_u$. The slow fading channel means the channel impulse response changes much slower than the transmitted symbol [1].

The relation between the various multipath parameters and the type of fading experienced by the signal are summarized in figure 2.4 [3].

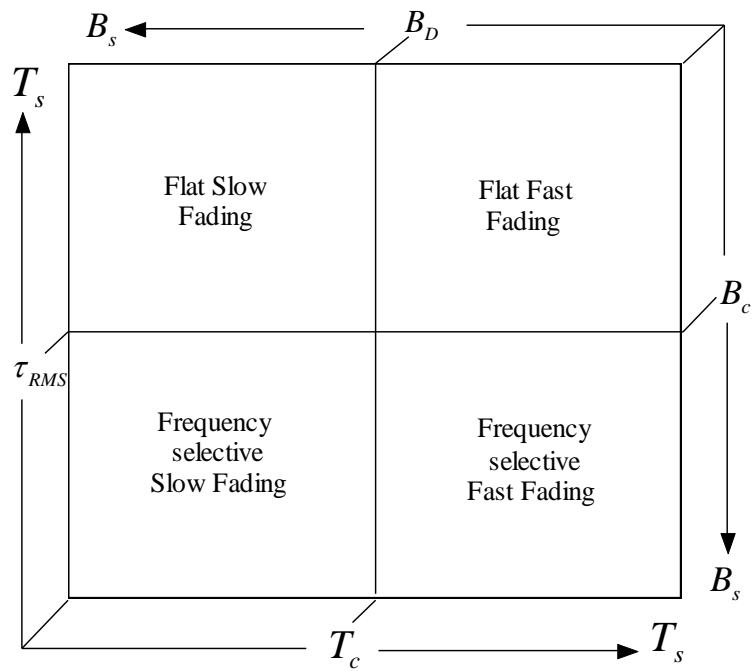

**Figure 2.4 : Fading illustration [3].**

## 2.2.3 Channel impulse response

The channel impulse response contains all information necessary to analyze and simulate any type of radio transmission through the channel. A wide sense stationary (WSS) model is assumed, such that the channel correlation is invariant over time [1]. A WSS channel has an impulse response given by

$$h(\tau) = \sum_{p=0}^{L_h-1} a_p \exp(-j\theta_p)\delta(\tau - \tau_p) \qquad (2.3)$$

Where $a_p, \tau_p$ and $\theta_p$ are the amplitude, the propagation delay and the phase shift of $p^{th}$ path respectively.

After the above discussions for wireless channel impairments; in dispersive channels the frequency selective fading channels are much more difficult to model than flat fading channels. When the frequency selective fading is occur the received signals are distorted and overlapped in time causing ISI and degrade system performance. There are several ways to minimize ISI. One is to reduce the symbol rate, but then the data rate is also reduced. Another technique is to utilize equalizers, equalizer is necessary for the optimum receiver to compensate or reduce the ISI in the received signal. In high data rate wireless communications, the transmitted symbol duration gets shorter and shorter. So, the implementation complexity of the equalization process becomes too high. It is therefore necessary to use alternative approaches that support high data rate transmission over multipath fading channels to combat ISI. In the following, alternative approach for transmitting over multipath fading channels using multi-carrier modulation (MCM) technique such as orthogonal frequency division multiplexing (OFDM) technique is discussed in the next section.

## 2.3  Basic principles of OFDM systems

The basic principle of OFDM is splitting a high data rate streams into a number of lower data rate streams and then transmitted these streams in parallel using several orthogonal subcarriers (parallel transmission). The available bandwidth $B_s$ is splitting into a large number of orthogonal subcarriers $N_u$. To achieve orthogonality between subcarriers; the subcarrier spacing $\Delta f$ is equal to the reciprocal of the useful symbol duration $B_s/N_u$. OFDM is considered as either a modulation technique or a multiplexing technique [4].

The choosing number of subcarriers $N_u$ depends on the coherence bandwidth and the coherence time. The subchannels bandwidth can be made small as compared to the coherence bandwidth $B_c$ of the channel i.e. $B_s/N_u \ll B_c$. To maintain the subchannels experience frequency flat fading which reduces equalization to a single complex multiplication per subcarrier. But the increasing of $N_u$ reduces the ISI but the useful symbol duration much long which degrades the system performance. If the useful OFDM symbol duration $T_u$ longer than the coherence time $T_c$ i.e. $T_u > T_c$, the channel frequency response changes significantly during the transmission of one symbol and a reliable detection of the transmitted information becomes impossible.

So, the coherence time of the channel determines the maximum number of the subcarriers during the OFDM symbol. Thus a reasonable range for $N_u$ can be derived as [5]:

$$B_s/B_c \ll N_u \ll \Re T_c \qquad (2.2)$$

### 2.3.1 SCM versus MCM

Single carrier modulation (SCM) uses a single carrier frequency to transmit all data symbols sequentially. Compared to multicarrier modulation, SC modulation has several advantages. It has low peak-to-average power ratio and it is less sensitive to frequency offsets and phase noise. The main disadvantage of SCM is that it is susceptible to multipath fading or interference because it uses only one carrier frequency [4] as shown in figure 2.5.

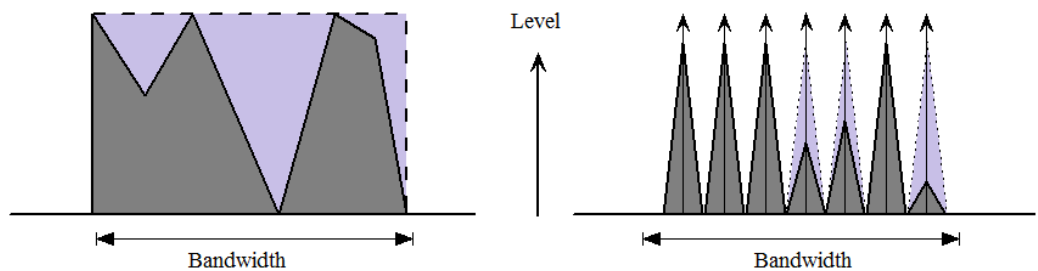

**Figure 2.5: SCM versus MCM.**

### 2.3.2 OFDM versus FDM

The OFDM technique is high spectral efficiency than frequency division multiplexing (FDM) technique. In FDM a guard band used to avoid inter-carrier interference (ICI), or crosstalk, from adjacent subcarriers but in OFDM the adjacent subcarriers overlapped without ICI between them due to the principle of orthogonality [4]. As shown in figure 2.6, OFDM saves almost 50 percent of the available bandwidth compared to FDM.

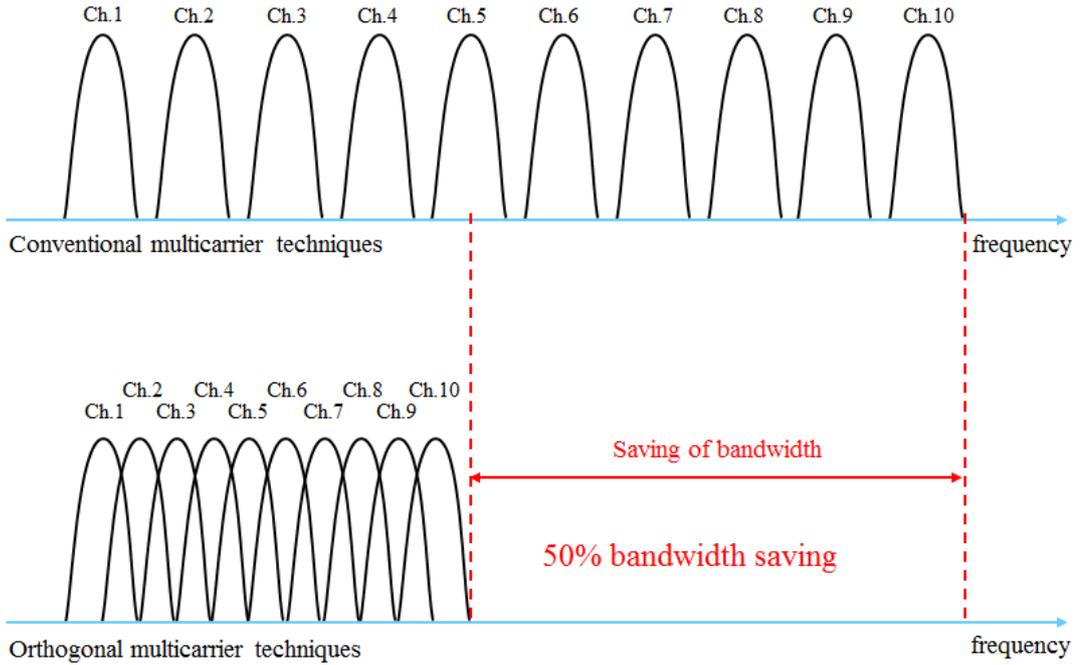

**Figure 2.6 : Comparison between FDM and OFDM [4].**

### 2.3.3 Orthogonality

In OFDM systems, the subcarriers can overlap without causing ICI due to the orthogonality between all subcarriers. In mathematics, two vectors perpendicular to each other are orthogonal and their dot product is equal to zero. In communications, orthogonality means two signals are uncorrelated, or independent, over one symbol duration. The set of subcarriers $\{\Phi_k(t)\}_{k=0}^{N_u-1}$ are mathematically orthogonal to each other under the following condition [3-5]:

$$\frac{1}{T_u}\int_0^{T_u} \Phi_k(t)\Phi_l^*(t)dt = \delta(k-l) \qquad (2.3)$$

Where $\delta(k-l)$ is the delta function defined as $\delta(k-l) = \begin{cases} 1 & k=l \\ 0 & k \neq l \end{cases}$. In OFDM in order to maintain all subcarriers to be orthogonal to each other, the peak of each subcarrier achieved when the other subcarriers are nulls, the subcarrier frequencies $f_k$ must be integer multiples of $1/T_u$ i.e. $f_k = k/T_u$ and minimum frequency separation between subcarriers must be $\Delta f = 1/T_u$ [6].

Figure 2.7 describes the waveform of three orthogonal subcarriers which together make up an OFDM signal in the time domain. The waveforms of all subcarriers have the same amplitude and same initial phase. However, in practice, the subcarriers are modulated in different amplitude and phase. The orthogonality within an OFDM symbol duration in the time domain means that all subcarriers have integer number of cycles and the number of cycles per symbol duration of adjacent subcarriers differs by exactly one [4].

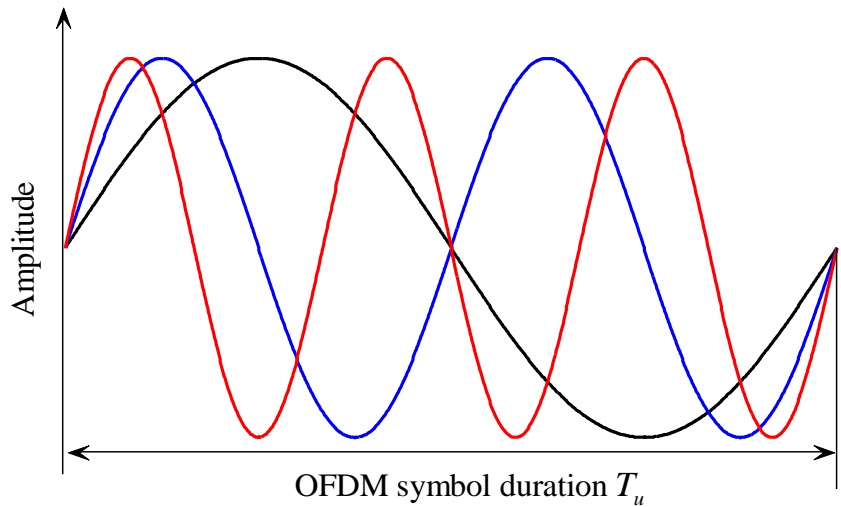

**Figure 2.7 : Orthogonal subcarriers in time domain [4].**

In the frequency domain each OFDM subcarrier has a $\text{sinc}(x) = \sin(x)/x$ frequency response as shown in figure 2.8. It shows the spectrum of an individual data subcarrier [4]. Figure 2.9 depicts the spectrum of an OFDM symbol consists of three subcarriers. It is observable from the plots that at the maximum of each subcarrier spectrum; all other subcarriers spectra are zero. It is also clearly from the plots that each subcarrier has maximum amplitude and nulls at the center frequency which are spaced equally by frequency difference equal to the carrier spacing. This verifies the orthogonal nature of the OFDM signal [4].

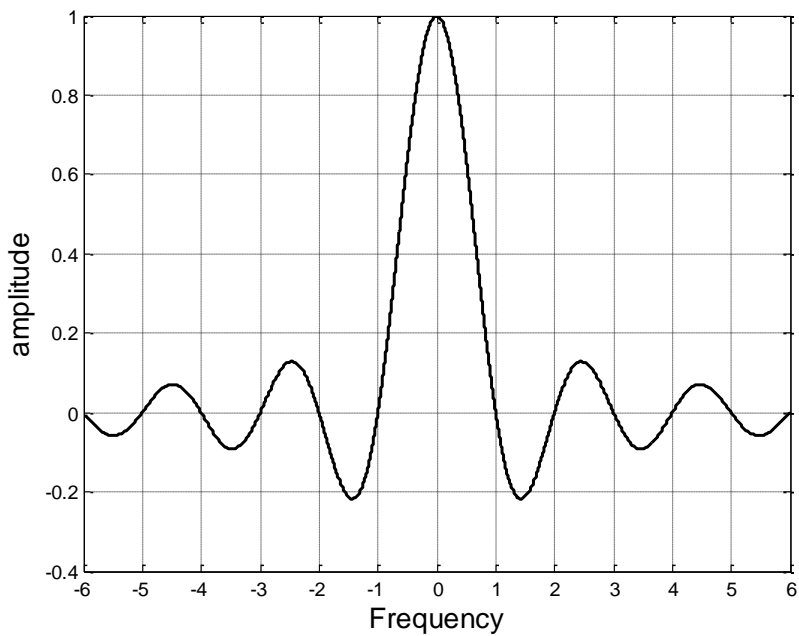

**Figure 2.8 : Spectrum of OFDM individual subcarrier.**

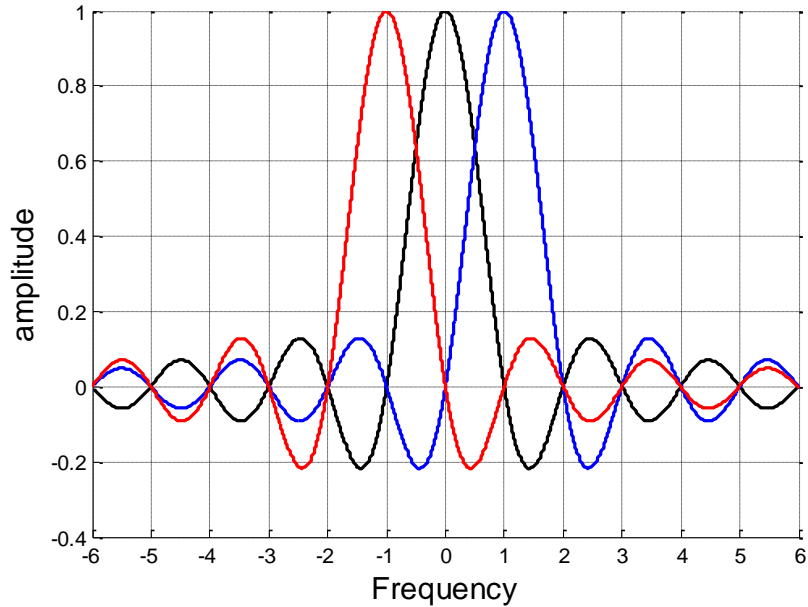

**Figure 2.9 : Spectrum of an OFDM symbol.**

## 2.3.4 Guard interval

One of the most important reasons for choosing OFDM transmission is its efficiency to deal with the delay spread of the multipath channel. By splitting a high rate data stream into a number of lower rate streams and transmitted simultaneously over a number of subcarriers. Since the symbol duration increases for lower rate parallel subcarriers, the relative amount of time dispersion caused by multipath delay spread is decreased. To eliminate ISI completely, a guard interval (GI) is introduced for each OFDM symbol. The GI is chosen larger than the channel delay spread, such that multipath components from one symbol cannot interfere with the next symbol. The GI does not contain any signal, in that case, the problem of inter-carrier interference (ICI) will appears which means that the subcarriers are no longer orthogonal [4].

As shown in figure 2.10, the OFDM symbol duration has three subcarriers and the guard interval is presented. If the receiver of the OFDM system tries to demodulate subcarrier 1, the interference from the other subcarriers is taken into account. Because of there is no integer number of cycles for each subcarrier over the OFDM symbol duration, thus there will be ICI, among subcarriers in the frequency domain. This problem can be removed via inserted the cyclic prefix (CP) in the GI this is explained in more details in the next section.

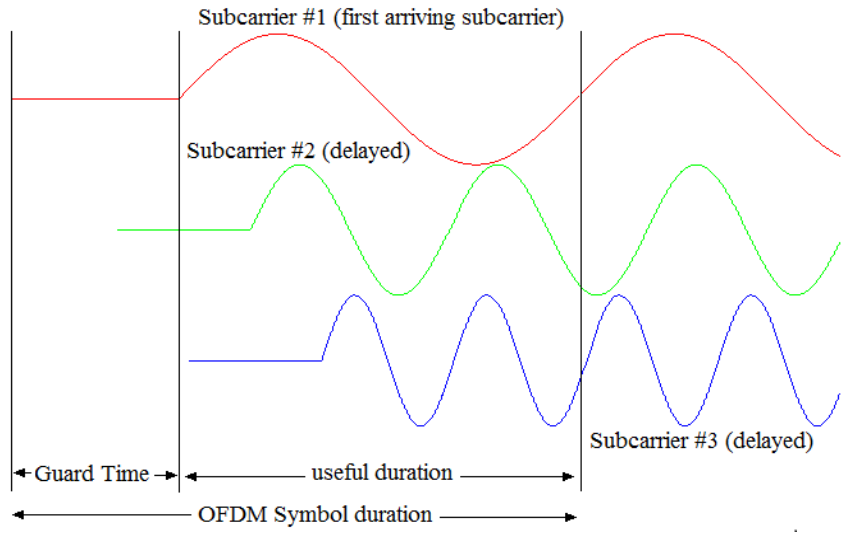

**Figure 2.10 Effect of multipath without cyclic extension in the GI.**

### 2.3.5 Cyclic prefix

The ICI is known by crosstalk between different subcarriers, which means they are no longer orthogonal [4]. To eliminate the ICI, the CP extends the OFDM signal into the GI. Generally, this can be done by copying the last $N_{cp}$ samples of the IDFT output and putting them at the beginning of the OFDM signal as shown in figure 2.11 and 2.12. Now, after inserting the CP, as long as the CP length is greater than or equal to channel delay spread, the delayed copies of the OFDM symbol always have the integer numbers of cycles within the OFDM useful duration. Thus, the ICI and ISI caused by multipath frequency selective fading are totally eliminated.

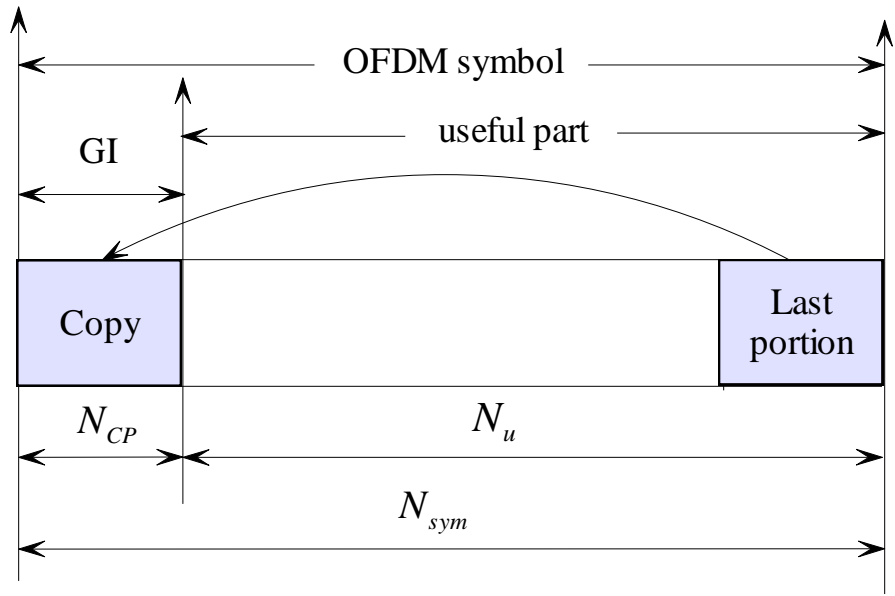

**Figure 2.11 : Cyclic prefix insertion.**

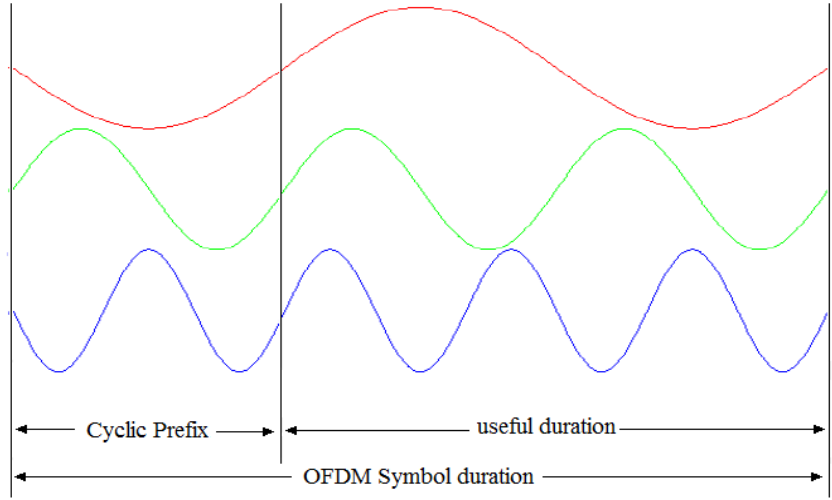

**Figure 2.12: Effect of adding CP in the GI.**

Choosing the GI is also important since a long time comparing to the symbol duration can reduce the power efficiency. That is because there is no useful information being sent during the GI. Normally, the GI is chosen to be two to four times of root mean square of the delay spread [4]. The transmitted energy required to transmit the signal increases with the length of cyclic prefix. The cost ($SNR_{loss}$) due to the insertion of CP is given by [5].

$$SNR_{loss} = -10 \log_{10}\left(1 - \frac{T_g}{T_{sym}}\right) \qquad (2.6)$$

Where $T_g$ is the CP length, $T_u$ is useful symbol duration of OFDM and $T_{sym}$ is the complete OFDM symbol duration and is equal to $T_{sym} = T_u + T_g$.

## 2.4 Advantages and disadvantages of OFDM

Compared to a single carrier wireless system, OFDM provides several advantages [3], [4]:

**Robustness to narrowband interference:** The duration of an OFDM symbol is much longer than that from an equivalent single carrier system, and narrowband interference will only affect a small fraction of the OFDM symbol. Channel coding and forward error correction (FEC) codes can be employed to recover the errors caused by narrowband interference. Thus OFDM is robust against narrowband interference.

**Resistance to frequency selective fading:** In an OFDM system, the frequency selective channel is divided into a number of frequency flat fading subchannels. The OFDM symbol duration is increased by converting the high rate data stream into several lower rate parallel data streams, which in turn reduces the relative channel delay spread.

**Simple equalization:** In an OFDM system, the channel bandwidth is divided into many narrow subchannels. The subchannel bandwidth is smaller than the channel coherence bandwidth, so the frequency response over individual subchannels is relatively flat. Thus, it is possible to have a simpler equalizer than that of an equivalent single carrier system. Furthermore, if the channel is time-invariant within OFDM symbol duration, one tap equalizer can be employed at the receiver, which is much simpler than the adaptive equalizer required in a single carrier system.

**Immunity to delay spread and multipath:** Generally, a CP is appended at the front of the OFDM symbol at the transmitter. It is typically a copy of the end of the OFDM symbol. The length of the cyclic prefix should be equal to or longer than that of the channel impulse response. Then the OFDM channel is converted from a linear circular channel into a cyclic circular channel so that the ISI can be eliminated.

**Efficient bandwidth usage:** In an OFDM system, the subcarriers are overlapped and no guard band is required, thus the spectrum efficiency can be close to the Nyquist limit.

**Computational efficiency:** In an OFDM system, an IFFT and FFT are implemented at the transmitter and receiver for modulation and demodulation, respectively, which significantly reduces the computational complexity of the system.

Although OFDM has been implemented in various applications, there are also some major drawbacks in OFDM systems [4]:

**High peak-to-average power ratio:** In an OFDM system, the transmitted symbol is the sum of the signals for all the subcarriers, which results in a high peak to- average power ratio (PAPR). In this case, the RF power amplifiers must operate over a wider linear region. Otherwise, the maximum power of the signals enters the non-linear region of the high power amplifier (HPA), which results in signal distortion, and induces intermodulation among the subcarriers and out of band radiation. However, a wider dynamic range linear power amplifier implies large power back-offs, which leads to inefficient amplification and expensive transmitter designs. Thus, it is desirable to reduce the PAPR in OFDM systems [7, 8].

**Symbol timing offset:** OFDM is highly sensitive to symbol timing offset (STO). Due to the use of IFFT and FFT for modulation and demodulation at transmitter and receiver respectively, correct timing (start of FFT window) is required at the receiver otherwise one FFT window will take sample from two transmitted OFDM symbol. This deteriorates the performance of OFDM system [9].

**Carrier frequency offset:** OFDM systems is also more sensitive to carrier frequency offset (CFO) between the oscillators of the transmitter and the receiver. The CFO leads to attenuation and phase rotation of the subcarriers, and ICI between subcarriers. The ICI decreases the SNR and degrades the BER performance of OFDM system [10, 11]. Thus, CFO estimation is needed in OFDM systems. A number of methods have been developed to reduce the sensitivity to frequency offset [4].

**Sensitivity to Doppler spread:** OFDM is sensitive to Doppler spread caused by user mobility [3], which results in loss of orthogonality among subcarriers. This in turn leads to inter-carrier interference and degrades system performance. While it is straightforward to estimate and reduce the ICI induced by phase noise, the ICI introduced by Doppler spread is a more challenging problem.

## 2.5 Applications of OFDM

OFDM is a promising MCM technique for high speed communication. It has been widely used in a number of communication systems such as: Wireless local area network (WLAN) such as: IEEE802.11a/g and HIPERLAN/2, Worldwide interoperability for microwave access (WiMAX); fixed WiMAX (IEEE802.16-2004) and mobile WiMAX (IEEE802.16e-2005). Its baseband version discrete multi-tone (DMT) has become the standard modulation technique for asymmetric digital subscriber line (ADSL; up to 6 Mbps) and very high speed digital subscriber line (VDSL; up to 100 Mbps), digital audio broadcasting (DAB) and digital video broadcasting (DVB) and Long-term evolution (LTE) [4, 12-13].

Finally, the impairments of OFDM will be discussed in detail in the next chapter. The impact of these impairments on the OFDM system performance will be analyzed.

# Chapter 3. OFDM Challenges

## 3.1 Introduction

The OFDM system model without impairments is derived first. Then OFDM impairments such as IQ imbalance, PAPR, symbol timing offset (STO) are introduced. The carrier frequency offset (CFO) are analyzed and discussed. The effects of CFO on the performance of OFDM system are explained.

## 3.2 Ideal OFDM system model

At the transmitter, the input bits streams are encoded by using modulation techniques such as M-PSK or M-QAM to complex data symbols. Let a block of complex data symbols $Z_m$ with $m = 0,1, \dots, N_u - 1$. The complex data symbols $Z_m$ are first in serial with duration $T_s$ which converted to parallel via serial to parallel (S/P) conversion and then modulated to $N_u$ subcarriers. Due to the S/P conversion, the duration of transmission time for $N_u$ symbols is extended to $N_u T_s$ which forms a single OFDM symbol with a length of $T_u$ (i.e. $T_u = N_u T_s$) where $T_s$ is the sampling time [14, 15].

To achieve the orthogonality between the $N_u$ sub-carriers, the sub-carriers spacing must be equal to $\Delta f = 1/T_u$. This is achieved by maintaining the subcarrier frequencies are integer multiples of $1/T_u$ i.e. $f_m = m/T_u$ , $m = 0,1, \dots, N_u - 1$. The baseband OFDM signals in the continuous-time domain can be expressed [14, 15] as:

$$z_l(t) = \frac{1}{N_u} \sum_{m=0}^{N_u-1} Z_{m,l} \, e^{j2\pi f_m(t-lT_u)/N_u}, \qquad lT_u < t < lT_u + nT_s \quad (3.1)$$

The main advantage of using OFDM is its modulator and demodulator is based on the IDFT and the DFT in the discrete domain respectively. The IDFT and the DFT replaced by IFFT and FFT for more computationally efficient [13]. The discrete-time OFDM symbol is obtained by sampled equation (3.1) at $t = lT_u + nT_s$ with $T_s = T_u/N_u$ and $f_m = m/T_u$ . Thus, the discrete-time OFDM symbol can be written [14, 15] as follows:

$$z_{n,l} = \frac{1}{N_u} \sum_{m=0}^{N_u-1} Z_{m,l} \, e^{j2\pi nm/N_u}, \qquad \begin{array}{l} n = 0,1 \dots \dots N_u - 1 \\ l = 0,1,2, \dots \dots \dots \dots \infty \end{array} \quad (3.2)$$

Where $z_{n,l}$ is defined as the $l^{th}$ transmitted OFDM symbol at $n^{th}$ samples. To avoid the effects of ISI and also to maintain the orthogonality between the on the sub-carriers, the GI between adjacent OFDM symbols must be $T_g \geq \tau_{max}$ [15]. The GI is a cyclic extension of each OFDM symbol which extends the duration of an OFDM symbol to $T_{sym} = T_u + T_g$. Let the CP length be $N_{CP} = GN_u$, where $G$ is the fractional overhead and $0 < G < 1$. The length of complete OFDM symbol is $N_{sym} = (1 + G) N_u$ [12]. The $l^{th}$ discrete transmitted OFDM symbol with CP is expressed as follows [16, 17]:

$$s_{n,l} = \begin{cases} Z_{n+N_u-N_{CP},l}, & n = 0, 1, \ldots\ldots, N_{CP} - 1 \\ Z_{n-N_{CP},l}, & n = N_{CP}, \ldots\ldots, N_u + N_{CP} - 1 \end{cases} \quad (3.3)$$

The block diagram of an OFDM modulator based on an IFFT and an OFDM demodulator based on a FFT is illustrated in figure 3.1.

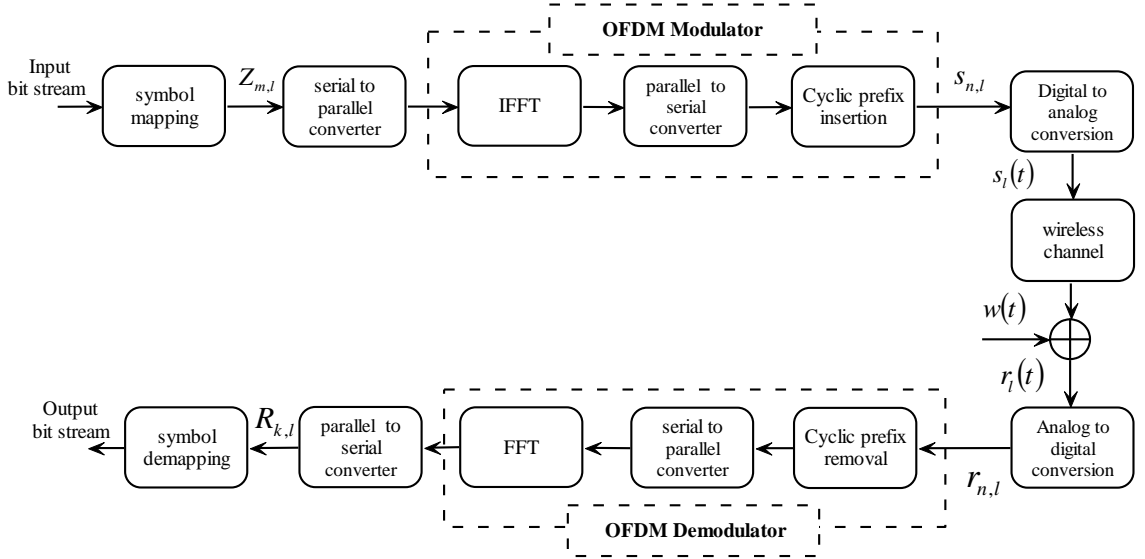

**Figure 3.1 : OFDM block diagram [15].**

The digital-to-analog converter (DAC) used to convert the discrete time $s_{n,l}$ into the analog version signal of $s_l(t)$. Then, the analog signal $s_l(t)$ is up-converted and passed to the wireless channel whose channel impulse response (CIR) $h(\tau)$. After the down conversion the received signal waveform $r_l(t)$ can be written as follows:

$$r_l(t) = s_l(t) \otimes h(\tau) + w_l(t) \quad (3.4)$$

Where $\otimes$, $r_l(t)$, $s_l(t)$, $h(\tau)$ and $w_l(t)$ denote the convolution operation, the received signal, the transmitted signal, the CIR and AWGN with zero mean and variance $\sigma_n^2$ respectively. It is assumed that the impulse response of the channel does not change during the symbol plus guard interval (i.e. the channel is slow fading behavior). The Analog-to-digital converter (ADC) converts the received signal $r_l(t)$ into the received sequence $r_{n,l}$ with $n = -N_{CP}, \ldots\ldots, N_u - 1$. Note that $r_{n,l}$ is the sampled version of $r_l(t)$ sampled at $t = nT_s$. After cyclic prefix removal the remaining sequence with index $n = 0,1, \ldots\ldots, N_u - 1$. The $l^{th}$ received OFDM symbol at the $n^{th}$ samples is written as follows:

$$r_{n,l} = \frac{1}{N_u} \sum_{m=0}^{N_u-1} H_{m,l} Z_{m,l} e^{-j2\pi mn/N_u} + w_{n,l}, \quad n = 0, \ldots\ldots, N_u - 1 \quad (3.5)$$

The using of CP leads to the linear convolution transformed into circular convolution [12]. If the CP length is greater than or equal to channel delay spread $(T_g \geq \tau_{max})$, ISI will be completely removed. This means that each OFDM symbol is free from interference from adjacent OFDM symbols and every subcarrier is also free from interference from adjacent subcarriers. Thus, the received OFDM symbols can be expressed in the frequency domain as [14]:

$$R_{k,l} = H_{k,l}Z_{k,l} + W_{k,l}, \qquad k = 0,1,\ldots\ldots,N_u - 1 \qquad (3.6)$$

Where $Z_{k,l}$, $H_{k,l}$, $W_{k,l}$ and $R_{k,l}$ denote the $k^{th}$ subcarrier of the $l^{th}$ transmitted symbol, channel frequency response, AWGN and received symbol in the frequency domain respectively. The received OFDM symbols in equation (3.6) can be represented as in figure 3.2.

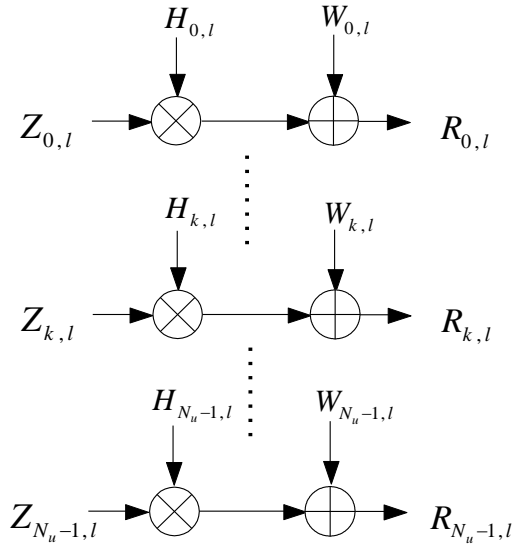

**Figure 3.2 : Frequency domain equivalent model of OFDM system [14].**

## 3.3 I/Q imbalance

Direct conversion, also is known by zero-intermediate frequency (IF) or homodyne, receiver structure. It translates the RF signal (i.e. passband) directly from the carrier frequency $f_c$ to baseband using one stage of mixing. The direct conversion receiver structure has several advantages: no IF stage, no image-rejection filter and easy integration due to low component count. However, a direction conversion RF front-end suffers from IQ imbalance [13]. The IQ imbalance refers to phase and gain imbalance between In-phase and quadrature-phase IQ paths; or equivalently, the real and imaginary parts of the complex signal. The IQ imbalance may be introduced at the transmitter only during frequency up-conversion or the receiver only during frequency down-conversion or both. Both the up-conversion and down-conversion are implemented in the analog domain by what is known as complex up-conversion and complex down-conversion [18]. To perform the complex frequency conversion; both the sine and cosine oscillating waveforms are required. The IQ imbalance is basically any mismatch between the I and Q branches from the ideal case (i.e. from the exact 90° phase difference and equal amplitudes) [18]. The gain and phase imbalance parameters at the direct conversion receiver denoted by $\mu$ and $\theta_r$ as shown in figure 3.3[19].

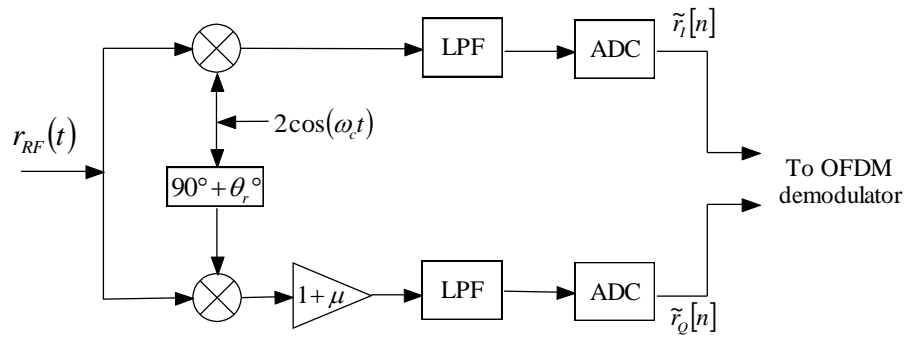

**Figure 3.3 : Direct conversion receiver with IQ imbalance [19].**

The IQ imbalance has two types: frequency independent IQ imbalance (frequency flat) and frequency dependent IQ imbalance (frequency selective) [21]. Figure 3.4 depicts the distortion due to the IQ imbalance for the QPSK baseband signals [13].

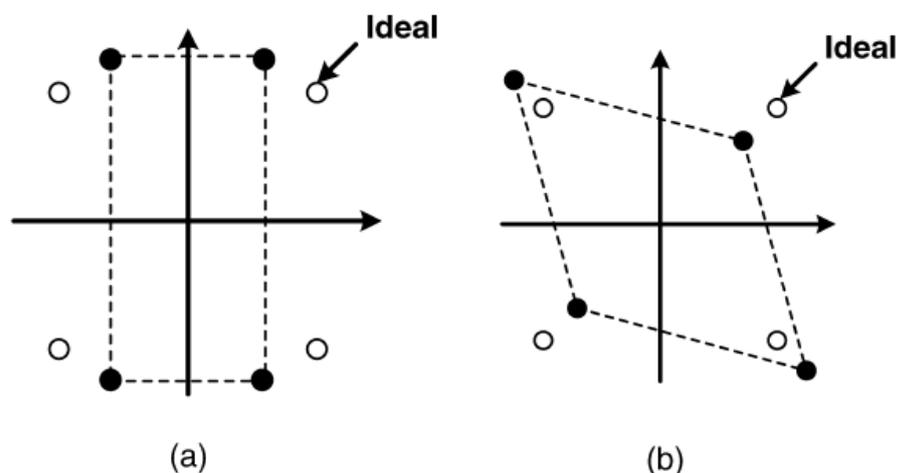

**Figure 3.4: IQ imbalance effect on QPSK signals: (a) gain error in LO signals and (b) phase error in LO signals [13].**

The performance degradation of the OFDM system caused by IQ imbalance has been investigated in [13]. The IQ signal is related to $\tilde{r}_I(t)$ and $\tilde{r}_Q(t)$ is expressed as follows [19]:

$$\hat{r}_{n,I} = r_{n,I} \tag{3.7}$$

$$\tilde{r}_{n,Q} = \frac{j}{2}\{(1+\mu)r_n^* e^{j\theta_r} - (1+\mu)r_n e^{-j\theta_r}\} \tag{3.8}$$

The complex envelope baseband of the received signal after ADC is given by:

$$\tilde{r}_n = \tilde{r}_{n,I} + j\tilde{r}_{n,Q} \tag{3.9}$$

The sample at the $k^{th}$ subcarrier after taken the FFT of the complex envelope baseband of the received signal $\tilde{r}_n$ in OFDM receiver is expressed as follows [19]:

$$\tilde{X}_k = \xi \tilde{X}_k + \sigma \tilde{X}^*_{-k} \tag{3.10}$$

Where

$$\xi = \frac{1}{2}\{1 + (1 + \mu)(\cos(\theta_r) - j\sin(\theta_r))\} \tag{3.11}$$

$$\sigma = \frac{1}{2}\{1 - (1 + \mu)(\cos(\theta_r) + j\sin(\theta_r))\} \tag{3.12}$$

Where $\xi$ and $\sigma$ are the signal gain (SG) and the mirror gain (MG) respectively [22]. The relation (3.10) shows that the gain and phase matches in the receiver mixer cause the received symbol $\tilde{X}_k$ at the subcarrier $(k)$ to be multiplied by the complex factor $\xi$ plus the complex conjugate of the received symbol $\tilde{X}^*_{-k}$ at $(-k)$ subcarrier which multiplied by another complex factor $\sigma$ [19]. The component $\tilde{X}^*_{-k}$ is often referred to as the mirror signal, since the subcarrier is located at the same distance from the DC-carrier [23].

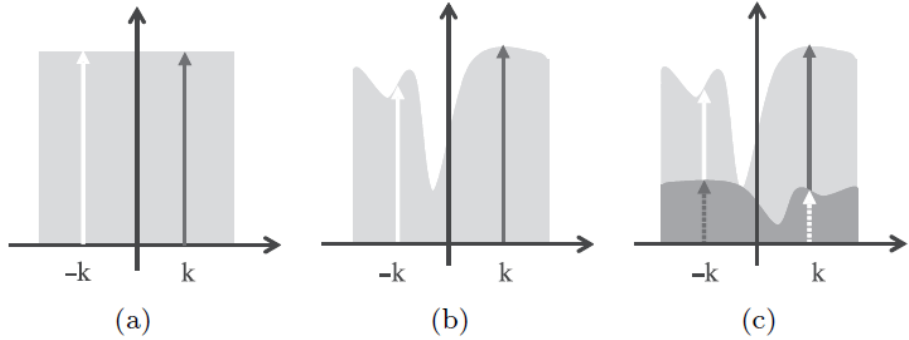

**Figure 3.5 : The effect of IQ imbalance on the reception of OFDM signal [23].**

The influence of IQ imbalance in a system experiencing a frequency selective channel is shown in figure 3.5, assume ideal up-conversion. The transmitted baseband signal is shown in figure 3.5 (a), where two subcarriers $(-k$ and $k)$. These subcarriers have the same separation from DC. The signal is up-converted to RF and transmitted through the frequency-selective channel, resulting in the received RF signal depicted in figure 3.5 (b). It is clear that subcarrier $(-k)$ is more attenuated by the channel than subcarrier $(k)$. Thus, the received signal is down-converted to baseband using the imperfect the direct conversion receiver with IQ mismatch. The mirror signal is not fully rejected, and mixes down into the regarded baseband channel. As shown in figure 3.5 (c), the subcarrier (k) experiences a contribution of the signal received on the mirror carrier $(-k)$ and vice versa [23]. Due to IQ imbalance power leaks from the signal on the mirror carrier to the carrier under consideration and thus causes ICI. As OFDM is very sensitive to ICI, IQ imbalance results in severe performance degradation [24].

## 3.4 Peak-to-average power ratio (PAPR)

When transmitted through a nonlinear device, such as a high-power amplifier (HPA) or DAC a high peak signal, generates out of band energy (spectral regrowth) and in-band distortion (constellation tilting and scattering). These degradations may affect the system performance severely [12]. Figure 3.6 shows a typical AM/AM response for an HPA, with the associated input back-off (IBO) and output back-off (OBO) [12]. To avoid such undesirable nonlinear effects, a waveform with high peak power must be transmitted in the linear region of the HPA by decreasing the average power of the input signal. The High back-off reduces the power efficiency of the HPA. The input back-off is defined as follows [12]:

$$IBO = 10 \log_{10} \left( \frac{P_{in_{sat}}}{\bar{P}_{in}} \right) \quad (3.13)$$

Where $P_{in_{sat}}$ is the saturation power above which is the nonlinear region and $\bar{P}_{in}$ is the average input power. The power efficiency of an HPA can be increased by reducing the PAR of the transmitted signal.

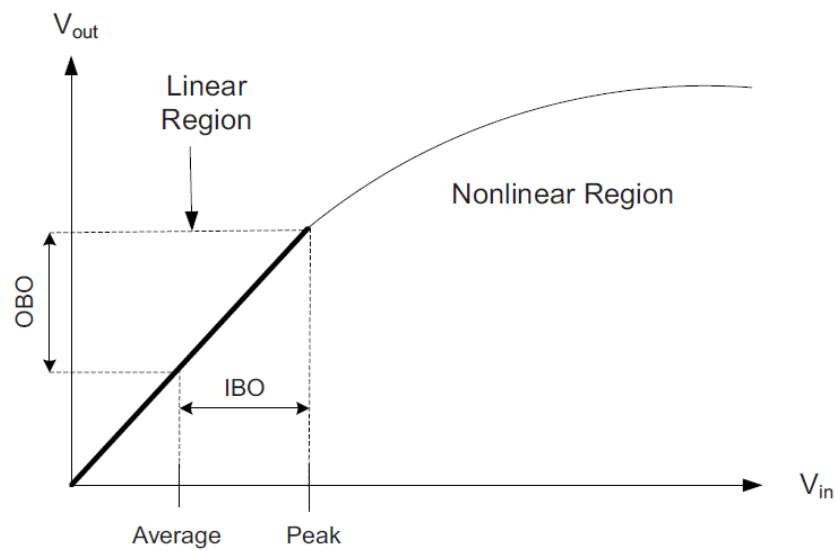

**Figure 3.6 : Input / Output curve of PA [12].**

One of the major drawbacks of OFDM system is high PAPR of transmitted signal. The large PAPR in OFDM is observed due to large dynamic range of its symbol waveforms. An OFDM signal consists of a number of independently modulated subcarriers, which can give a large PAPR, when added up coherently. When $N_u$ signals are added with the same phase, they produce a peak power that is $N_u$ times the average power. Most RF communication systems have a high power amplifier (HPA) in the transmitter to obtain sufficient transmit power. This high PAPR forces the HPA to have a large back-off in order to get linear amplification of the signal, which significantly reduces the efficiency of the amplifier. On the other hand, if an amplifier work with nonlinear characteristics, it will cause undesired distortion of OFDM signal and degrades the BER performance of the system [7]. For a continuous time baseband OFDM signal, the PAPR is defined as the ratio between the maximum instantaneous power and its average power. If $z(t)$ is a transmitted baseband OFDM signal, then PAPR is defined as [7]:

$$PAPR[z(t)] = \frac{\max_{0 \leq t \leq T_u} [|z(t)|^2]}{\bar{P}_{in}} \tag{3.14}$$

Where $\bar{P}_{in}$ is defined as the average power of $z(t)$.

## 3.5 Symbol timing offset (STO)

In OFDM systems, to avoid the ISI, the DFT window should include sample only from one OFDM symbol as shown in figure 3.7.

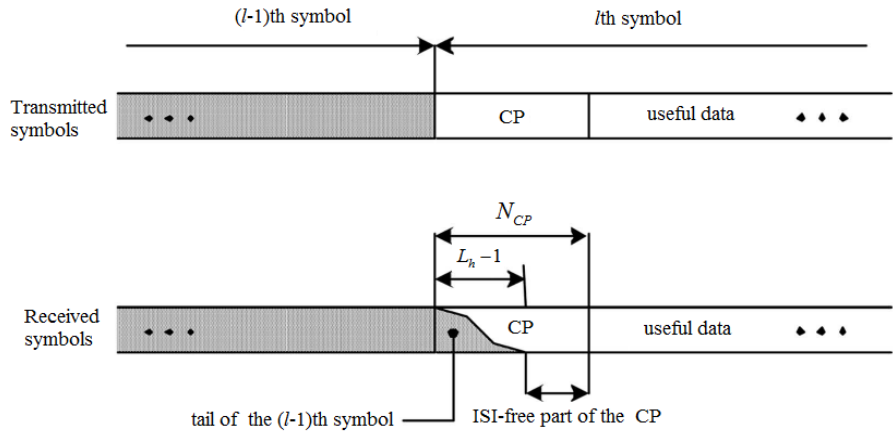

**Figure 3.7 : Partial overlapping between received symbols due to multipath dispersion [9].**

The tail of each received symbol extends over the first $L_h - 1$ samples of the successive symbol due to multipath dispersion. If the length of the CP must be greater than the CIR duration thus a certain range of the GI is not affected by the previous symbol at the receiver [9]. As long as the timing error satisfies the condition of $L_h - N_{cp} - 1 \leq \delta \leq 0$ where $L_h$ is the length of channel impulse response, thus there is no ISI at the DFT output. The output of the DFT of the $k^{th}$ subcarrier of the $l^{th}$ symbol is given by [9]:

$$\tilde{Y}_{k,l} = e^{j2\pi k\delta/N} H_k Z_{k,l} + W_{k,l} \tag{3.15}$$

The above equation indicates that the timing error $\delta$ appears as a linear phase across subcarriers. It can be compensated by the channel equalizer, which cannot distinguish between phase shifts introduced by the channel and those caused by timing offset [9].

If the value of timing offset out of the interval $L_h - N_{cp} - 1 \leq \delta \leq 0$ leads to ISI between adjacent OFDM symbols at the DFT output. And also the orthogonality between adjacent subcarriers will destroy thus ICI will occur. The output of the DFT of the $k^{th}$ subcarrier of the $l^{th}$ symbol is given by [9]:

$$\tilde{Y}_{k,l} = e^{j2\pi k\delta/N}\alpha(\delta)H_k Z_{k,l} + \beta_{k,l} + W_{k,l} \qquad (3.16)$$

Where $\alpha(\delta)$ is an attenuation factor while $\beta_{k,l}$ accounts for ISI and ICI [9].

## 3.6 Carrier frequency offset (CFO)

The CFO is the mismatch between the received carrier frequency and the carrier frequency generated by the local oscillator at the receiver. The OFDM systems are more sensitive to CFO than single carrier frequency systems. The CFO phenomenon is introduced from two sources: first, the mismatch between the local oscillator at transmitter and receiver. The other source, the Doppler shifts caused by the relative motion between the transmitter and receiver [25].

The accuracy of an oscillator is measured in terms of parts per million (ppm). In IEEE 802.11a WLAN, the oscillator precision tolerance is specified to be less than ±20 ppm, so that the CFO is in the range from -40 ppm to 40 ppm. If the transmitter oscillator generates frequency with offset +20 ppm and the receiver oscillator also is generating with offset -20, thus the total CFO in the received baseband signal is 40 ppm. In the IEEE 802.11a standard the carrier frequency is 5.2 GHz thus, the CFO is up to ±208 kHz [13]. This CFO is large compared to the subcarrier spacing of 312.5 kHz (i.e. the offset in each subcarrier larger than one half the subcarrier spacing) which leads to more deterioration the system performance. The frequency offset due to the relative mobility between the transmitter and the receiver in the range of some hundreds of Hz. For example, if the carrier frequency is 5.2 GHz and a velocity of 100 km/h, the offset value is 481.48 Hz, which is very small compared to the subcarrier spacing [13].

### 3.6.1 OFDM system model with CFO

Let us define the normalized CFO $\varepsilon$ as a ratio of the CFO $\delta f$ to subcarrier spacing $\Delta f$ is given by:

$$\varepsilon = \frac{\delta f}{\Delta f} \qquad (3.17)$$

The CFO can be modeled as a complex multiplicative distortion of received signal in the time domain. Thus, the received signal in time domain, equation (3.5), can be modified as follows [14]:

$$y_{n,l} = 1/N_u \sum_{m=0}^{N_u-1} H_{m,l} Z_{m,l} e^{j2\pi(m+\varepsilon)n/N_u} + w_{n,l} \qquad (3.18)$$

The effect of CFO only on the received signal in time domain without the effect of channel and noise can be written as follows [14]:

$$y_{n,l} = z_{n,l} e^{j2\pi n\varepsilon/N_u} \qquad (3.19)$$

Equation (3.19) indicates that the CFO causes a phase rotation by $2\pi n\varepsilon$ to the signal $z_{n,l}$. Note that the phase rotation is proportional to the normalized CFO $\varepsilon$ and the time index $n$ [14]. The effect of CFO only on the received signal in frequency domain without the effect of channel and noise also can be expressed as follows [14]:

$$Y_{k,l} = Z_{k-\varepsilon,l} \qquad (3.20)$$

Equation (3.20) indicates that the CFO leads to shift by the same frequency shift $\varepsilon$ for all subcarriers to the signal $Z_{k,l}$ in the frequency domain. The normalized CFO can be divided into two parts: integer CFO (IFO) $\varepsilon_i$ and fractional CFO (FFO) $\varepsilon_f$ $(i.e.\, \varepsilon = \varepsilon_i + \varepsilon_f)$ [14]. The effects of the IFO and FFO on the received signal are illustrated in figure 3.8. These effects on the OFDM systems will be discussed in more detail in the next sections. The CFO impairment in OFDM systems is modeled as shown in figure 3.9.

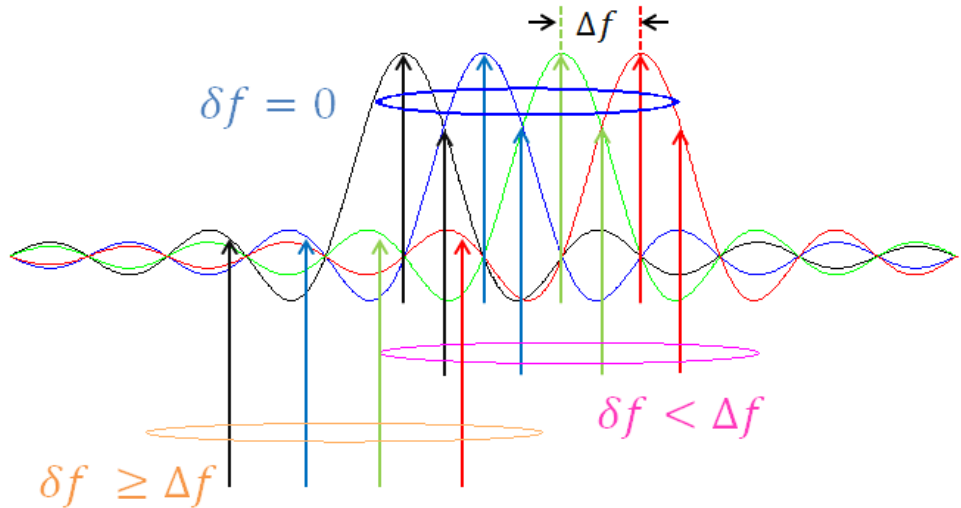

**Figure 3.8 : Impact of carrier frequency offset.**

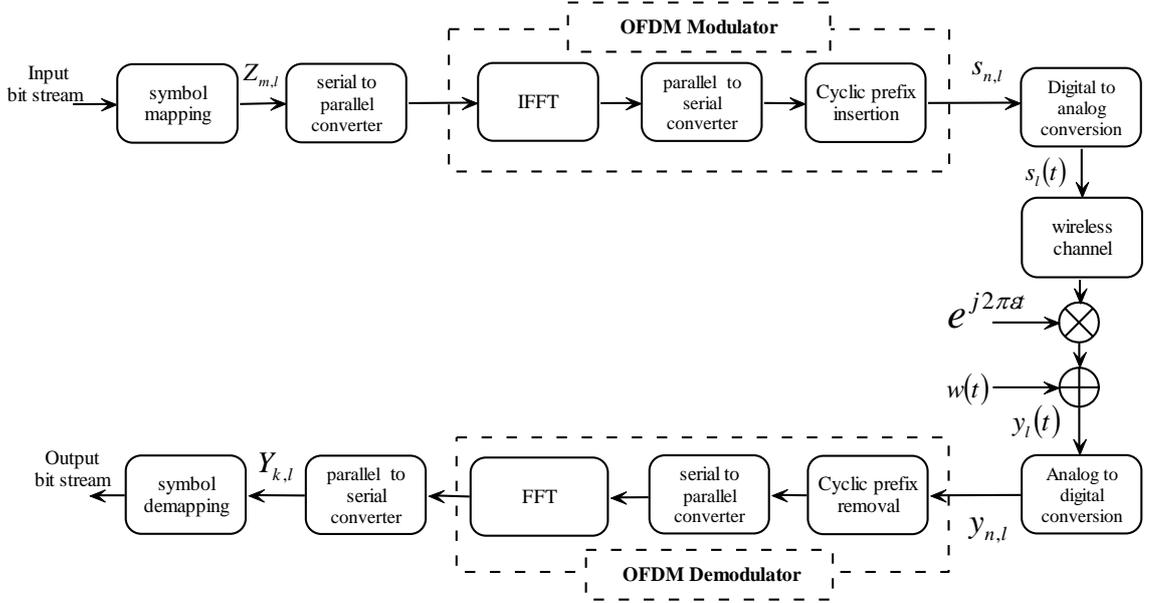

**Figure 3.9 : OFDM system with CFO modeling.**

### 3.6.2 IFO problem modelling

The IFO $(\varepsilon_i)$ is an integer multiple of the subcarrier spacing $(\Delta f)$ [9]. The IFO is considered in equation (3.18); the $l^{th}$ received OFDM symbol at $n^{th}$ sample can be expressed as:

$$y_{n,l} = 1/N_u \sum_{n=0}^{N_u-1} H_{m,l} Z_{m,l} e^{j2\pi(m+\varepsilon_i)n/N_u} + w_{n,l}, n = 0,1 \ldots \ldots N_u - 1 \quad (3.21)$$

After some manipulations, the $l^{th}$ received OFDM symbol at the $k^{th}$ subcarrier after FFT can be expressed as follows [9]:

$$Y_{k,l} = e^{j2\pi\varepsilon_i l N_{sym}/N_u} H(|k - \varepsilon_i|_{N_u}) Z_{k,l}(|k - \varepsilon_i|_{N_u}) + W_{k,l} \quad (3.22)$$

Where $|k - \varepsilon_i|_{N_u}$ denotes $(m - \varepsilon_i)$ modulo to $N_u$ which means that the value of $(k - \varepsilon_i)$ reduced to interval $[0, N_u - 1]$ and $W_{k,l}$ is the AWGN at the $k^{th}$ subcarrier. Equation (3.22) indicates that the IFO $\varepsilon_i$ leads to cyclic shift for the transmitted OFDM symbols i.e. the received symbols which were mapped on the subcarriers are in the wrong positions in the demodulated spectrum, resulting in a high BER. But orthogonality between all subcarriers still maintained so there is no ICI [9].

### 3.6.3 FFO problem modelling

The FFO $(\varepsilon_f)$ is a fraction of one subcarrier spacing [9]. The FFO is considered, in equation (3.18), the $l^{th}$ received OFDM symbol after CP removal at $n^{th}$ sample of can be written as follows [14]:

$$y_{n,l} = 1/N_u \sum_{m=0}^{N_u-1} H_{m,l} Z_{m,l} e^{j2\pi(m+\varepsilon_f)n/N_u} + w_{n,l}, n = 0,1 \ldots \ldots N_u - 1 \quad (3.23)$$

The $l^{th}$ received OFDM symbol at the $k^{th}$ subcarrier after FFT can be expressed as follows [14]:

$$Y_{k,l} = \sum_{n=0}^{N_u-1} r_{n,l} e^{-j2\pi nk/N_u} + W_{k,l}, \qquad k = 0,1 \ldots \ldots N_u - 1 \qquad (3.24)$$

Where $W_{k,l} = \sum_{n=0}^{N_u-1} w_{n,l} e^{-j2\pi nk/N_u}$, after manipulation, the received $l^{th}$ OFDM symbol at the $k^{th}$ subcarrier $Y_{k,l}$ can be expressed as follows [14]:

$$\begin{aligned} Y_{k,l} &= \sum_{n=0}^{N_u-1} \left\{ 1/N_u \sum_{m=0}^{N_u-1} H_{m,l} Z_{m,l} e^{j2\pi(m+\varepsilon_f)n/N_u} e^{-j2\pi kn/N_u} \right\} + W_{k,l} \\ &= \sum_{m=0}^{N_u-1} H_{m,l} Z_{m,l} \left\{ 1/N_u \sum_{n=0}^{N_u-1} e^{j2\pi(m-k+\varepsilon_f)n/N_u} \right\} + W_{k,l} \qquad (3.25) \end{aligned}$$

Let $\Lambda_{m-k} = 1/N_u \sum_{n=0}^{N_u-1} e^{j2\pi(m-k+\varepsilon_f)n/N_u}$ which describes the ICI effect. If $\varepsilon_f = 0$ thus, the component $\Lambda_0 = 1$ for $m = k$ and $\Lambda_{k-m} = 0 \; \forall \; m - k \neq 0$ this means that there is no ICI between all subcarriers [20]. If $\varepsilon_f \neq 0$, the component of $\Lambda_0$ can be getting at $k = m$. After some manipulation the component of $\Lambda_0$ can be expressed as follows:

$$\Lambda_0 = 1/N_u \sum_{n=0}^{N_u-1} e^{j2\pi n\varepsilon_f/N_u} = \left\{ \frac{\sin(\pi\varepsilon_f)}{N_u \sin(\pi\varepsilon_f/N_u)} \right\} e^{j\pi\varepsilon_f(N_u-1)/N_u} \qquad (3.26)$$

Where $\Lambda_0$ is defined as the attenuation and phase rotation factor to the $k^{th}$ desired subcarrier due to the FFO. It is depends on the normalized frequency offset $\varepsilon_f$ but is independent of k. This means that all subcarriers experience the same degree of attenuation and rotation of the wanted component [26]. Also, the $k^{th}$ desired subcarrier is subject to ICI. This is the sum of components dependent on each of values $Z_{m,l}$, $m \neq k$. The contribution of each $Z_{m,l}$ depends on $(m - k) \bmod N_u$ and on the normalized frequency offset $\varepsilon_f$. It does not depend directly on $k$ [26]. If $m \neq k$, the component of $\Lambda_{m-k}$ is defined as the ICI coefficient between $m^{th}$ and $k^{th}$ subcarriers [27]. After some manipulation the component of $\Lambda_{m-k}$ is expressed as follows [14]:

$$\Lambda_{m-k} = \left\{ \frac{\sin(\pi\varepsilon_f) e^{-j\pi(m-k)/N_u}}{N_u \sin(\pi(m-k+\varepsilon_f)/N_u)} \right\} e^{j\pi\varepsilon_f(N_u-1)/N_u} \qquad (3.27)$$

Substitute from equations (3.26) and (3.27) into equation (3.25), the frequency domain received signal with FFO is expressed as follows [14, 20]:

$$Y_{k,l} = e^{j2\pi\varepsilon_f[G+(1+G)l]} \sum_{k=0}^{N_u-1} \{H_{m,l}Z_{m,,l}(\Lambda_0 + \Lambda_{m-k})\} + W_{k,l} \quad (3.28)$$

$$Y_{k,l} = \underbrace{e^{j2\pi\varepsilon_f[G+(1+G)l]}\Lambda_0 H_{k,l}Z_{k,l}}_{signal} + \underbrace{e^{j2\pi\varepsilon_f[G+(1+G)l]} \sum_{m=0,m\neq k}^{N_u-1} \Lambda_{m-k}H_{m,l}Z_{m,l}}_{ICI}$$

$$+ \underbrace{W_{k,l}}_{AWGN}, \quad k = 0,1 \ldots \ldots N_u - 1 \quad (3.29)$$

Let the ICI component is defined as $I_{k,l} = \sum_{m=0,m\neq k}^{N_u-1} \Lambda_{m-k}H_{m,l}Z_{m,l}$ [14]. Thus, the $l^{th}$ received OFDM symbol at the $k^{th}$ subcarrier after FFT can be rewritten as follows [14]:

$$Y_{k,l} = e^{j2\pi\varepsilon_f[G+(1+G)l]}H_{k,l}Z_{k,l}\Lambda_0 + e^{j2\pi\varepsilon_f[G+(1+G)l]}I_{k,l} + W_{k,l} \quad (3.30)$$

Equation (3.30) consists of three components in frequency domain. The first component represents the amplitude reduction and phase rotation of the $k^{th}$ desired subcarrier, the second component represents the ICI from other subcarriers into $k^{th}$ desired subcarrier due to loss of orthogonality among subcarrier frequency components and the third component represents AWGN [14].

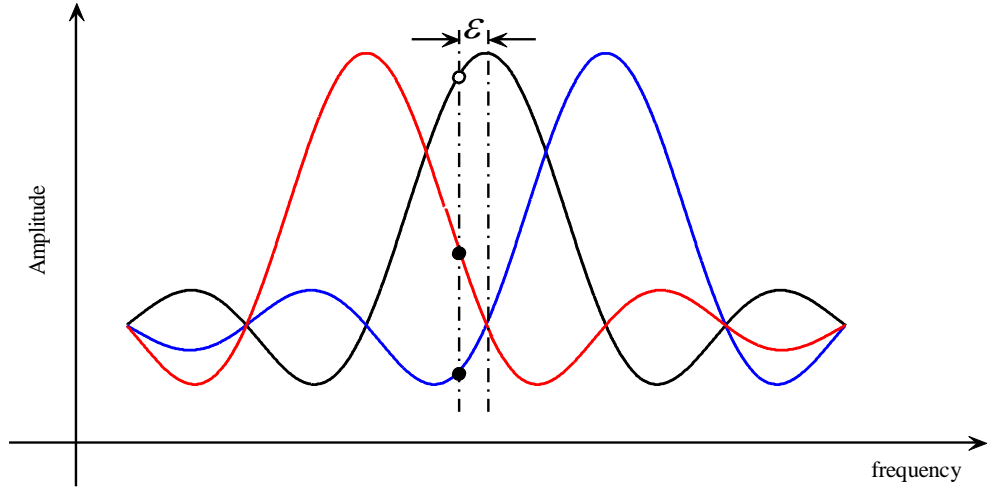

**Figure 3.10 : Inter-carrier interference (ICI) due to CFO.**

Finally, when $\varepsilon = \varepsilon_i + \varepsilon_f$ with $\varepsilon_f \neq 0$, the CFO is referred to as FFO. In that case, all the subcarriers are shifted by an integer multiple of subcarrier spacing $\Delta f$ (depending upon the $\varepsilon_i$ value) and a fraction of subcarrier spacing (depending upon the $\varepsilon_f$ value). There are two destructive effects caused by a FFO as seen from equation (3.30). First, the amplitude reduction and phase distortion to the desired subcarrier. Note that the amplitude reduction occurs because of the desired subcarrier is no longer sampled at the peak of $\text{sinc}(.)$ function of the FFT. Second, destroys the orthogonality between all subcarriers which introduce the ICI form the neighboring subcarriers. The adjacent subcarriers cause interference because they are not sampled at their zero crossings [14, 25]. In figure 3.10, the amplitude reduction is shown by a maker " ○ " and ICI by adjacent subcarriers is denoted by " ● ".

The phase rotation caused by CFO is depicted in figure 3.11. The constellation of 16QAM modulation under the following simulation parameters: five OFDM symbols, $N_u = 64$ subcarriers, $N_{CP} = 16$ in the absence of noise, for $\varepsilon = 0$ and $\varepsilon = 0.01$ [25]. The CFO is given by the multiplication distortion with factor of $e^{j2\pi\varepsilon n/N_u}$. This means that the signal $z_n$ is rotated by a phase angle of $2\pi\varepsilon(N_u + N_{cp})/N_u$ between two consecutive OFDM symbols with time indices $l$ and $l + 1$ [28].

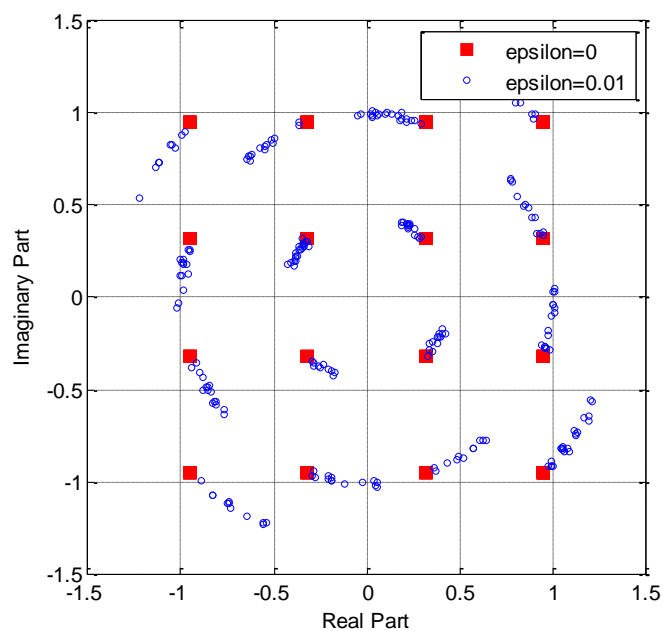

**Figure 3.11 : Phase rotation caused by CFO for 16QAM modulation.**

The BER performance versus $E_b/N_0$ of the OFDM system for 16QAM modulation under the following simulation parameters: five OFDM symbols, $N_u = 64$ subcarriers, $N_{CP} = 16$ in AWGN for different CFO can be investigated as shown in figure 3.12. Figure 3.12 depicts for large CFOs, the errors caused by ICI deteriorate the BER performance even at high SNR. For small CFOs the BER performance is dominated by SNR. For normalized CFO smaller than 0.01, we find the BER performance degradation due to ICI can be neglect. These results are consistent with previous published results [29].

Note that the modulation scheme with large constellation points is more sensitive to frequency offset than a small constellation modulation scheme. This is the reason for the SNR requirements for the higher constellation modulation scheme are larger for the same BER performance [25]. Thus, the performance of the OFDM systems depends on modulation type.

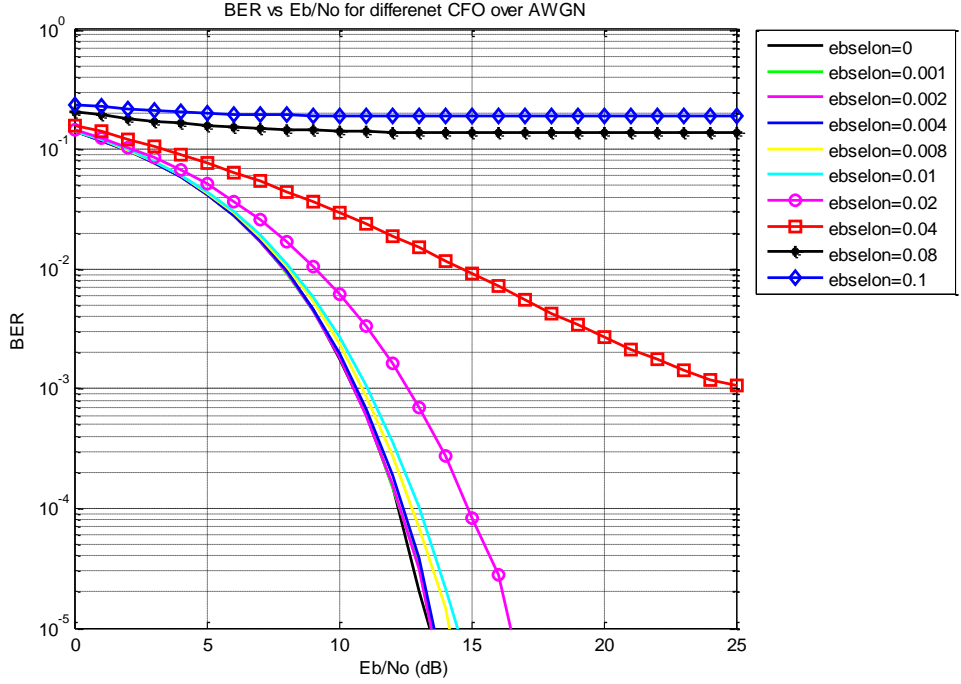

**Figure 3.12 : BER versus $E_b/N_0$ for OFDM system for 16QAM modulation under different CFO [29].**

The effects of frequency offset on the performance of the system can be evaluated by the loss in the SNR [9] is expressed as follows:

$$\gamma(\varepsilon) = \frac{\text{SNR}^{(ideal)}}{\text{SNR}^{(real)}} \quad (3.31)$$

Where $\text{SNR}^{(ideal)}$ is the SNR without CFO and $\text{SNR}^{(real)}$ is the SNR in the presence of frequency offset $\varepsilon$. For OFDM systems the SNR loss in dB can be expressed as follows [9]:

$$\gamma(\varepsilon) = \frac{1}{|\Lambda_0|^2}\left\{1 + \frac{E_s}{N_0}[1 - |\Lambda_0|^2]\right\} \quad (3.32)$$

We conclude that the CFO has significant effects on the performance of OFDM systems. Thus, CFO estimation is crucial in OFDM systems. In the next chapter several schemes for frequency synchronization will be discussed in detail. The proposed dual bandwidth scheme for estimate and correct CFO will be developed.

# Chapter 4. Frequency Synchronization Schemes

## 4.1 Introduction

The OFDM systems are very much sensitive to synchronization impairments such as CFO. Thus, frequency synchronization has been one of the most important researches in OFDM systems. In this chapter, a survey of frequency synchronization techniques to overcome the CFO problem in OFDM systems is introduced. Data-aided (DA) schemes and non-data-aided (NDA) schemes are presented. The definitions and the main parameters of the PLL are introduced. The CFO estimation using conventional pilot tones versus the CFO estimation using clustered pilot tones are discussed. The dual bandwidth scheme is proposed. The improved dual bandwidth scheme using clustered pilot tones is investigated.

Frequency synchronization in OFDM is performed via two stages: acquisition and tracking. Acquisition stage means IFO ($\varepsilon_i$) is estimated and is also known by initial or coarse CFO estimation parameters (i.e. finds the correct tone numbering). Tracking stage is a fine tuning for CFO to get better estimation (i.e. finds the FFO ($\varepsilon_f$) from the received subcarriers) and is also known by fine CFO estimation. Acquisition parameter estimation schemes generally have a wide range but low accuracy. Tracking algorithms have a narrower range and finer accuracy [30].

The frequency synchronization is used to get an accurate CFO estimate from the received signal and then use the estimate value to compensate the signal. However, estimation error is inevitable, and the residue frequency offset usually exists in OFDM systems. Several CFO estimators have been proposed for OFDM frequency synchronization. These schemes are classified as follows: DA and NDA (i.e. blind) schemes. DA schemes yield much better performance than NDA schemes at the cost of bandwidth efficiency [31]. The classification of CFO compensation schemes are illustrated in figure 4.1.

## 4.2 Non-data-aided (NDA) schemes

In NDA schemes the transmitted data are used without any other additional information (i.e. do not use of known symbols) for CFO estimation [32].

### 4.2.1 CFO estimation using cyclic prefix

The CFO estimation using CP exploiting the redundancy created by the CP can estimate time and frequency parameters. The CP is an identical copy of the last $N_{CP}$ samples of every OFDM symbol which is appended in front of the symbol. Although the loss of transmission power and bandwidth associated with the cyclic, it can be used for CFO estimation.

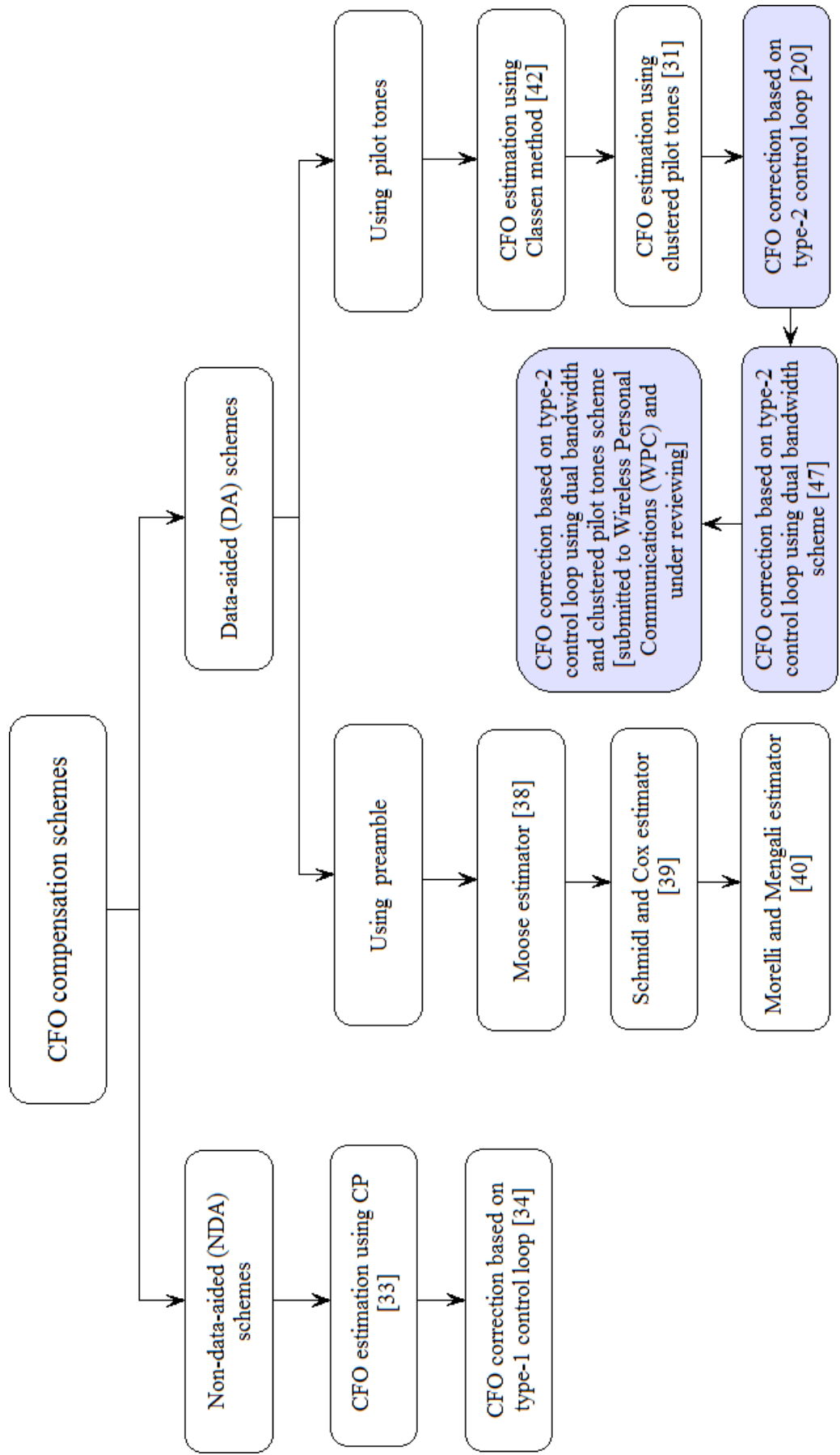

**Figure 4.1: Classification of CFO compensation schemes.**

This is most commonly done by averaging the correlation between the CP and the end of the OFDM symbol as shown in figure 4.2 [33].

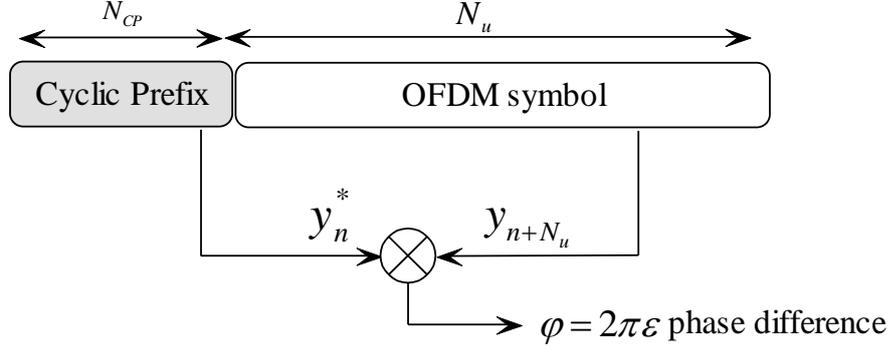

**Figure 4.2: Structure of observed OFDM signal with CP.**

Assuming that there is no timing offset, the CFO $\varepsilon$ leads to phase rotation of $2\pi n\varepsilon/N_u$ in the received signal. The phase of the correlation product $y_{n,l}^* \, y_{n+N_u,l}$ contains information about CFO $\varepsilon$ [4, 30]. Assuming that the channel effect is neglected, thus the phase difference between CP and the end part of an OFDM symbol due to CFO is $2\pi N_u \varepsilon / N_u = 2\pi\varepsilon$ as shown in figure 4.2. Therefore, the estimation of CFO can be obtained by dividing the correlation phase with $2\pi$ [4].

Then, the CFO can be found from the phase angle of the product of CP and the end part of an OFDM symbol, for example $\hat{\varepsilon} = 1/2\pi \, arg\{y_{n,l}^* \, y_{n+N_u,l}\}$, $n = -1, -2, \ldots\ldots, N_{CP}$. The noise effect is reducing via the averaging over the samples in a CP interval [14]. Thus, the estimation of normalized CFO can be expressed as follows:

$$\hat{\varepsilon} = \frac{1}{2\pi} arg \left\{ \sum_{n=-N_{CP}}^{-1} y_{n,l}^* \, y_{n+N_u,l} \right\} \tag{4.1}$$

The argument operation $arg(.)$ returns values in the range of $[-\pi, +\pi)$ and is performed by using $\tan^{-1}(.)$ [9]; the estimation range of CFO in equation (4.1) is $[-\pi, +\pi)/2\pi = [-0.5, +0.5)$ which is less than one half of the subcarrier spacing i.e. $|\hat{\varepsilon}| < 0.5$. Consequently, integral CFO cannot be estimated by this technique. Generally, NDA schemes are used for fine CFO estimation only [30]. The CFO estimation using CP is investigated as shown in figure 4.3 [33].

The value of $[y_{n,l}^* \, y_{n+N_u,l}]$ becomes real only when there is no frequency offset. This implies that it becomes imaginary as long as the CFO exists. Thus, the imaginary part of $[y_{n,l}^* \, y_{n+N_u,l}]$ can be used for CFO estimation [30]. In this case, the estimation error is defined as follows [14]:

$$e_\varepsilon = \frac{1}{L_{av}} \sum_{n=1}^{L_{av}} Im\{y_{n,l}^* \, y_{n+N_u,l}\} \qquad (4.2)$$

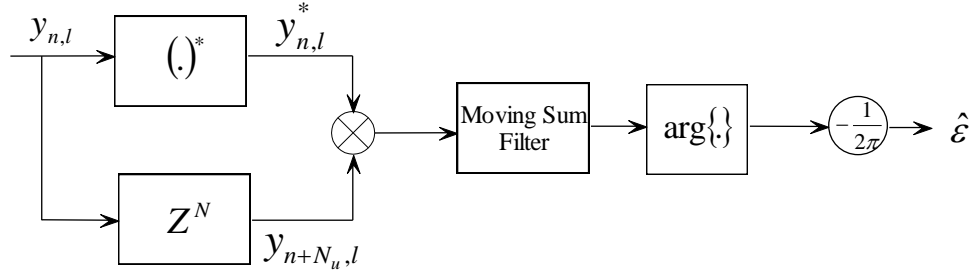

**Figure 4.3 : Structure of the estimator using cyclic prefix [33].**

Where $L_{av}$ denotes the number of samples used for averaging. The CFO estimator described in equation (4.1) was originally derived for AWGN scenario; it can be used under multipath environment. But the performance of the estimator in multipath dispersive channel will be reduced. This is due to fact that the multipath induced ISI and ICI can be considered as an additional noise in $y_{n,l}$ for $n = [-N_{CP}, -N_{CP}+1, \ldots\ldots, N_{CP}+L_h]$ where $L_h$ is the dispersive channel length in number of samples [30].

### 4.2.2 CFO correction based on type-1 control loop

In 2005, Linling Kuang et.al [34], a time-frequency decision-feedback loop (TF-DFL) for CFO tracking in OFDM systems without using of the pilot subcarriers was proposed. The residual CFO appears to be a linearly increasing phase offset in time domain and a fixed phase offset within one OFDM symbol in frequency domain. The residual CFO up to 10% of the subcarrier spacing can be tracked via two first order phase lock loops (PLL) in both time and frequency loops. The linearized first order PLL is shown in figure 4.4.

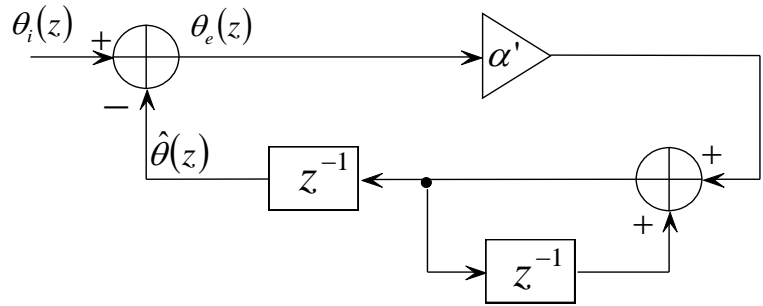

**Figure 4.4 : Linearized first order PLL.**

Therefore, the phase transfer function of the first order PLL is expressed as follows [34]:

$$\mathcal{H}(z) = \frac{\hat{\theta}(z)}{\theta_i(z)} = \frac{\alpha'.z^{-1}}{1-(1-\alpha')z^{-1}} \qquad (4.3)$$

Likewise, the phase error transfer function is

$$\mathcal{H}_e(z) = \frac{\theta_e(z)}{\theta_i(z)} = \frac{1 - z^{-1}}{1 - (1 - \alpha')z^{-1}} \tag{4.4}$$

Where $\theta_i(z)$, $\hat{\theta}(z)$, $\theta_e(z)$ and $\alpha'$ are the Z-transforms of input phase sequence $\theta_i(n)$, output phase sequence $\hat{\theta}(n)$, phase error sequence $\theta_e(n)$ and proportional gain of the loop filter of first order PLL respectively. For stability $|1 - \alpha'| < 1$, or $0 \leq \alpha' < 2$. If the input phase is constant $\theta_i(n) = \theta_0 \ \forall \ n \geq 0$, its Z-transforms is $\theta_i(z) = \frac{z^{-1}}{1-z^{-1}}$. Then, the steady state phase error is given by [34]:

$$\lim_{n \to \infty} \theta_e(n) = \lim_{z \to 1} (1 - z^{-1})\theta_e(z) = 0 \tag{4.5}$$

Thus, the first order PLL can be able to converge the steady state phase error to zero for phase offset. If the input phase is a linear increased one such that $\theta_i(n) = \Delta_0.n \ \forall \ n \geq 0$, its Z-transforms is $\theta_i(z) = \frac{z^{-1}}{(1-z^{-1})^2}$. Then, the steady state phase error is given by:

$$\lim_{n \to \infty} \theta_e(n) = \lim_{z \to 1} (1 - z^{-1})\theta_e(z) = \frac{\Delta_0}{\alpha'} \tag{4.6}$$

Thus, the first order PLL cannot be able to converge the steady state phase error to zero to frequency offset. According to equation (4.6), for very small frequency offsets the first order PLL only suffers a small loss with a reasonable ($\alpha'$). In this case, the first order PLL may be sufficient in tracking the residual CFO. Thus, the first order PLL used in both time and frequency loops to track the CFO in the received OFDM signal. One PLL in time loop (TL), to track the linear increasing phase offset; another PLL in frequency loop (FL), to track the constant residual phase offset as shown in figure 4.5 [34]. The TL output sequence, $\hat{\varphi}_l$ only updates at the $l^{th}$ OFDM symbol. Thus, the pre-FFT sample $y_{n,l}$ is rotated by [34]:

$$y'_{n,l} = y_{n,l}.e^{-j(n+N_{CP}+lN_{sym})\hat{\varphi}_l}, \quad n = 0,1,\ldots,N_u - 1 \tag{4.7}$$

The FL output sequence $\hat{\psi}_{k,l}$ updates at every subcarrier. The post-FFT data symbol $Y_{k,l}$ is rotated by [34]:

$$Y'_{k,l} = Y_{k,l}.e^{-j\hat{\psi}_{k,l}}, \quad k = 0,1,\ldots,N_u - 1 \tag{4.8}$$

Then, the corrected data symbol $Y'_{k,l}$ may be demapped to bit stream. In the phase error detector (PED), $Y'_{k,l}$ are used for extracting the error increment $\varepsilon_{k,l}$ as follows [34]:

$$\text{DFL2}: \varepsilon_{k,l} = Im\{\text{sgn}(Y'^*_{k,l}).e_{k,l}\} = e^Q_{k,l}.\text{sgn}(a_{k,l}) - e^I_{k,l}.\text{sgn}(b_{k,l}) \tag{4.9}$$

Where $e_{k,l}^I = a_{k,l} - I_{k,l}$, $e_{k,l}^Q = b_{k,l} - Q_{k,l}$, $a_{k,l}$ and $b_{k,l}$ are the real and image parts of $Y'_{k,l}$, respectively, $\text{sgn}(.)$ denotes the signum function, and $I_{k,l}$ and $Q_{k,l}$ are the real and image parts of the decision output, respectively. The error increment $\varepsilon_{k,l}$ enters into the FL filter directly. Then, it is averaged before going to the TL filter as follows:

$$\bar{\varepsilon}_l = \frac{1}{N_c} \sum_{k=0}^{N_c-1} \varepsilon_{k,l} \qquad (4.10)$$

The DFL2 has the same phase-detect characteristics as that of the DFL1 [35], but avoids using multiplier. The TF-DFL is faster than the CFO estimation technique using CP over multipath fading channel. The pilot subcarriers in each OFDM symbol are removed thus, the TF-DFL reduces the pilot subcarriers overhead in the tracking stage. The TF-DFL utilized the all data subcarriers for CFO estimation and compensation. Thus, it is robust to selective fading channel compared to that of the CFO estimation technique using CP. It is implemented in DSP chip. It has small computational load compared to [33, 36]. Simulation results are achieved in Rayleigh fading channels, with the BER performance close to that of an offset-free system for QPSK, 16QAM, and 64QAM modulation schemes. It can only handle very small CFO where its estimation range is $|\hat{\varepsilon}| \leq 0.1$.

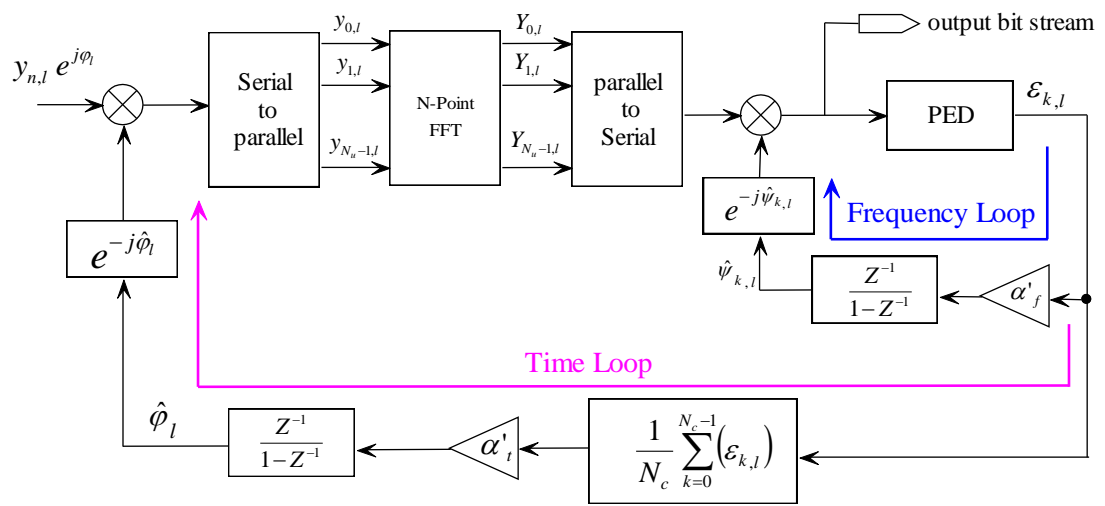

**Figure 4.5 : TF-DFL structure.**

## 4.3 Data aided (DA) schemes

In DA schemes, additional training symbols are inserted into the transmitted data such as preamble before data blocks pilot tones pilot tones in each OFDM symbol or a combination of the two. The DA schemes lead to lose some amount of bandwidth for frequency synchronization. Thus, the DA schemes is not bandwidth efficient than NDA schemes [31, 37].

### 4.3.1 CFO estimation using preamble

The training symbol (preamble) is known to the receiver and it is transmitted before the data blocks, and then CFO is estimated by detecting the phase rotation of preamble during a block period [37]. As for the applications of OFDM, many CFO estimators are designed for burst transfer mode [33, 38-39] where the short distance environment such as WLAN applications is regarded as a slow-varying channel model, and the preamble is, thus, designed for CFO estimation under this scenario [31].

### 4.3.2 Moose CFO estimator

In 1994, P. H. Moose [38] proposed to use two identical OFDM symbols without guard interval in between to perform CFO estimation as shown in figure 4.6.

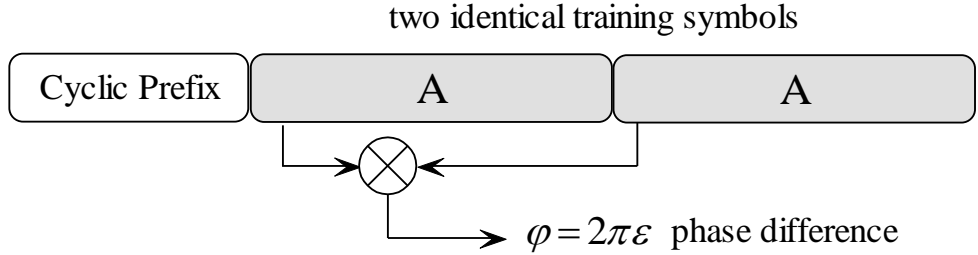

**Figure 4.6 : The OFDM training symbol structures proposed for frequency estimation by Moose.**

From equation (3.18), the $l^{th}$ received OFDM symbol in the presence of CFO in the time-domain is given by [38]:

$$y_{n,l} = (1/N_u) \sum_{k=0}^{N_u-1} H_{k,l} Z_{k,l} e^{j2\pi n(k+\varepsilon)/N_u} + w_{n,l}, n = 0,1,\dots,N_u - 1 \quad (4.11)$$

Let us drop the subcarrier $l$ and the time index $n$ is extended to $2N_u - 1$. If an OFDM transmission symbol is repeated without guard interval. The corresponding received signal in time domain can be expressed as [38]:

$$y_n = (1/N_u) \sum_{k=0}^{N_u-1} H_k Z_k e^{j2\pi n(k+\varepsilon)/N_u} + w_n, n = 0,1,\dots,2N_u - 1 \quad (4.12)$$

Now the $k^{th}$ element after the $N_u$-point FFT of the first $N_u$ can be written as [38]:

$$Y_{1k} = \sum_{n=0}^{N_u-1} y_n e^{-j2\pi nk/N_u} + W_{1k}, \quad k = 0,1,\dots,N_u - 1 \quad (4.13)$$

And the $k^{th}$ element of the $N_u$-point FFT of the second half of the sequence is given by [38]:

$$Y_{2k} = \sum_{n=N_u}^{2N_u-1} y_n e^{-j2\pi nk/N_u} + W_{2k} = \sum_{n=0}^{N_u-1} y_{n+N_u} e^{-j2\pi nk/N_u} + W_{2k} \quad (4.14)$$

But from equation (4.12) $y_{n+N_u} = y_n e^{j2\pi\varepsilon}$

$$Y_{2k} = \sum_{n=0}^{N_u-1} (y_n e^{2\pi j\varepsilon}) e^{-j2\pi nk/N_u} + W_{2k}$$

$$= e^{j2\pi\varepsilon} \sum_{n=0}^{N_u-1} y_n e^{-j2\pi nk/N_u} + W_{2k} \quad (4.15)$$

$$Y_{2k} = Y_{1k} e^{j2\pi\varepsilon} + W_{2k}, \quad k = 0, 1, \ldots N_u - 1 \quad (4.16)$$

Note that both the ICI and the desired signal components are phase rotated equally by the CFO (by a phase shift proportional to frequency offset), from first DFT to the second one. Therefore, equation (4.16) can be used to estimate the CFO accurately even if the offset is too large to demodulate the data satisfactorily [38]. Moose estimator was originally derived in the frequency domain as:

$$\hat{\varepsilon} = (1/2\pi) \tan^{-1} \left\{ \left( \sum_{k=0}^{N_u-1} Im[Y_{2k} Y_{1k}^*] \right) \bigg/ \left( \sum_{k=0}^{N_u-1} Re[Y_{2k} Y_{1k}^*] \right) \right\} \quad (4.17)$$

By using equation (4.16) in the absence of noise we find that the angle of $Y_{2k} \cdot Y_{1k}^*$ is $2\pi\varepsilon$ for each $k$ as follows:

$$Y_{2k} \cdot Y_{1k}^* = (Y_{1k} e^{j2\pi\varepsilon}) \cdot Y_{1k}^* = |Y_{1k}|^2 e^{j2\pi\varepsilon} \quad (4.18)$$

The variance of estimation error for the estimator in equation (4.17) is derived in [38] as:

$$var[\hat{\varepsilon}|\varepsilon, \{Y_k\}] = \frac{N_0}{(2\pi)^2 (T_u/N_u) \sum_{n=0}^{N_u-1} |z_n|^2} \quad (4.19)$$

If the number of subcarriers increases the variance gets smaller which is an indication of performance improvement. Let the multipath channel which remains static during the two training symbols. Moreover, the training symbol pair is received over the dispersive channel proceeded by a cyclic prefix of length $N_{cp} > L_h$. Since the modulation phase values are repeated, the phase shift of all the subcarriers between consecutive training symbols is due to CFO only and hence remains unchanged. Therefore, no guard interval is required between the two identical training symbols and the equation (4.17) can be used for multipath environment as well [38].

The estimation range of equation (4.17) is one half of the subcarrier spacing of the repeated symbol $|\hat{\varepsilon}| \leq 0.5$. Therefore, this technique is not suitable for acquisition mode where estimation range is of primary focus. The main drawback of this scheme is the relatively short acquisition range. Moose also describes how to increase this range by using shorter training symbols to find the carrier frequency offset. By shortening the training symbols by a factor of two would double the range of carrier frequency acquisition. However the performance of estimator gets worse as the estimation range increases. The reason for this performance degradation is that as the training symbols get shorter, fewer samples are available for correlation average [38].

### 4.3.3 Schmidl and Cox CFO estimator

In 1997, Schmidl and Cox [39] suggested two training symbols. The first training symbol contains two identical halves of length $N_u/2$ that are generated by transmitting a pseudonoise (PN) sequence on the even subcarriers while zeros are used on the odd frequencies in the time domain. The first training symbol can be used for two purposes. First the correlation between the two identical halves can be used for frame/symbol timing. Second, the two identical halves will remain identical after passing through the channel, except that there will be a phase difference between them caused by the carrier frequency offset. This means that it is used for a coarse estimate of the CFO. The second training symbol contains a PN sequence; PN1 on the even subcarriers and another PN2 on the odd subcarriers. The PN sequence can be chosen so that it copes well with the high peak to average ratio problem of OFDM as shown in figure 4.7.

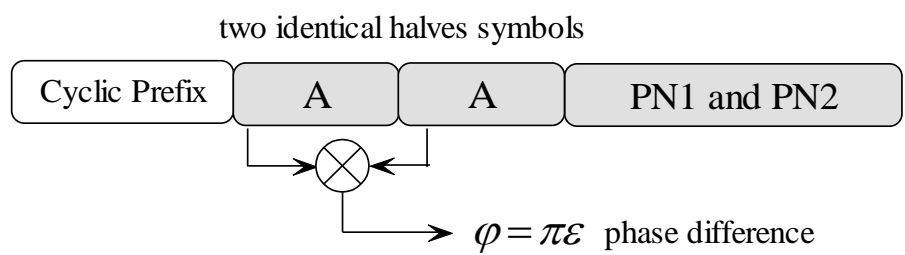

**Figure 4.7 : The OFDM training symbol structures proposed for frequency estimation by Schmidl and Cox.**

Assuming the first half of the first training symbol is identical to the second half in time domain at the transmitter. Assuming that there is no timing offset existing and ignoring the effect of the channel. Thus, the phase rotation due to the carrier frequency offset $\delta f$ only exists at the receiver [39]. If the received complex envelope at time $t$ in the first half is $y(t) = u(t)e^{-j2\pi\delta f t}$, after $T_u/2$ seconds the signal is $y(t + T_u/2) = u(t + T_u/2)e^{-j(2\pi\delta f t + \pi\delta f T_u)}$. Since $u(t) = u(t + T_u/2)$ by design, we have [5]:

$$y^*(t)\, y(t + T_u/2) \approx |u(t)|^2 e^{j\pi\delta f T_u} \tag{4.20}$$

The result has a phase of $\varphi = \pi\delta f T_u = \pi\varepsilon$ that is proportional to the frequency offset [6]. Thus, the phase difference $\varphi$ between two halves is obtained by calculating their correlation is approximately equal to $\pi\varepsilon$ [39]. This phase difference $\hat{\varphi}$ can be estimated as [39]:

$$\hat{\varphi} = angle(P_{S\&C}(d)) \tag{4.21}$$

Where $P_{S\&C}(d)$ is defined as the correlation between the two halves of training symbol. If $|\hat{\varphi}|$ is less than $\pi$, then the FFO estimate is given by

$$\hat{\varepsilon}_f = \hat{\varphi}/\pi \tag{4.22}$$

But the value of $angle(P_{S\&C}(d))$ cannot reflect the total $\varphi$ since the angle of a complex number is limited to the range of $[-\pi, \pi]$. In this case, the actual $angle(P_{S\&C}(d))$ would be $\varphi + 2\pi\upsilon$, where $\upsilon$ is defined as an integer and the actual normalized CFO would be [9, 39]:

$$\varepsilon = \varepsilon_f + \varepsilon_i \tag{4.23}$$

Where $\varepsilon_i = 2\upsilon$, the fractional part is estimated using (4.22). The integer part can be found using the second training symbol. First, the two training symbols are partially frequency corrected by multiplying the samples by $e^{-j2\pi n\hat{\varepsilon}_f/N_u}$. Thus, the FFT outputs of these two symbols will not have ICI since the remaining frequency offset is an integer multiple of $1/T_u$, and the orthogonallity between subcarriers is restored. However we still need to find out this integer part of the frequency offset and corrects it. Since the IFO degrades the SNR [6].

For IFO estimation, the second training symbol contains a differentially encoded PN1 sequence on even subcarriers and PN2 sequence on odd subcarriers. Let the FFT's of received first and second training symbols is $F_{1,k}$ and $F_{2,k}$ and the differentially modulated even frequencies of second training symbol be $v_k$. At the output of the FFT unit, because of the uncompensated frequency shift of $2\upsilon/T_u$, the training symbols $F_{1,k}$ and $F_{2,k}$ will be the same as the transmitted data except that they will be shifted by $2\upsilon$ positions (frequency shifting property of FFT). The number of even positions shifted can be calculated by finding $\hat{g}$ to maximize the metric [39]:

$$B(g) = \frac{\left|\sum_{k \in X_E} F^*_{1,k+2g} v^*(k) F_{2,k+2g}\right|^2}{2\left(\sum_{k \in X_E} |F_{2,k}|^2\right)^2} \tag{4.24}$$

With integer $g$ spanning the range of possible frequency offsets. Where $X_E = \{-W_E, -W_E + 2, \dots\dots, -4, -2, 2, 4, \dots\dots, W_E - 2, W_E\}$ be the set of indices for even frequency components and $W_E$ is the number of even frequencies with the

PN sequence [39]. Then, the normalized CFO estimate is expressed as follows [9, 39]:

$$\hat{\varepsilon} = \hat{\varepsilon}_f + \hat{\varepsilon}_i \qquad (4.25)$$

The main advantage of the Schmidl and Cox estimator is its acquisition range larger than the Moose estimator but with large overhead.

### 4.3.4 Morelli and Mengali CFO estimator

Morelli and Mengali [40] proposed an extension of Schmidl and Cox algorithm by using only one training symbol composed of $Q$ identical parts where $Q > 2$ in the time domain as shown in figure 4.8.

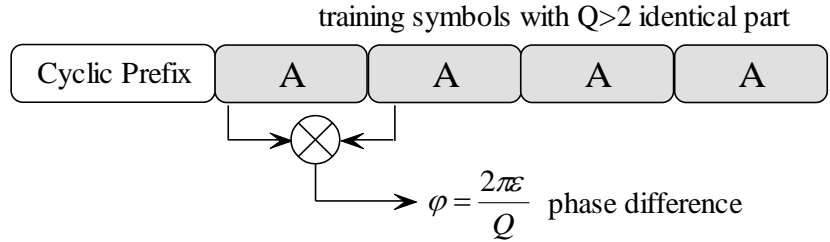

**Figure 4.8 : The OFDM training symbol structures proposed for frequency estimation by Morelli and Mengali.**

They are generated by transmitting a PN sequence on the frequency multiples of $Q/T_u$ and setting zeros for the remaining frequencies. The Morelli and Mengali exploit the correlation between identical parts of the training symbol defined as [40]:

$$\Psi(q) = \frac{1}{N_u - qM} \sum_{n=qM}^{N_u-1} y_n\, y^*_{n-qM}, \qquad 0 \leq q \leq P \qquad (4.26)$$

Where $M = N_u/Q$, is the length in sampling interval of each part of the training symbol and $P$ is denoted by a design parameter and is less than or equal to $Q - 1$. Substituting equation (3.18) into (4.26), we get the following [40]:

$$\Psi(q) = e^{j2\pi n\varepsilon/N_u}\, D(n)[1 + \alpha(q)] \qquad (4.27)$$

Where $D(n) \triangleq \frac{1}{N_u - qM}\sum_{n=qM}^{N_u-1}|r_n|^2$ is the real-valued random envelope and α(q) is the noise term defined in [40]. By considering the phase differences of the correlation estimate is given by:

$$\varphi(q) \triangleq [arg\{\Psi(q)\} - arg\{\Psi(q-1)\}]_{2\pi}$$
$$\approx 2\pi\varepsilon/Q + \alpha_I(q) - \alpha_I(q-1) \qquad 0 \leq q \leq P \qquad (4.28)$$

Where $[r]_{2\pi}$ denotes modulo-$2\pi$ operation (it reduce $r$ to the interval $[-\pi, \pi)$), arg$\{\Psi(q)\}$ is the argument of $\Psi(q)$ and $\alpha_I(q)$ is the imaginary part of $\alpha(q)$. The best linear unbiased estimator (BLUE) of the CFO estimator can be expressed as follows [40]:

$$\hat{\varepsilon} = \frac{1}{(2\pi/Q)} \sum_{q=1}^{P} w_{M\&M}(q)\varphi(q) \qquad (4.29)$$

Where the weighting function $w_{M\&M}(q)$ is given by:

$$w_{M\&M} = \frac{3(Q-q)(Q-q+1) - P(Q-P)}{P(4P^2 - 6QP + 3Q^2 - 1)} \qquad (4.30)$$

The weighting function in equation (4.30) is channel independent. Thus, the Morelli and Mengali scheme is the BLUE for any channel. From [40] we conclude that, the estimation range of the Morelli and Mengali scheme is $|\hat{\varepsilon}| \leq Q/2$. The Morelli and Mengali scheme can estimate CFO by using a single training symbol. Thus, the training overhead is reduced by a factor of two than the Schmidl and Cox scheme with accuracy comparable to Schmidl and Cox's estimator. The cost to be paid is an increase in computational complexity.

### 4.3.5 CFO estimation using pilot tones

The transmission of a preamble in the front of the frame transmission and there is no data symbols are permitted during the preamble period. The preamble method is not suitable for fast fading channels. Where for every many data blocks, a few synchronization blocks are transmitted, thus, the channel may change a lot during transmitting data blocks. For continual transfer mode like broadcasting applications [41] synchronization subcarriers (pilots) are embedded in every OFDM symbol to track the variations of the CFO by measuring the phase shift of pilot symbols in two consecutive OFDM symbols [37, 42-43]. The estimation performance of the continue pilots method is better than the non-data-aided estimator. In this method some a special pilots are required for every OFDM symbol. Thus, this method decreases the spectral efficiency of the systems and scattered pilots are used [37].

### 4.3.5.1 CFO estimation using Classen method

In 1994, Classen [42] proposed a known pilot symbols are spread uniformly in the frequency domain. The pilot subcarriers known as sync-subchannels can be transmitted in every OFDM symbol for CFO tracking. The CFO estimation scheme proposed in [42] is performed via two stages: an acquisition stage and a tracking stage. In the acquisition stage, large frequency offsets (multiples of subcarrier spacing) are estimated. In the tracking stage the small fractional frequency offsets are estimated. Figure 4.8 depicts the two stages of the synchronization structure.

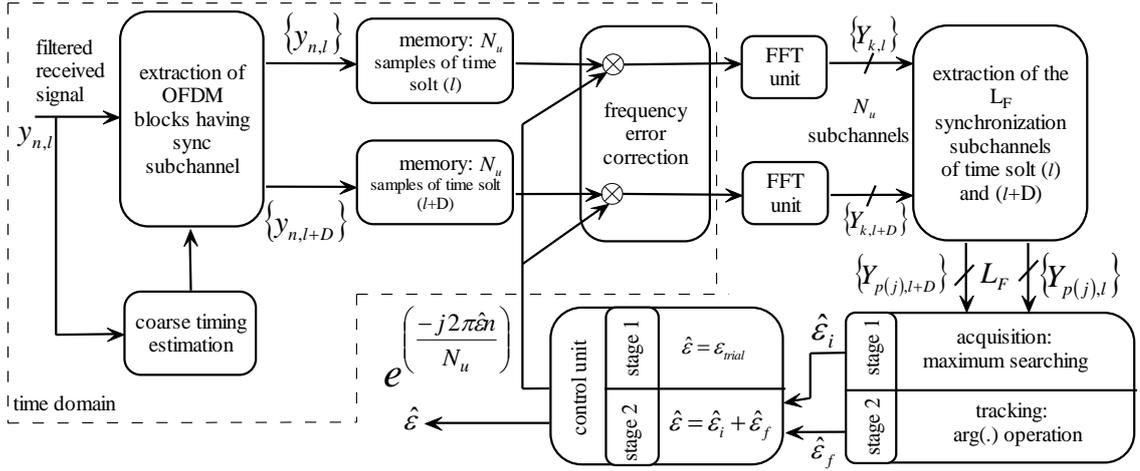

**Figure 4.9 : CFO synchronization scheme using pilot tones.**

After time synchronization, $N_u$ samples of two OFDM symbols $y_{n,l}$ and $y_{n,l+D}$ are saved in the memory. The FFT unit transforms these symbols into frequency domain (FD) signals $Y_{k,l}$ and $Y_{k,l+D}$ with $k = 0,1,\ldots\ldots N_u - 1$. Then pilots at known locations are extracted from $l^{th}$ and $(l+D)^{th}$ OFDM symbols [14, 42]. These pilot subcarriers are then utilized for CFO estimation in the frequency domain and then, the frequency correction is done in the time domain. The function of stage 1 is to obtain a coarse CFO estimate as fast as possible. Stage 2 uses the coarse frequency estimate delivered from stage 1 and performs the tracking process. The Splitting of the frequency synchronization task into two stages allows a large amount of degree of freedom where each stage has a specific algorithm which is tailored to perform a specific task (i.e. the two algorithms are independent). Therefore, stage 1 can be optimized for higher estimation range and speed regardless of the tracking performance while stage 2 can be used for higher accuracy where the large acquisition is no longer required [14, 42].

**Tracking mode algorithm:**

After the estimation and correction of the IFO which is achieved by the acquisition stage. The FFO still exists; the FFO is smaller than one half of the subcarrier spacing. The pilot tones are transmitted over a subset of available subcarriers. Let $L_F$ uniformly spaced subcarriers are reserved for pilot tones, and then the fine CFO estimator can be expressed as follows [14, 42]:

$$\hat{\varepsilon}_f = \frac{1}{2\pi D T_u} \arg\left\{ \sum_{j=0}^{L_F-1} \left(Y_{l+D,p(j)}[\hat{\varepsilon}_i] \cdot Y^*_{l,p(j)}[\hat{\varepsilon}_i]\right) \cdot \left(Z^*_{l+D,p(j)} Z_{l,p(j)}\right) \right\} \quad (4.31)$$

Where $p(j)$, $Z_{l,p(j)}$, $L_F$ and $D$ denotes the location of $j^{th}$ pilot tone, the pilot tone at location $p(j)$ in FD at the $l^{th}$ OFDM symbol, the number of pilot tones and an integer respectively.

**Acquisition mode algorithm:**

The acquisition task should be done as fast as possible and the estimation range should be higher as well. The main objective of the acquisition stage is the estimation range regardless of accuracy where it is followed by a more accurate tracking stage. The acquisition algorithm proposed in [42] achieved by a search operation for the training symbols transmitted over $L_F$ pilot subcarriers. Note that the magnitude of the argument of the arg(.) function of equation (4.31) is maximum when $\hat{\varepsilon}_i$ coincides with true value of CFO ($\varepsilon$). Therefore, the coarse CFO estimation can be calculated as follows [42]:

$$\hat{\varepsilon}_i = \frac{1}{2\pi T_u} \max_{\varepsilon_{trial}} \left\{ \left| \sum_{j=0}^{L_F-1} \left( Y_{l+D,p(j)}[\hat{\varepsilon}_{trial}] \cdot Y^*_{l,p(j)}[\hat{\varepsilon}_{trial}] \right) \cdot \left( Z^*_{l+D,p(j)} Z_{l,p(j)} \right) \right| \right\} \quad (4.32)$$

Where $\varepsilon_{trial}$ is the trial frequency value and $Y^*_{l,p(j)}[\hat{\varepsilon}_{trial}]$ is the FFT output with $\varepsilon_{trial}$ -offset-corrected input. In practice, $0.1/T_u$ spaced trial values are sufficient [42]. In tracking stage, only FFO $\hat{\varepsilon}_f$ is estimated and corrected. But in acquisition mode, both coarse and fine CFOs (i.e. IFO and FFO) are estimated and overall correction is done according to $\hat{\varepsilon} = \hat{\varepsilon}_i + \hat{\varepsilon}_f$ [42]. Since the frequency correction is achieved by counter rotating the received samples in time domain (i.e. before FFT unit) [14, 42].

Finally, Classen method introduces a method which jointly finds both the symbol timing and carrier frequency offset. However, it is very computationally complex because it uses a trial and error method where the carrier frequency is incremented in small steps over the entire acquisition range until the correct carrier frequency is found. It is impractical to do the complete search to estimate CFO.

### 4.3.5.2 CFO correction based on type-2 control loop

In 2012, Xiaofei Chen et.al, [20], CFO tracking scheme for OFDM system based on type-2 control loop was proposed. The problem is solved as a sequence of estimation and correction steps. There are two major contributors to the CFO measurement uncertainty are the additive white noise and the ICI. The noise variance can be reduced via averaging but the ICI introduced uncertainty can be decreased by iterative CFO compensations. These considerations lead to CFO correction based on type-2 control loop using conventional pilot tones.

The CFO estimator is obtained by using the one pilot tone; an unbiased estimate of the normalized CFO is obtained from a block of $L$ OFDM symbols as [20, 31]:

$$\hat{\varepsilon}_k = (1/2\pi(1+G)) \tan^{-1} \left\{ \frac{Im\left(\frac{1}{L}\sum_{l=1}^{L} Y^*_{k,l-1} Y_{k,l}\right)}{Re\left(\frac{1}{L}\sum_{l=1}^{L} Y^*_{k,l-1} Y_{k,l}\right)} \right\} \quad (4.33)$$

The performance of the phase estimation $\hat{\varphi}_k$ can be get by computed the variance of $Y^*_{k,l-1}Y_{k,l}$ [20]. Let the $k^{th}$ FFT subcarrier be the pilot tone subcarrier and the pilot tone $Z_{k,l} = +1$, $\forall l$. Assuming the transmitted data are i.i.d, zero mean with variance $\sigma_s^2$, and the data symbols are uncorrelated with noise. Thus, the expected value of the conjugate product $\{Y^*_{k,l-1}Y_{k,l}\}$ can be written as [20]:

$$E\{Y^*_{k,l-1}Y_{k,l}\} = |\Lambda_0|^2|H_k|^2 e^{j\varphi(1+G)N_u} \quad (4.34)$$

And the variance of $\{Y^*_{k,l-1}Y_{k,l}\}$ can be computed as:

$$var[Y^*_{k,l-1}Y_{k,l}] = 2|\Lambda_0|^2|H_k|^2\sigma_s^2 C + \sigma_s^4 C^2 + 2\sigma_s^2\sigma_n^2 C + 2|\Lambda_0|^2|H_k|^2\sigma_n^2 + \sigma_n^4 \quad (4.35)$$

Where $C = \sum_{m=0, m\neq k}^{N-1}|\Lambda_{m-k}|^2|H_k|^2$ thus, the component of $var[Y^*_{k,l-1}Y_{k,l}]$ is composed by four sources: signal, ICI, noise and channel gain. Note that in the case of a moderate frequency selective channel, the channel gain coefficients $H_k$ have very limited. Let the channel is flat fading and set $H_k = 1$ the component of $var[Y^*_{k,l-1}Y_{k,l}]$ is plotted. Under $\sigma_s^2\sigma_n^2 = 10$ for different values of SNR the estimation variance is plotted as shown in figure 4.10 [20]. We conclude that if the SNR is decreased, the estimation variance will increase and the curve will change slowly with the CFO this mean that the AWGN is more dominant than the ICI.

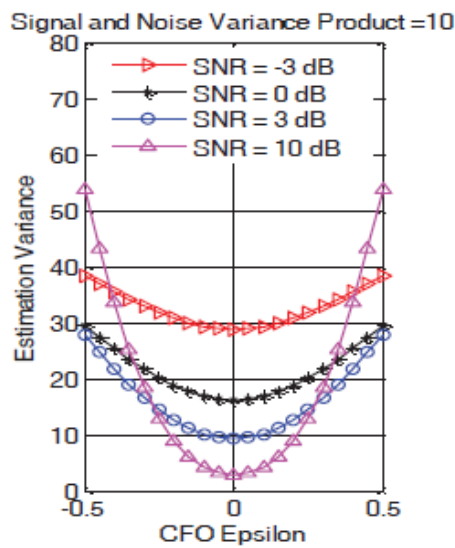

**Figure 4.10 : $var[Y^*_{k,l-1}Y_{k,l}]$ versus $\varepsilon$ under $\sigma_s^2\sigma_n^2 = 10$ for different values of SNR [20].**

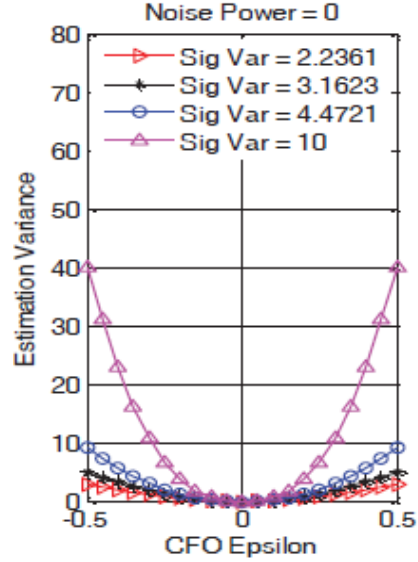

**Figure 4.11 :** $var[Y^*_{k,l-1}Y_{k,l}]$ **versus $\varepsilon$ under $\sigma_n^2 = 0$ for different values of signal power [20].**

Under $\sigma_n^2 = 0$ for different values of signal power the estimation variance can be plotted as shown in figure 4.11 [20]. We conclude that the ICI decreases as $\varepsilon$ approaches to zero.

From equation (4.35) and figure 4.10 and 4.11, the AWGN is an important contributor for estimation variance [20]. Thus, the estimation variance can be reduced by using the combination of the following two approaches: first by making $L$ independent measurements as equation (4.33) reduces the estimation variance by a factor of $L$. Second, iteratively correct the CFO, since smaller $\varepsilon$ gives lower estimation variance [20]. The CFO correction loop is based on type-2 control loop is shown in figure 4.12.

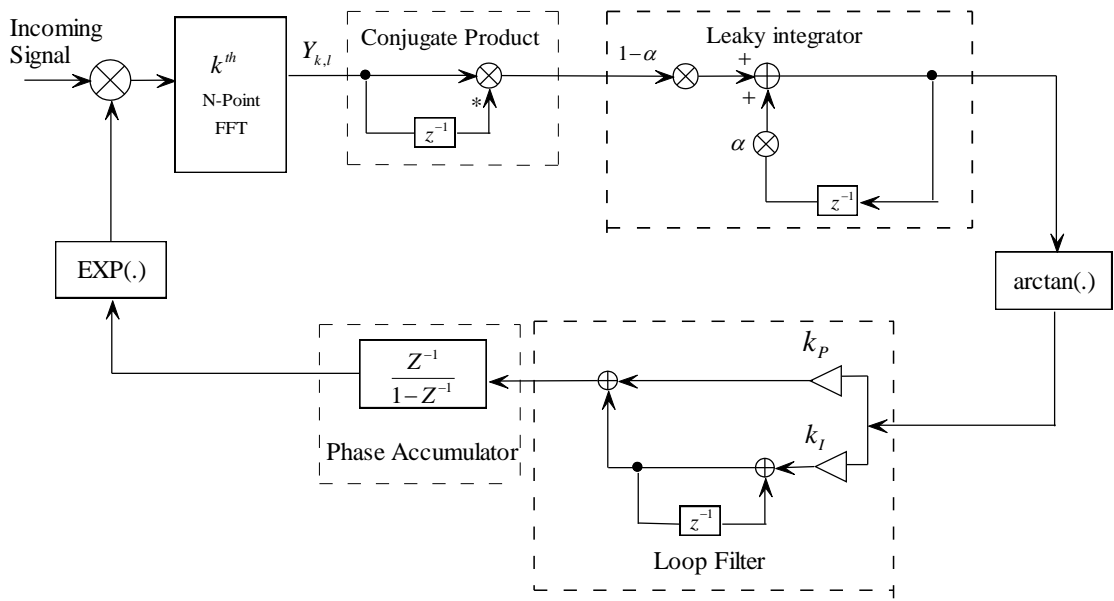

**Figure 4.12 : OFDM CFO estimation and correction block diagram [20].**

The phase locked loop (PLL) consists three components the phase detector (PD), the loop filter (LF) and the voltage controlled oscillator (VCO). In the CFO correction, the PD (or the CFO estimator) consists of three components the conjugate product block, the leaky integrator and the ATAN operator. The rest components of the PLL are the PI filter, the PI stands for "Proportional + Integral" action and the phase accumulator [20]. The leaky integrator is used for averaging to reduce the estimation variance. The parameter $\alpha$ is the forgetting factor of the leaky integrator, which is a design constant close to unity [20].

The phase transfer function of linearized discrete PLL shown in figure 4.12 in the absence of the phase noise $N(z)$ can be expressed as follows [20, 34]:

$$H(z) = \frac{\hat{\theta}(z)}{\theta_i(z)} = \frac{(k_P + k_I)\left(z - \frac{k_P}{k_P + k_I}\right)}{z^2 - 2\left(1 - \frac{k_P + k_I}{2}\right)z + (1 - k_P)} \quad (4.36)$$

The phase error transfer function is expressed as follows:

$$H_e(z) = \frac{\theta_e(z)}{\theta_i(z)} = \frac{z^2 - 2z + 1}{z^2 - 2\left(1 - \frac{k_P + k_I}{2}\right)z + (1 - k_P)} \quad (4.37)$$

The proportional gain $(k_p)$ and the integral gain $(k_I)$ of the loop filter of type-2 control loop coefficients can be computed via the following equations [20].

$$k_p = \frac{4\zeta\theta_n}{1 + 4\zeta\theta_n + \theta_n^2} \quad , \quad k_I = \frac{4\theta_n^2}{1 + 4\zeta\theta_n + \theta_n^2} \quad (4.38)$$

Where $\theta_n$ is the loop bandwidth and the parameter $\zeta$ is the damping factor.

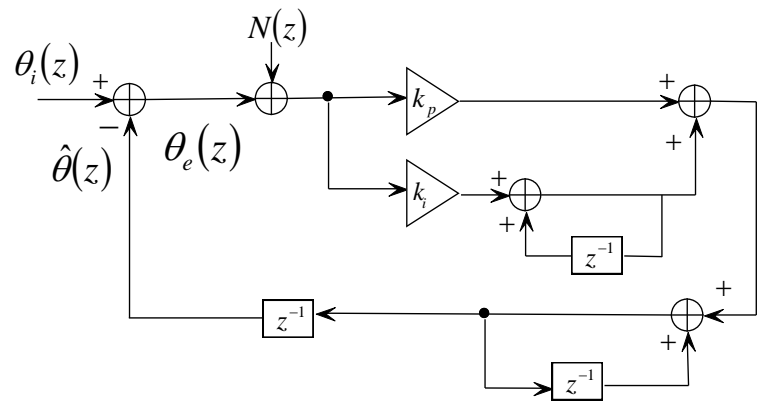

**Figure 4.13 : Linearized discrete PLL phase model.**

The CFO correction based on type-2 control loop can handle CFO larger than the type-1 control loop [34]. But the scheme [20] have large acquisition time of the PLL which make the system is slow. In the next sections, the methodology of a reduction of PLL acquisition time is introduced. The dual bandwidth scheme is used for rapid acquisition and good tracking PLL. The proposed dual bandwidth scheme is based on the switched capacitor of the loop filter of the PLL. The dual bandwidth scheme is improved using clustered pilot tones is proposed.

## 4.4 CFO estimation using conventional pilot tones

The CFO estimation using conventional pilot tones is based on a few of isolated pilot tones embedded in the received data stream. On each pilot tone, identical pilot symbols are transmitted for all OFDM symbols. The principle of these algorithms is that the frequency estimation problem can be converted into a phase estimation problem by considering the phase shift between subsequent subchannel samples $Y_{k,l}$ and $Y_{k,l-1}$, where the subscripts $k$ and $l$ denote the index of pilot tones and the index of OFDM symbol respectively. The subsequent subchannel samples $Y_{k,l}$ and $Y_{k,l-1}$ are given by [20, 31]:

$$Y_{k,l} = e^{j2\pi\varepsilon_f(G+(1+G)l)} \left[\Lambda_0 H_{k,l} Z_{k,l} + I_{k,l}\right] \quad (4.39)$$

$$Y_{k,l-1} = e^{j2\pi\varepsilon_f(G+(1+G)l)} \left[\Lambda_0 H_{k,l-1} Z_{k,l-1} + I_{k,l-1}\right] \quad (4.40)$$

Let $\emptyset = 2\pi\varepsilon_f(1+G)$, from equations (4.39) and (4.40), the received component $Y_{k,l}$ consists of a desired component due to $Z_{k,l}$ and another term due to the interferences $I_{k,l}$ where $I_{k,l} = \sum_{m=0, m \neq k}^{N_u-1} \Lambda_{m-k} H_{m,l} Z_{m,l}$ [14].

Assuming the $k^{th}$ subchannel is a pilot tone, the pilot symbols on the $k^{th}$ subchannel should have $Z_{k,1} = Z_{k,2} \ldots \ldots = Z_{k,L}$ [31]. Using the one pilot tone, an unbiased estimate of the phase is obtained from a block of $L$ OFDM symbols as follows [20, 31]:

$$\widehat{\emptyset}_k = tan^{-1} \left\{ \frac{Im\left(\frac{1}{L}\sum_{l=1}^{L} Y_{k,l-1}^* Y_{k,l}\right)}{Re\left(\frac{1}{L}\sum_{l=1}^{L} Y_{k,l-1}^* Y_{k,l}\right)} \right\} \quad (4.41)$$

Where $Re(.)$ and $Im(.)$ represent the real part and imaginary parts, respectively. The superscript * stands for conjugate. Subsequently, the normalized CFO can be obtained as follows [31]:

$$\hat{\varepsilon}_k = \frac{\hat{\varphi}_k}{2\pi(1+G)} \quad (4.42)$$

The performance of phase estimation is given by [31]:

$$var(\widehat{\emptyset}_k) = \frac{1}{(L-1)^2} \left[\frac{1}{SIR_k} + \frac{(L-1)}{2SIR_k^2}\right] \quad (4.43)$$

Where $var(.)$ represents the variance, $SIR_k$ denotes signal to interference ratio (SIR) of $Y_{k,l}$ and is derived from equation (4.39) as follows [31]:

$$SIR_k = \frac{|c_0|^2|Z_{k,l}|^2}{\sum_{m=1}^{N_u-1}|\Lambda_m|^2} \quad (4.44)$$

The variance of the phase estimate in equation (4.43) reveals that it is inversely proportional to the SIR of the pilot tone. Therefore, it is interest of designing a pilot tone with a high SIR, so that a more accurate CFO estimate can be derived by a clustered pilot tones design [31].

## 4.5 CFO estimation using clustered pilot tones

In 2007, Wei Zhang [31], a clustered pilot tones design is proposed. The pilot tones are organized into clusters. Every two pilot tones together as a group and these tone groups are equally spaced along the subcarrier axis. In each cluster, the pilot symbols are transmitted always antipodal to each other. The pilot symbol on the left side $(a)$ in each cluster is made antipodal to the one on the right $(-a)$. The idea is that the ICI of consecutive pilot tones can cancel each other out, thus gives a much better SIR. Let $S_f$, $S_r$ and $S$ denote the sets of the left, right and all pilot positions in one OFDM symbol respectively. In one pilot tone cluster, it has $Z_{k,l} = -Z_{k+1,l}$ for $k \in S_f$ and $(k+1) \in S_r$. The decoded value on the left pilot tone can be expressed as [31]:

$$Y_{k,l} = e^{j2\pi\varepsilon_f(G+(1+G)l)} \cdot \left[ (\Lambda_0 - \Lambda_1)Z_{k,l} + \sum_{m \in S_f, m \neq k} (\Lambda_{m-k} - \Lambda_{m-k+1})Z_{m,l} \right.$$
$$\left. + \sum_{m \notin S} \Lambda_{m-k} Z_{m,l} \right] \quad (4.45)$$

Also, the value on the right pilot tone can be written as

$$Y_{k+1,l} = e^{j2\pi\varepsilon_f(G+(1+G)l)} \cdot \left[ (\Lambda_{-1} - \Lambda_0)Z_{k,l} + \sum_{m \in S_r, m \neq k} (\Lambda_{m-k-1} - \Lambda_{m-k})Z_{m,l} \right.$$
$$\left. + \sum_{m \notin S} \Lambda_{m-k-1} Z_{m,l} \right] \quad (4.46)$$

Thus, the clustered pilot tones are subtracted in pairs and expressed as follows:

$$\mathcal{R}_{k,l} = Y_{k,l} - Y_{k+1,l} \quad (4.47)$$

Substituting from equations (4.45) and (4.46) into equation (4.47) thus $\mathcal{R}_{k,l}$ can be rewritten as follows [31]:

$$\mathcal{R}_{k,l} = e^{j2\pi\varepsilon_f(G+(1+G)l)} \cdot \left[(2\Lambda_0 - \Lambda_1 - \Lambda_{-1})Z_{k,l} + \sum_{m \notin S}(\Lambda_{m-k} - \Lambda_{m-k-1})Z_{m,l}\right] \quad (4.48)$$

From equation (4.48), find that $\mathcal{R}_{k,l}$ consists of the desired signal part and ICI part. The signal power of $\mathcal{R}_{k,l}$ is given by [31]:

$$\mathcal{P}_s = |(2\Lambda_0 - \Lambda_1 - \Lambda_{-1})|^2 |Z_{k,l}|^2 \quad (4.49)$$

And the ICI power of $\mathcal{R}_{k,l}$ is given by [31]:

$$\mathcal{P}_{ICI} = \sum_{m \notin S} |(\Lambda_m - \Lambda_{m-1})|^2 \quad (4.50)$$

By using equations (4.49) and (4.50), the SIR of the clustered pilot tones can be denoted by [31]:

$$SIR_k = \frac{|(2\Lambda_0 - \Lambda_1 - \Lambda_{-1})|^2 |Z_{k,l}|^2}{\sum_{m \notin S}|(\Lambda_m - \Lambda_{m-1})|^2} \quad (4.51)$$

Finally, the SIR is significantly improved by the clustered pilot tones than isolated pilot tones [31]. Using the one clustered pilot tone, the normalized CFO is obtained from a block of $L$ OFDM symbols as [20, 31]:

$$\hat{\varepsilon}_k = \left(\frac{1}{2\pi(1+G)}\right) tan^{-1} \left\{\frac{Im\left(\frac{1}{L}\sum_{l=1}^{L}\mathcal{R}^*_{k,l-1}\mathcal{R}_{k,l}\right)}{Re\left(\frac{1}{L}\sum_{l=1}^{L}\mathcal{R}^*_{k,l-1}\mathcal{R}_{k,l}\right)}\right\} \quad (4.52)$$

And its variance has the similar form with equation (4.43)

$$var(\hat{\emptyset}_k) = \frac{1}{(L-1)^2}\left(\frac{1}{SIR_k} + \frac{L-1}{2\,SIR_k^2}\right) \quad (4.53)$$

The CFO correction scheme is based on second order type-2 phase locked loop (PLL) [20]. The order of the PLL is defined as the highest order of the denominator of the loop transfer function. The type of the loop refers to the number of perfect integrators within the loop [44, 45].

## 4.6 Phase locked loop (PLL) basics

The PLL is a feedback control circuit. A PLL is a control loop which synchronizes its output signal (generated by VCO) with a reference or input signal in frequency as well as in phase. In the synchronized (often called locked) state the output frequency of a PLL is exactly same as the input signal and the phase error between the oscillator's output signal and the reference signal is zero, or remains constant. In the unlocked state, the PLL generates a control signal which is related to the phase error. This signal acts on the oscillator in such a way that the phase error is again reduced to a minimum. Thus, the main objective of the PLL always adjusts the phase of the output signal to lock to the phase of the reference signal [44]. A basic PLL block diagram is shown in figure 4.14.

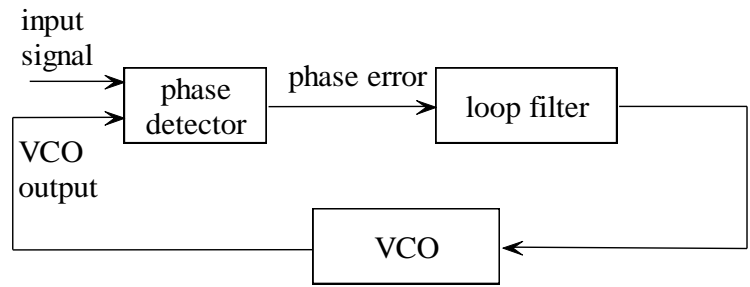

**Figure 4.14 : A basic PLL block diagram.**

The linearized PLL is shown in figure 4.15; the output of the phase detector $\theta_e(s)$ represents the difference between the phase of the input signal $\theta_i(s)$ and the phase of the VCO output $\hat{\theta}(s)$ [45].

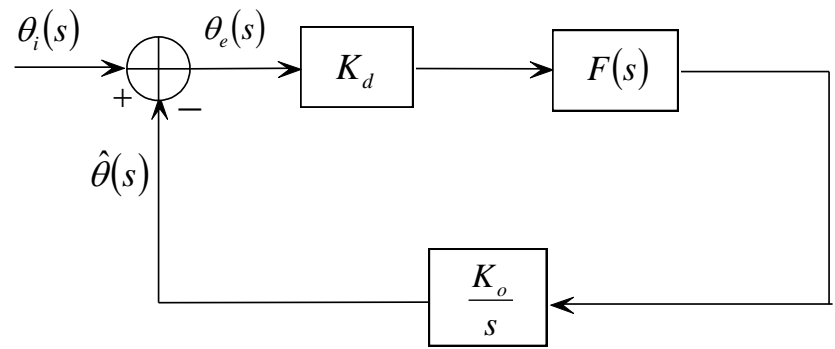

**Figure 4.15 : Linearized phase locked loop.**

The basic transfer function of PLL is given by:

$$H(s) \equiv \frac{\hat{\theta}(s)}{\theta_i(s)} = \frac{K_d K_o F(s)}{s + K_d K_o F(s)} \tag{4.54}$$

The order of the PLL is defined as the highest order of $s$ in the denominator of the loop transfer function. The type of the loop refers to the number of perfect integrators in the loop. A PLL has an implicit perfect integrator with the VCO, so the first order loop is a first order, type1 loop. A filter $F(s)$ with a perfect integrator would yield a type-2 loop [44, 45]. The second order type-2 PLL is obtained by using the active low pass filter (or active PI filter) as shown in figure 4.16. The transfer function of the PI filter is given by [44, 43] is expressed as follows:

$$F(s) = \frac{\tau_2}{\tau_1} + \frac{1}{\tau_1 s} \tag{4.55}$$

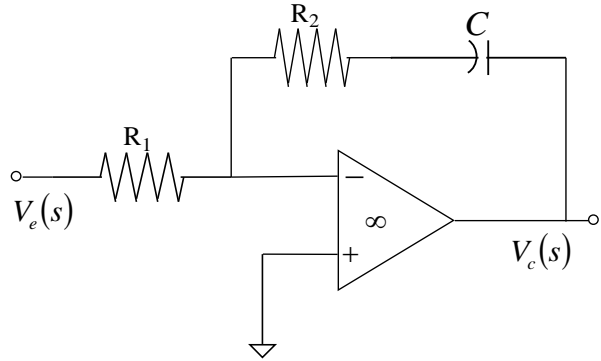

**Figure 4.16 : active PI filter.**

Where $\tau_1 = R_1 C$ and $\tau_2 = R_2 C$. The active PI filter in figure 4.16 is specified by another configuration. This configuration is proportional plus integral configuration as shown in figure 4.17.

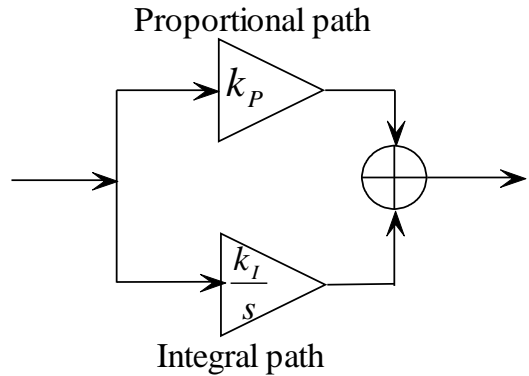

**Figure 4.17 : proportional plus integral configuration [45].**

The transfer function of the PI configuration shown in figure 5.4 is given by:

$$F(s) = K_P + \frac{K_I}{s} \qquad (4.56)$$

Where $K_P$ is the gain coefficient of the proportional path and $K_I$ is the coefficient of the integral path of the loop filter. The two configurations in figures 4.16 and 4.17 are electrically equivalent if $K_P = \tau_2/\tau_1$ and $K_I = 1/\tau_1$ [45]. Thus, the phase transfer function for the second order type-2 PLL is given by:

$$H(s) = \frac{K_d K_o (s\tau_2 + 1)/\tau_1}{s^2 + s\, K_d K_o \tau_2/\tau_1 + K_d K_o/\tau_1} \qquad (4.57)$$

Also, the loop parameters are given by:

$$\omega_n = \sqrt{\frac{K_d K_o}{R_1 C}} \qquad (4.58)$$

$$\zeta = \frac{\omega_n R_2 C}{2} \qquad (4.59)$$

The phase transfer function of the second order PLL is performing a low-pass filter operation for input phase signals $\theta_i(t)$. Thus, the second order PLL tracks the input phase that is within the loop bandwidth and fails to track phase that is outside the loop bandwidth. Thus, input phase within the loop bandwidth is transferred to the VCO's phase output $\hat{\theta}(s)$, but input phase outside the loop bandwidth is attenuated [44, 45]. There are two key parameters of a PLL system are its tracking and acquisition ranges. They can be defined as follows:

Tracking range $(B_T)$: is the range of frequencies over which the PLL can maintain lock with an input signal (i.e. phase lock). The tracking range is given by [46].

$$B_T = f_{max} - f_{min} \qquad (4.60)$$

Where $f_{max}$ and $f_{min}$ are the maximum and minimum frequencies over which phase lock can be maintained. The factors limiting the tracking range include the maximum frequency deviation of the VCO and the dc voltage range of the PD output. The tracking range is independent on the loop filter because of when the PLL is in phase lock the difference frequency $f_i - f_o$ almost zero [46].

Acquisition range $(B_{ac})$: is the range of frequencies over which the PLL can acquire lock with an input signal as long as phase lock does not exist i.e. $f_i \neq f_o$. Once the phase lock has been acquired, the VCO output frequency $f_o$ can track or follow the input frequency $f_i$ over the entire tracking range. The Acquisition range is always within the tracking range. The Acquisition range of the PLL is given by [46]:

$$B_{ac} = f_2 - f_1 \qquad (4.61)$$

Where $f_1$ and $f_2$ are the lowest and the highest frequency which the PLL can lock onto [46]. The acquisition range is also known as the capture range or pull-in range [44]. The acquisition and tracking ranges of the PLL can be illustrated in figure 4.18.

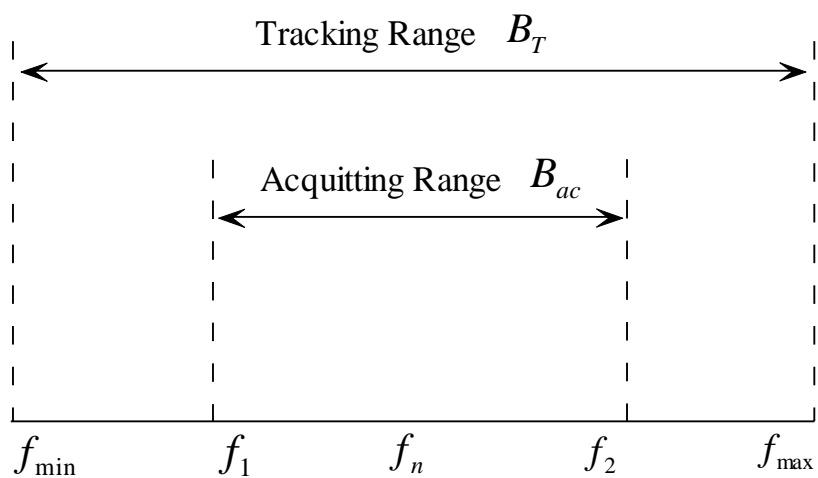

**Figure 4.18 : Tracking and acquisition range.**

If the bandwidth of the loop filter is greater than the difference frequency $f_i - f_n$ thus, the phase lock can be achieved. But if the bandwidth of the loop filter is less than the difference frequency $f_i - f_n$ thus, the phase lock cannot be achieved as shown in figure 4.19.

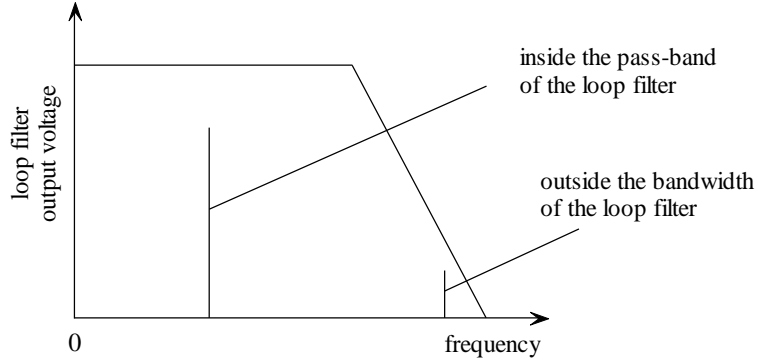

**Figure 4.19: Transfer function of the loop filter.**

Noise bandwidth ($B_L$) of the second order type-2 PLL can be calculated as expressed as follows [44, 45]:

$$B_L = \frac{\omega_n}{2}(\zeta + 1/4\zeta) \tag{4.62}$$

The acquisition time is given by [44]

$$T_p = \frac{(\Delta\omega_0)^2(\zeta + 1/4\zeta)^3}{16{B_L}^3} \tag{4.63}$$

Where $\Delta\omega_0$ is initial frequency offset, thus from equations (4.62) and (4.63) we find that the acquisition time can be decreased via increasing the noise bandwidth ($B_L$). A PLL has the ability to suppress both the noise superimposed on the input signal and noise generated by the VCO. The narrower bandwidth of the PLL can achieve more effective jitter filtering of the PLL. However, the error of the VCO frequency cannot be reduced rapidly. Although a narrow bandwidth is better for rejecting large amounts of input noise. Furthermore, the frequency acquisition is slow and impractical. So, there is a tradeoff between jitter filtering and fast acquisition. Therefore, there are aided frequency-acquisition techniques to solve this problem, such as the variable-bandwidth methods [45].

## 4.7 Proposed dual bandwidth scheme

The proposed dual bandwidth scheme can be built to have a large bandwidth for rapid acquisition and a much narrower bandwidth for good tracking in the presence of noise. The signal to command switching of bandwidth can be the lock indication voltage from the quadrature phase detector. When the loop is out of lock, the output voltage $Q'$ of the lock indication is zero and sets the switches to their wideband position. When the loop locks, the output voltage $Q'$ of the lock indication appears and sets the switches into their narrowband position.

The lock indication is a quadrature phase detector which has two inputs: the received input signal and the VCO output with 90° phase shift. The main phase detector has an output voltage proportional to $\sin(\theta_e)$ and quadrature output proportional to $\cos(\theta_e)$. In the locked condition $\theta_e$ small, so $\cos(\theta_e) \approx 1$. In the unlocked state of the PLL, the outputs from both phase detectors are beat notes at the difference frequency, so the DC output is almost zero. Thus, the filtered output of the quadrature detector gives indication of lock [44, 45]. The proposed dual bandwidth scheme is realized by switching capacitor of the loop filter as shown in figure 4.20.

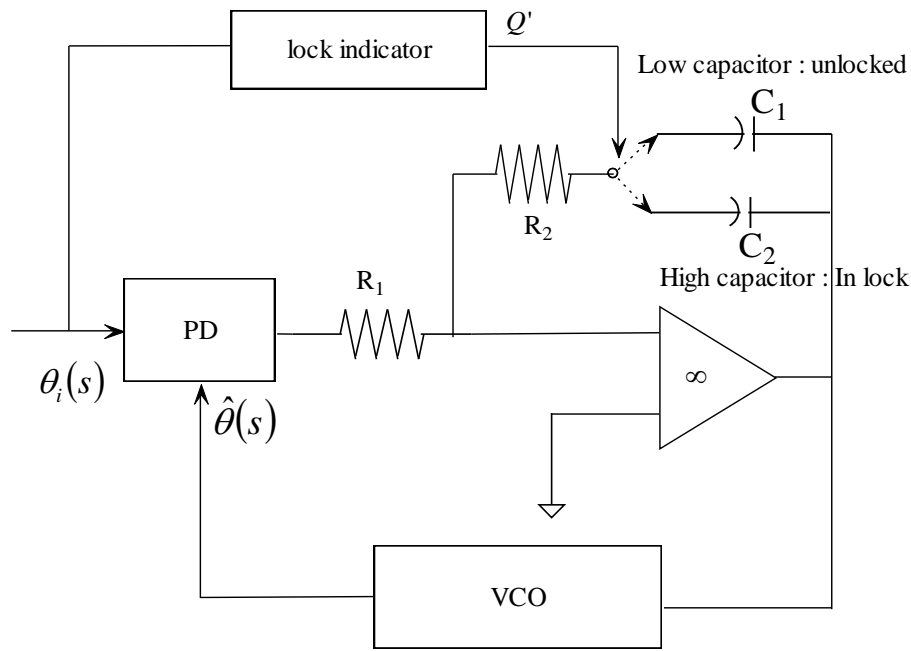

**Figure 4.20 : PLL block diagram using switching capacitors [44].**

The CFO correction based on type-2 control loop [20] for OFDM systems can be improved via the proposed dual bandwidth scheme. Thus, the linearized discrete PLL in figure 4.13 can be modified using dual bandwidth scheme as shown in figure 4.21. The loop filter parameters can be modified and expressed as follows:

$$k_p = \frac{4\zeta\theta_{nj}}{1 + 4\zeta\,\theta_{nj} + \theta_{nj}^2}\,, \qquad k_I = \frac{4\theta_{nj}^2}{1 + 4\zeta\,\theta_{nj} + \theta_{nj}^2}\,, \qquad j = 1,2 \qquad (4.64)$$

Where $\theta_{nj}$ denotes the variable loop bandwidth. If $j = 1$, the switches set into $k_{p\,wide}$ and $k_{I\,wide}$ but if $j = 2$, the switches set into $k_{p\,narrow}$ and $k_{I\,narrow}$. Note that we will use the same phase and phase error transfer function in equations (4.51) and (4.52) in two different cases of bandwidth.

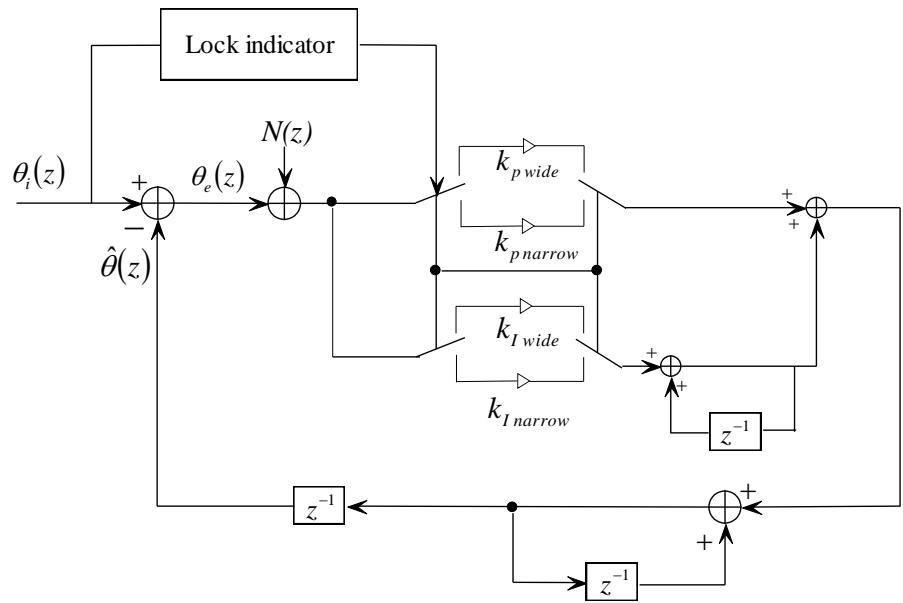

**Figure 4.21 : Linearized discrete PLL using dual bandwidth scheme.**

Now the linearized discrete PLL using dual bandwidth scheme is used to propose FFT and CFO estimation and correction for OFDM system as shown in figure 4.22.

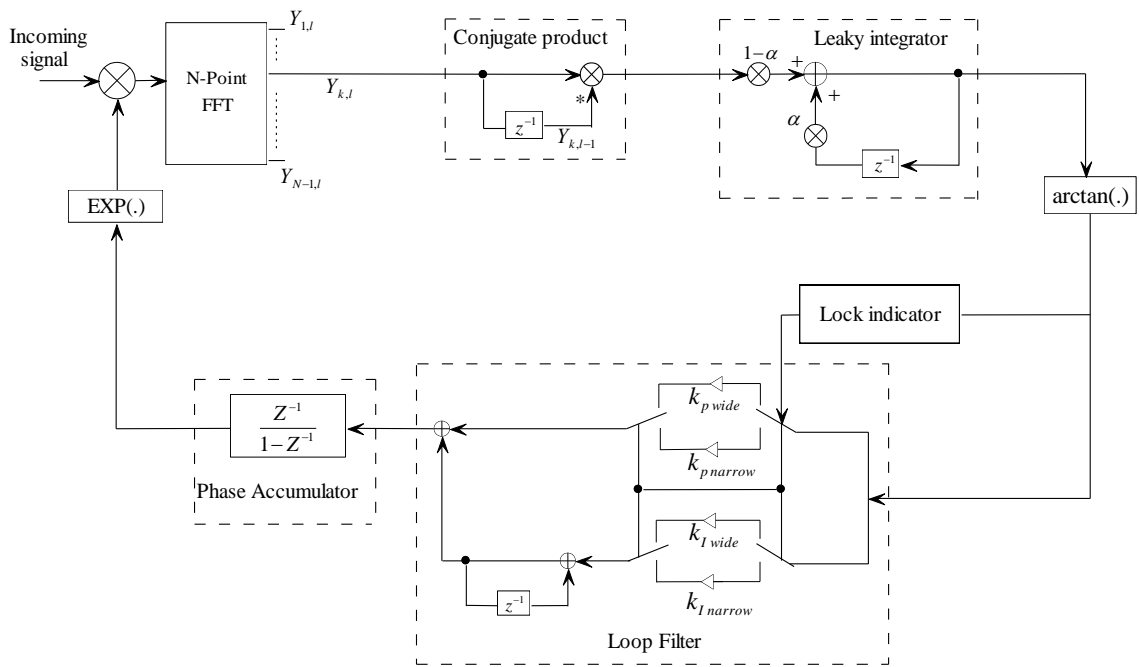

**Figure 4.22 : OFDM CFO estimation and correction using dual bandwidth scheme.**

Finally, the flow chart for dual bandwidth scheme is designed as shown figure 4.23.

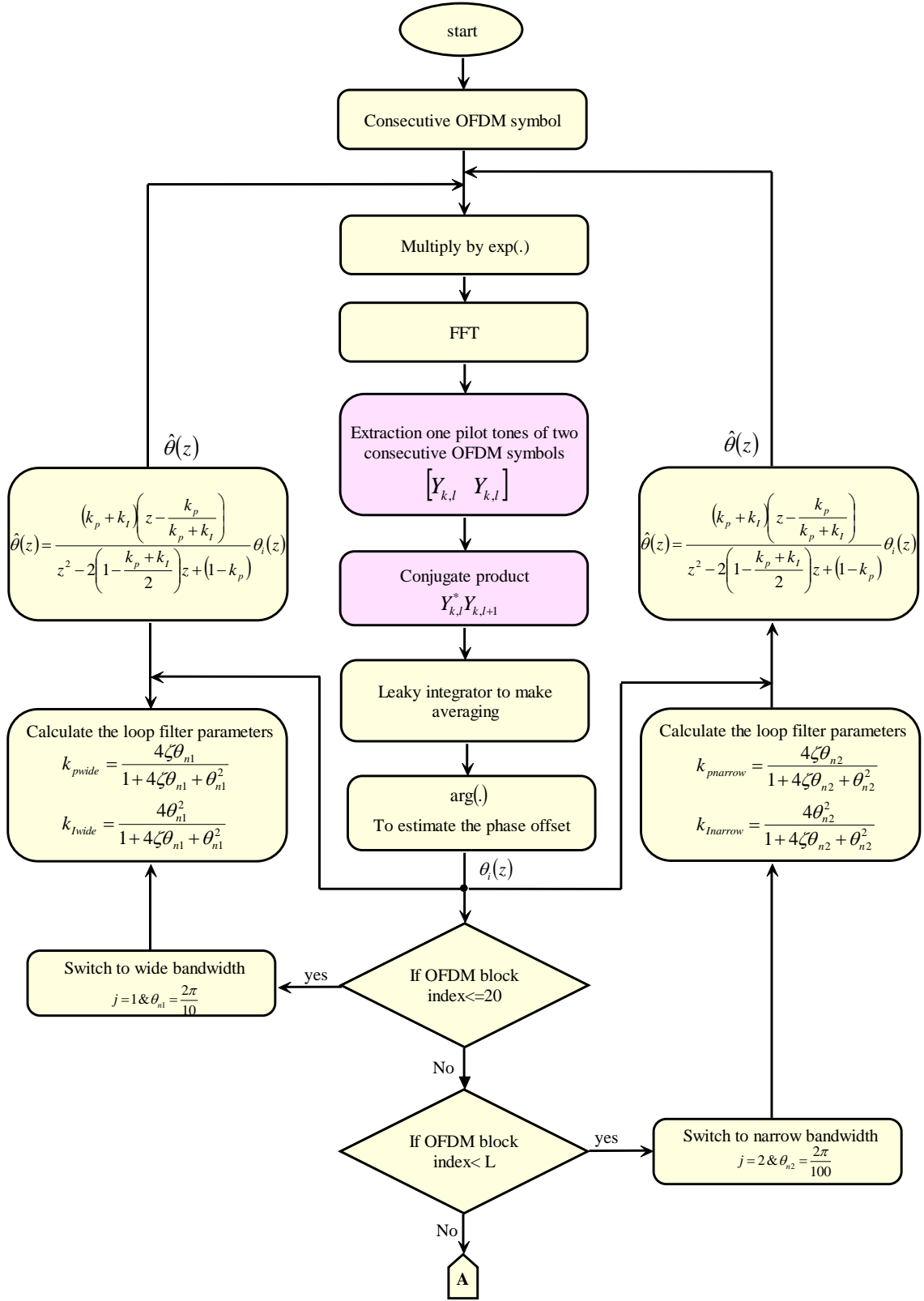

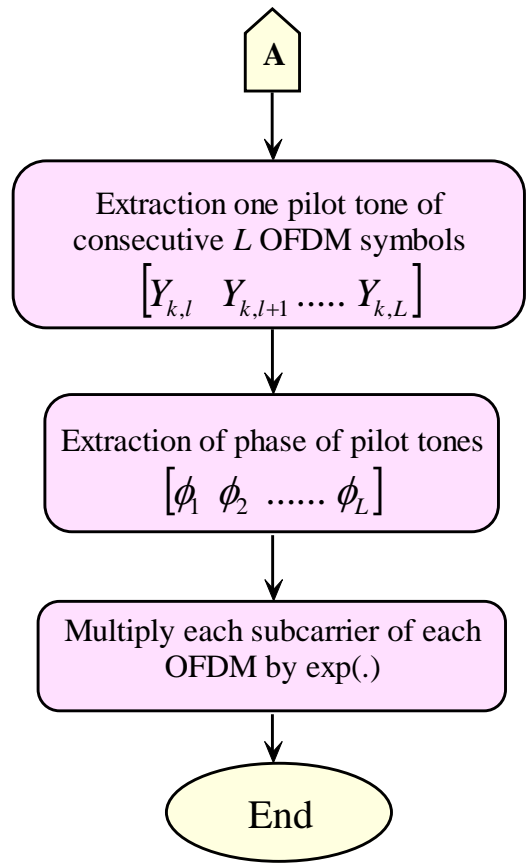

**Figure 4.23 : Flow chart for dual bandwidth scheme.**

## 4.8 Improved dual bandwidth scheme

The dual bandwidth scheme [47] is improved via enhancement the CFO estimator (i.e. PD of the PLL) by using clustered pilot tones instead of conventional pilot tones. The pilot tones are organized into clusters. In each cluster, the adjacent pilot symbols to be transmitted are always set to be antipodal $a$ is transmitted at the left pilot tone and $-a$ at the right pilot tone in each cluster [31]. The CFO estimation and correction using dual bandwidth and clustered pilot tones scheme for OFDM systems is realized as shown in figure 4.24.

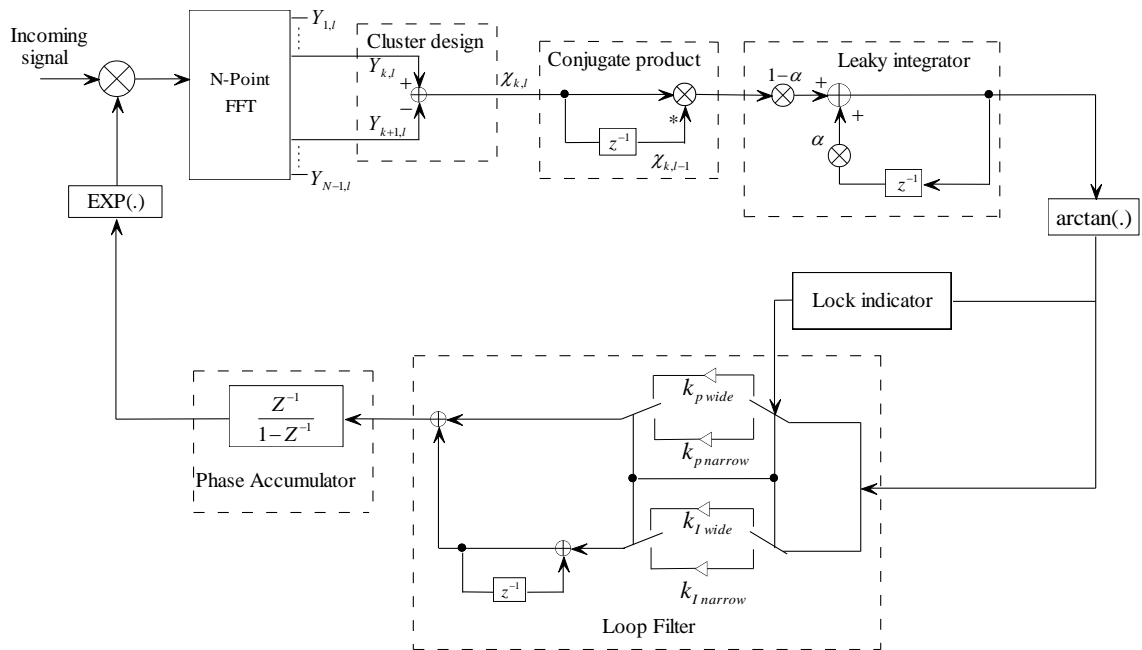

**Figure 4.24 : OFDM CFO estimation and correction using dual bandwidth scheme and clustered pilot tones.**

The flow chart for improved dual bandwidth using clustered pilot tones scheme is designed as shown in figure 4.25.

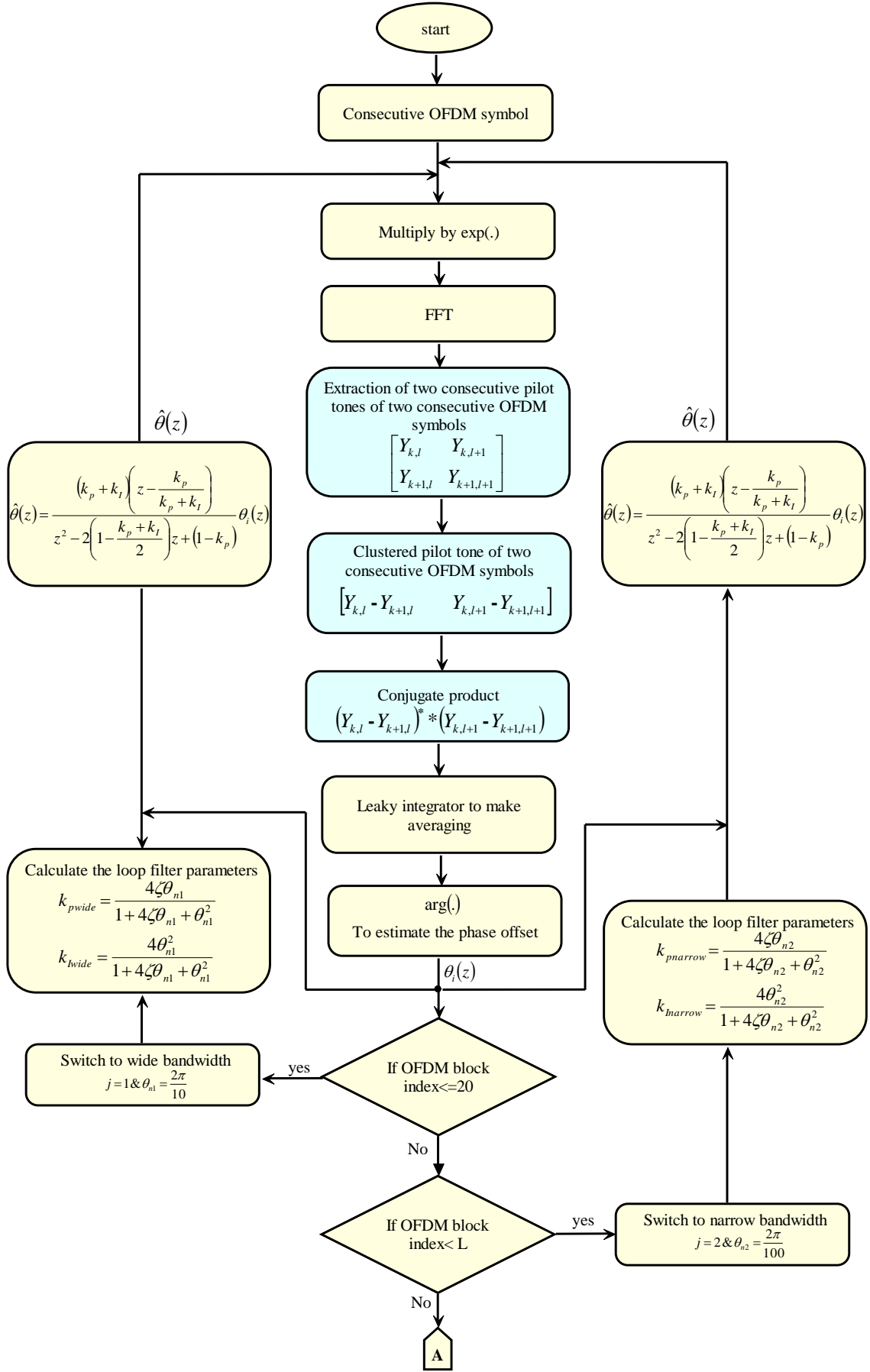

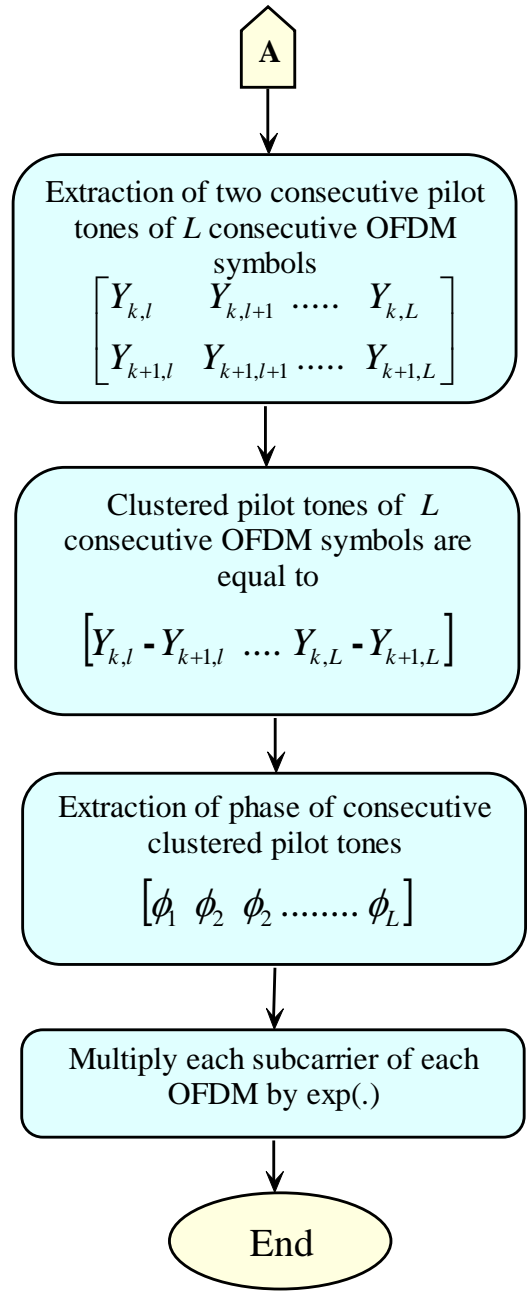

**Figure 4.25 : Flow chart for dual bandwidth using clustered pilot tones scheme.**

Finally, the proposed dual bandwidth scheme for CFO correction based on type-2 control loop becomes faster than scheme [20] which improves the BER performance. Improved dual bandwidth scheme using clustered pilot tones improves the BER performance compared to the dual bandwidth scheme [47]. In the next chapter, we will simulate and investigate the proposed dual bandwidth scheme and the improved dual bandwidth scheme using clustered pilot tones for CFO correction in OFDM systems for different modulation types over Rayleigh multipath fading channel.

# Chapter 5. Numerical Results

## 5.1 Introduction

In this chapter the performance of OFDM for different types of modulation schemes in both AWGN and multipath Rayleigh fading channels is evaluated. The performance of the proposed dual bandwidth scheme for CFO estimation and correction is evaluated. The proposed dual bandwidth scheme is assessed via computer simulations using different modulation schemes over multipath Rayleigh fading channels.

## 5.2 OFDM system parameters

The WLAN OFDM system parameters utilized in the simulation are listed in table 5-1 [48].

Table 5-1: WLAN OFDM system parameters [47].

| Parameter | Value |
| --- | --- |
| System bandwidth ($BW$) | 20MHz |
| Number of data subcarriers | 48 |
| Number of pilot subcarriers | 4 |
| Total number of subcarriers | 52 |
| IFFT/FFT size ($N_u$) | 64 |
| Subcarrier frequency spacing ($\Delta f$) | $312.5 kHz\ (BW/N_u)$ |
| IFFT/FFT period ($T_u$) | $3.2 \mu s\ (1/\Delta f)$ |
| GI duration ($T_g$) | $0.8 \mu s\ (T_u/4)$ |
| Symbol duration ($T_{sym}$) | $4\ \mu s$ |
| Carrier frequency offset ($\delta f$) | $125\ kHz$ |
| Relative frequency offset | $\varepsilon = 125 * 1000/312.5 * 10\text{^}3 = 0.4$ |

## 5.3 Offset free OFDM system performance

The BER performance versus $E_b/N_0$ for offset free OFDM system was investigated for quadrature phase shift keying (QPSK) and 16-quadrature amplitude modulation (16QAM).

### 5.3.1 OFDM system performance over AWGN channel

Figure 5.1 shows the BER versus $E_b/N_0$ for OFDM system with BPSK, QPSK, 16QAM and 64QAM modulation schemes over AWGN channel.

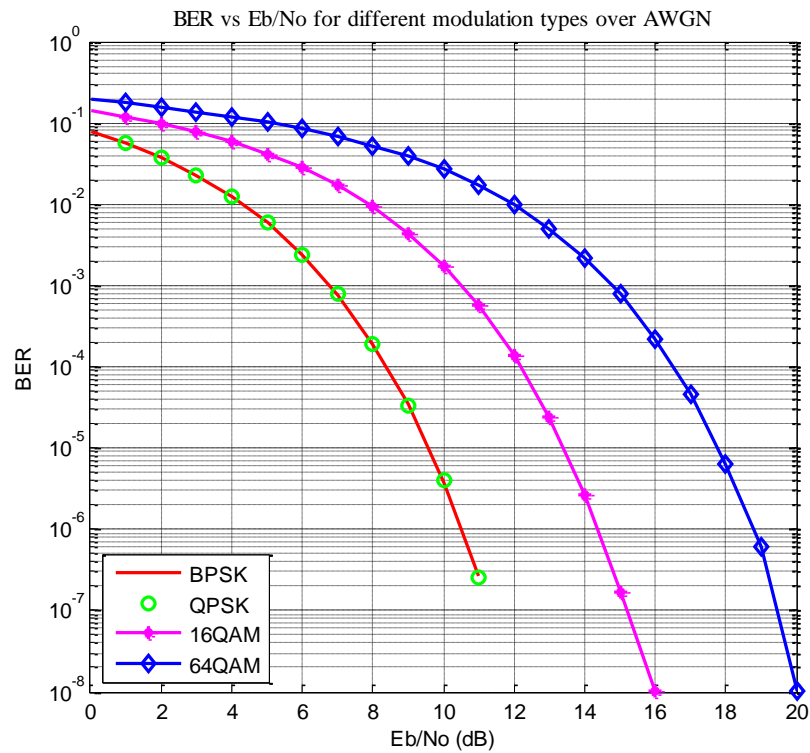

**Figure 5.1 : BER versus $E_b/N_0$ for OFDM system with different types of modulations over AWGN channel.**

Table 5-2 shows $E_b/N_0$ for BPSK, QPSK, 16QAM and 64QAM modulation schemes over AWGN channel at BER=$10^{-3}$.

**Table 5-2: $E_b/N_0$ for different modulation schemes over AWGN channel.**

| Modulation scheme | $E_b/N_0$ dB |
|---|---|
| BPSK | 7 |
| QPSK | 7 |
| 16QAM | 11 |
| 64QAM | 15 |

### 5.3.2 OFDM system performance over multipath Rayleigh fading channel

Figure 5.2 shows the BER versus $E_b/N_0$ for OFDM system with BPSK, QPSK, 16QAM and 64QAM modulation schemes over multipath Rayleigh fading channel with channel delay spread $0.5 \mu sec$.

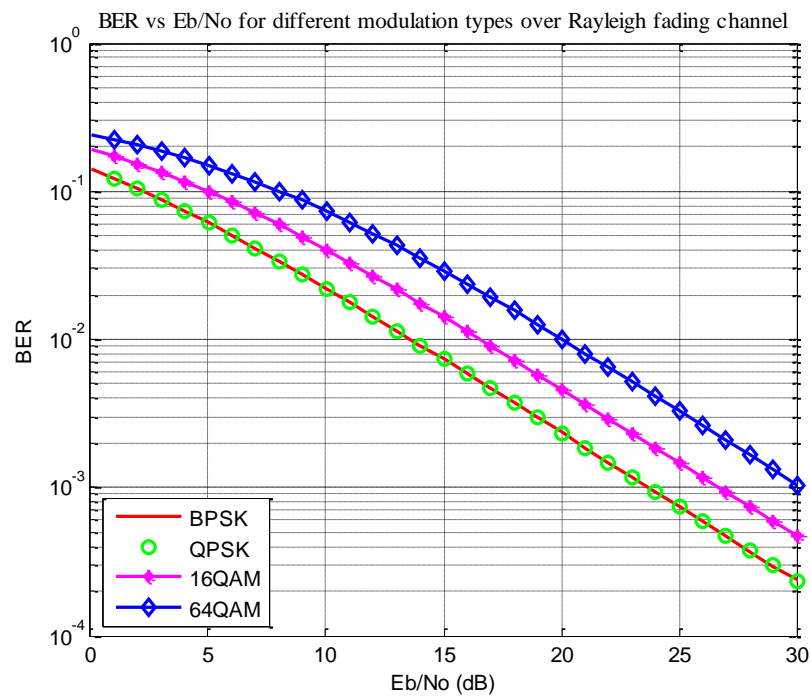

**Figure 5.2 : BER versus $E_b/N_0$ for OFDM systems with different types of modulations over multipath Rayleigh fading channel.**

Table 5-3 shows $E_b/N_0$ for BPSK, QPSK, 16QAM and 64QAM modulation schemes over multipath Rayleigh fading channel at BER=$10^{-3}$.

**Table 5-3: $E_b/N_0$ for different modulation schemes over multipath Rayleigh fading channel.**

| Modulation scheme | $E_b/N_0$ dB |
|---|---|
| BPSK | 24 |
| QPSK | 24 |
| 16QAM | 27 |
| 64QAM | 30 |

We conclude that, the BPSK has the lowest BER while the 64QAM has highest BER than other modulation schemes. The BPSK and QPSK modulation schemes have the same performance. The 16QAM and 64QAM modulation schemes require higher power as compared to BPSK and QPSK modulation schemes to get the same BER performance.

### 5.3.3 OFDM system performance with and without CP

Figures 5.3 and 5.4 show the BER versus $E_b/N_0$ for OFDM systems with QPSK and 16QAM modulation schemes respectively without and with CP over AWGN channel.

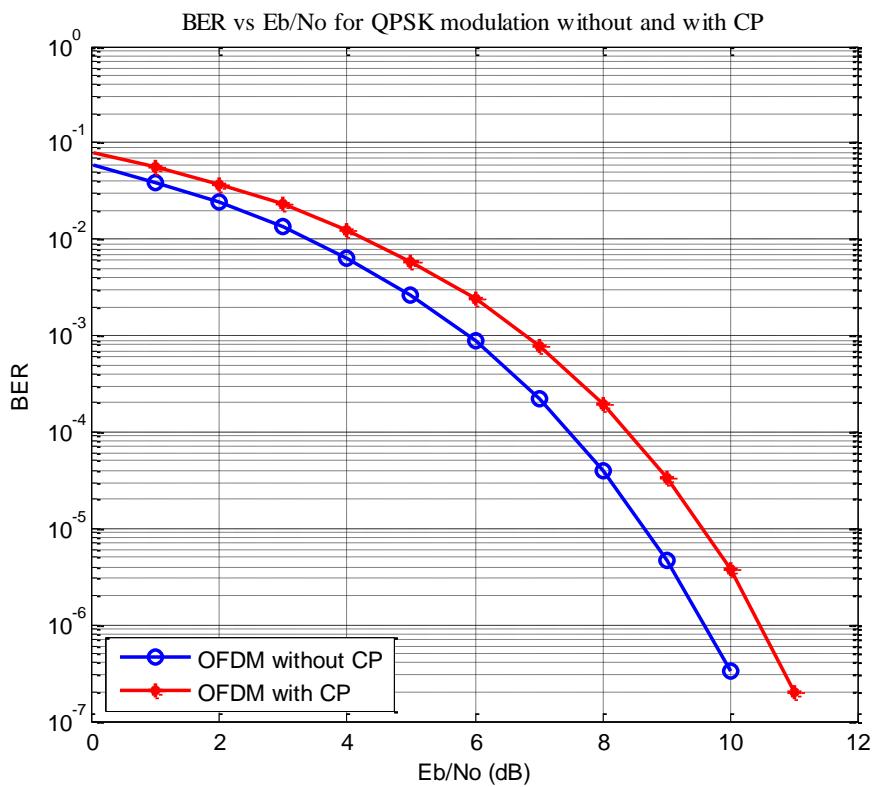

**Figure 5.3 : BER versus $E_b/N_0$ for OFDM systems without and with CP for QPSK over AWGN channel.**

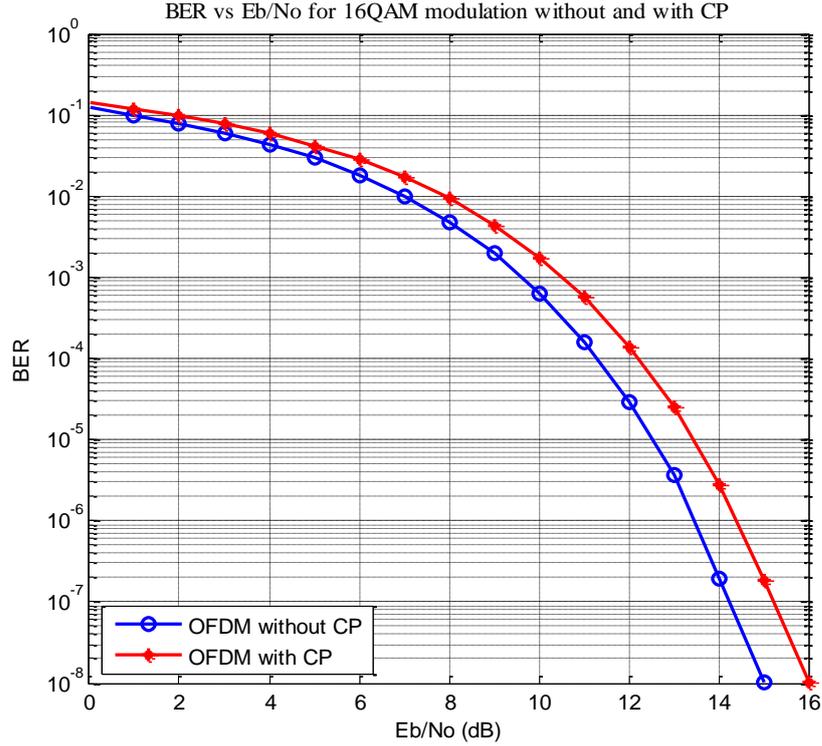

**Figure 5.4 : BER versus $E_b/N_0$ for OFDM systems without and with CP for 16QAM over AWGN channel.**

Table 5-4 shows $E_b/N_0$ for QPSK and 16QAM modulation schemes for OFDM system with and without CP over AWGN channel at BER=$10^{-3}$.

**Table 5-4 : $E_b/N_0$ for different modulation schemes for OFDM system with and without CP over AWGN channel.**

| Modulation scheme | $E_b/N_0$ dB | |
|---|---|---|
| | Without CP | With CP |
| QPSK | 6 | 7 |
| 16QAM | 10 | 11 |

We conclude that the OFDM system with CP requires higher power than the OFDM system without CP. The $SNR_{loss}$ due to the CP can be calculated theoretically and by simulation for $G = 1/4$ in OFDM system. From Theoretically analysis the loss in SNR due to the CP in OFDM system is calculated as follows:

$$SNR_{loss} = -10\log_{10}\left(1 - \frac{T_g}{T_{sym}}\right) = 10\log_{10}(1 + G) = 0.969 \text{ dB} \approx 1\text{dB}$$

From simulation analysis the loss in SNR due to the CP in OFDM system is equal $SNR_{loss} = 1\text{dB}$.

### 5.3.4 OFDM system performance for different CP length

Figures 5.5 and 5.6 show the BER versus $E_b/N_0$ for OFDM systems with QPSK and 16QAM modulation schemes respectively for different CP length over multipath Rayleigh fading channel.

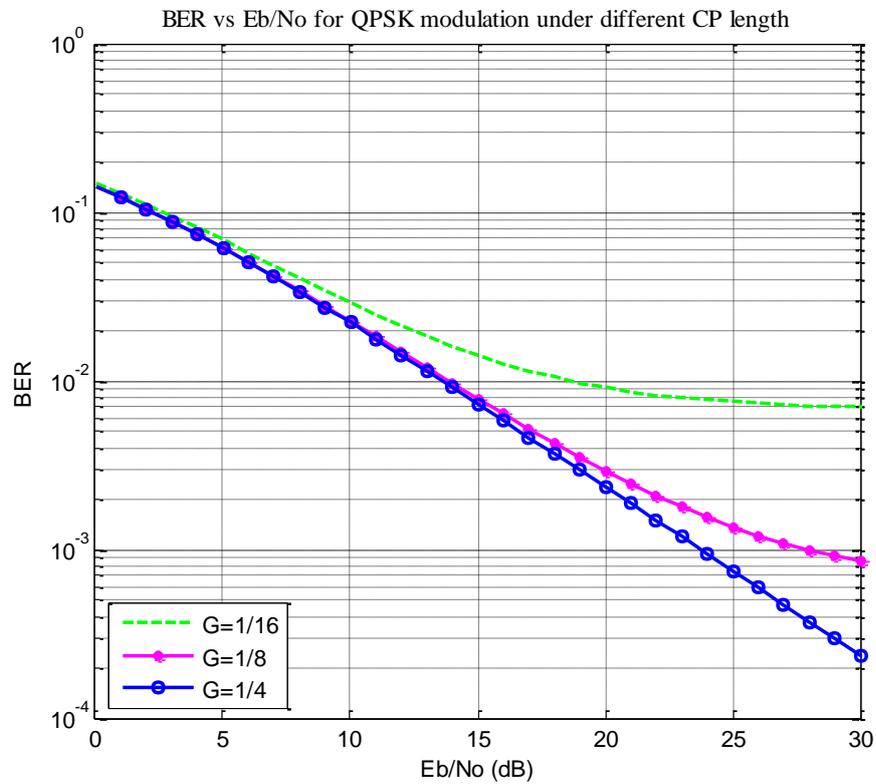

**Figure 5.5 : BER versus $E_b/N_0$ for OFDM system for different CP length for QPSK modulation over multipath Rayleigh fading channel.**

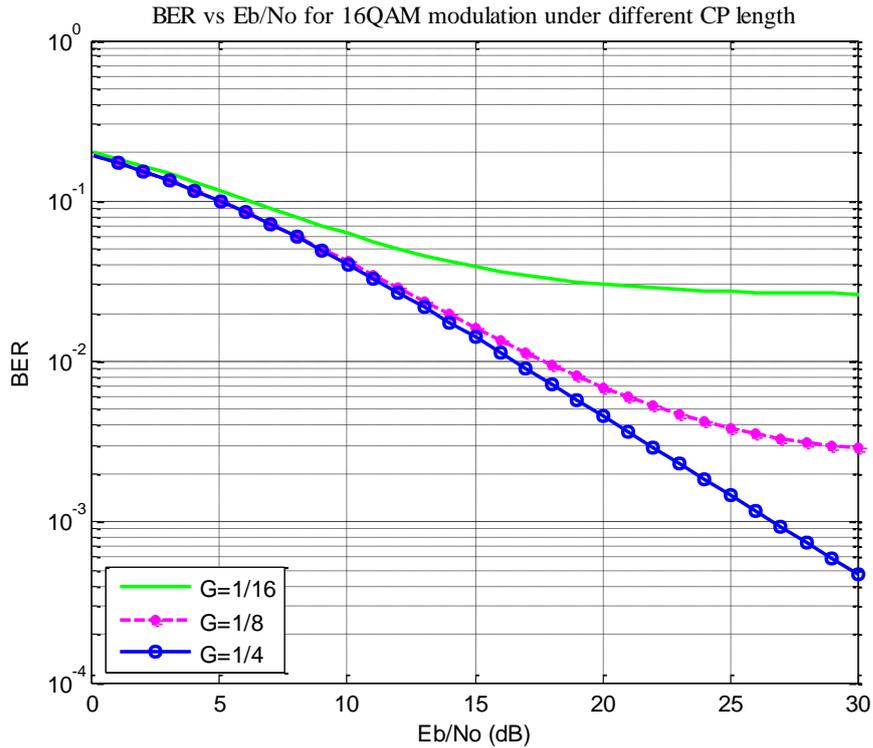

**Figure 5.6 : BER versus $E_b/N_0$ for OFDM system for different CP length for 16QAM modulation over multipath Rayleigh fading channel.**

Table 5-5 shows the BER for QPSK and 16QAM modulation schemes for OFDM system for different CP length over multipath Rayleigh fading channel at $E_b/N_0 = 25\ dB$.

**Table 5-5: BER performance for different modulation schemes for OFDM system for different CP length over multipath Rayleigh fading channel.**

| Modulation scheme | BER | | |
| --- | --- | --- | --- |
| | $G = 1/16$ | $G = 1/8$ | $G = 1/4$ |
| QPSK | 0.007502 | 0.001355 | 0.000748 |
| 16QAM | 0.02733 | 0.003861 | 0.001478 |

We conclude that if the CP length is increased than channel delay spread the BER is decreased. But increased the CP length reduces the throughput. If the CP length is decreased than channel delay spread the BER is increased due to ISI.

### 5.3.5 OFDM system performance for different channel delay

Table 5-6 shows the mean delay spread for different kinds of environment [49].

**Table 5-6 : The mean delay spread for different kinds of environment [49].**

| Type of environment | Delay spread |
|---|---|
| Rural | 0.5 |
| Suburban | 1.5 |
| Urban | 3 |
| Dense urban | 4.5 |

Figures 5.7 and 5.8 show the BER versus $E_b/N_0$ for OFDM systems with QPSK and 16QAM modulation schemes respectively under different channel delay spread over multipath Rayleigh fading channel.

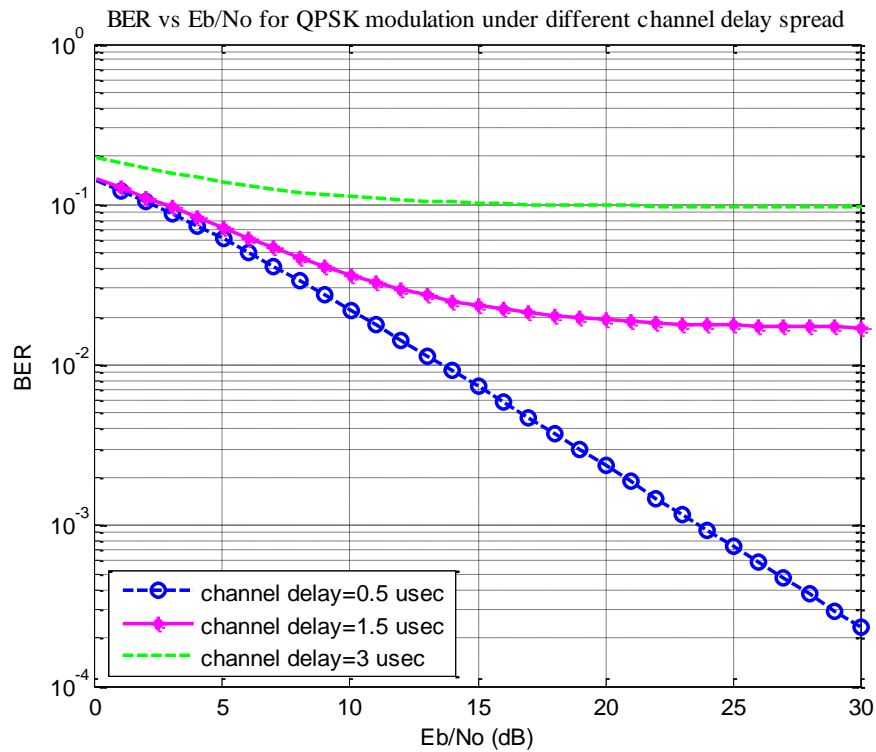

**Figure 5.7 : BER versus $E_b/N_0$ for OFDM system under different channel delay spread for QPSK modulation over multipath Rayleigh fading channel.**

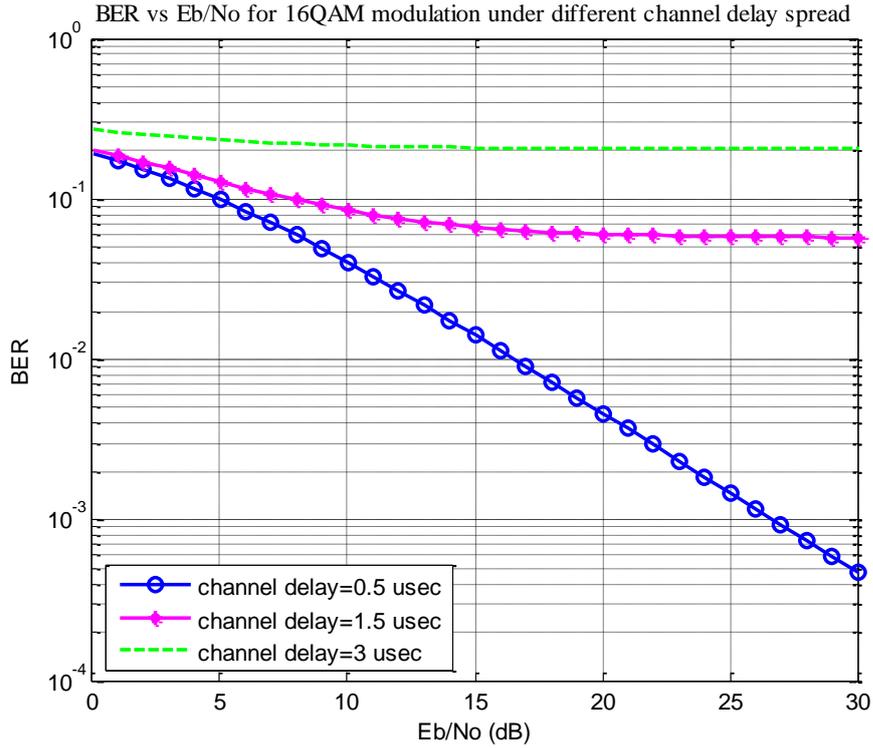

**Figure 5.8 : BER versus $E_b/N_0$ for OFDM system under different channel delay spread for 16QAM modulation over multipath Rayleigh fading channel.**

Table 5-7 shows the BER for QPSK and 16QAM modulation schemes for OFDM system under different channel delay spread over multipath Rayleigh fading channel at $E_b/N_0 = 30\ dB$.

**Table 5-7 : BER performance for different modulation schemes for OFDM system under different channel delay spread over multipath Rayleigh fading channel.**

| Modulation | channel delay spread $\tau_{max}\ \mu sec$ | | |
|---|---|---|---|
| BER | 0.5 | 1.5 | 3 |
| QPSK | 0.0002355 | 0.0172 | 0.09722 |
| 16QAM | 0.0004702 | 0.05765 | 0.2054 |

We conclude that if the CP length is greater than or equal to channel delay spread the transmitted OFDM symbols exposed to flat fading channel as shown in the case of $\tau_{max} = 0.5\ \mu sec$. If the CP length is less than channel delay spread the transmitted OFDM symbols exposed to frequency selective fading and the BER is increased as shown in the case $\tau_{max} = 1.5\ \mu sec$ and $3\ \mu sec$.

### 5.3.6 OFDM system performance for different FFT size

Figures 5.9 and 5.10 show the BER versus $E_b/N_0$ for OFDM systems with QPSK and 16QAM modulation schemes respectively for different FFT size but for the same bandwidth $(BW) = 20MHz$, $G = 1/4$ and over multipath Rayleigh fading channel with channel delay spread $\tau_{max} = 4.5\ \mu sec$.

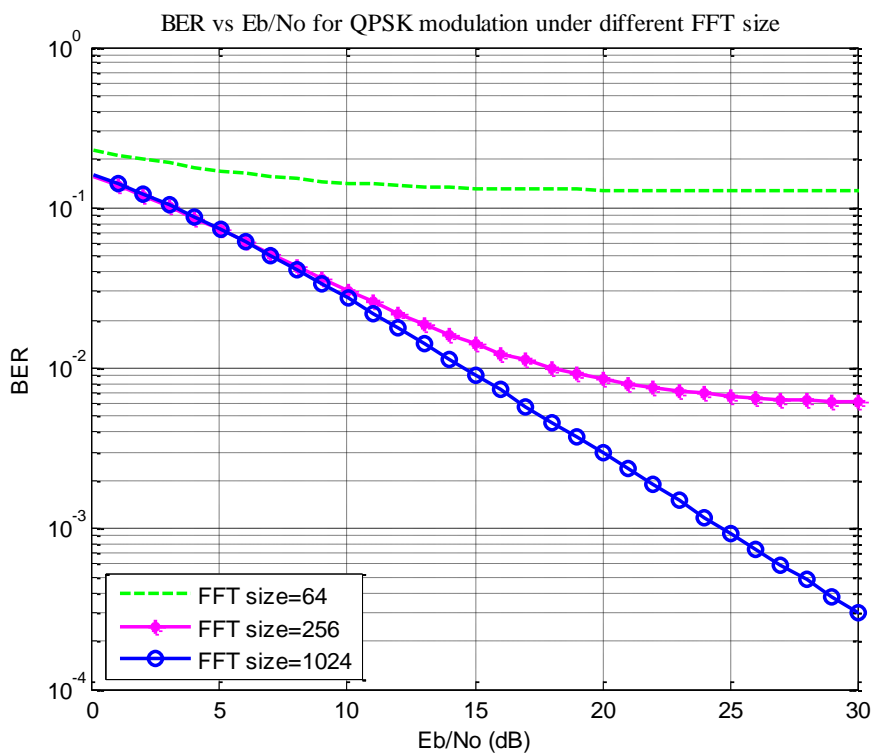

**Figure 5.9 : BER versus $E_b/N_0$ for OFDM system for different FFT size for QPSK over multipath Rayleigh fading channel with $\tau_{max} = 4.5\ \mu sec$.**

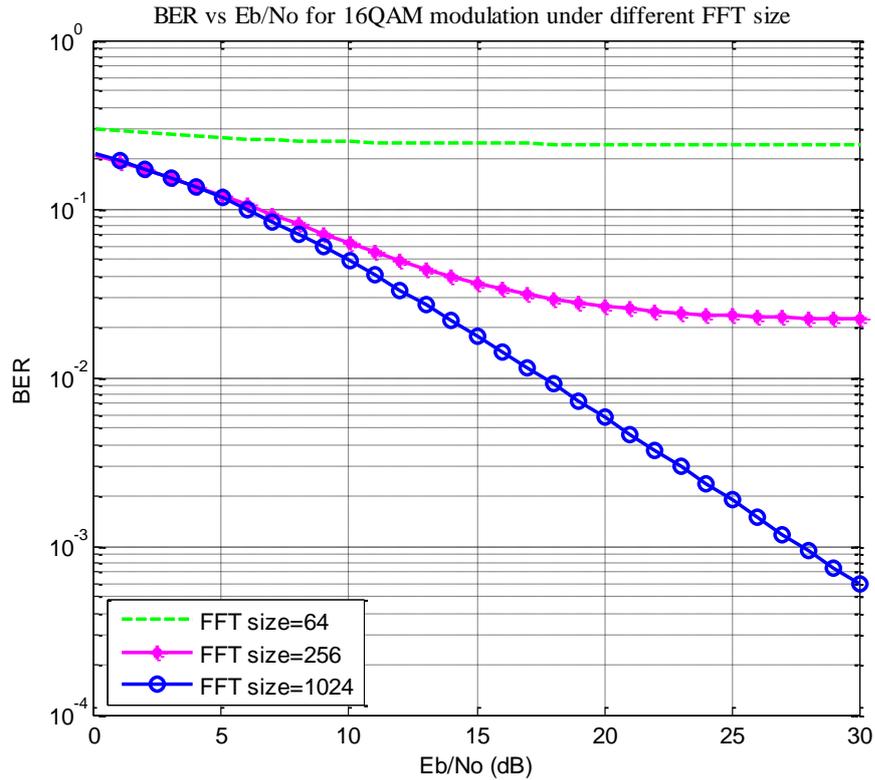

**Figure 5.10 : BER versus $E_b/N_0$ for OFDM system for different FFT size for 16QAM over multipath Rayleigh fading channel with $\tau_{max} = 4.5 \mu sec$.**

We conclude that for $G = 1/4$, the guard interval $(T_g)$ for $N_u = 64$, $N_u = 256$ and $N_u = 1024$ is equal to $0.8~\mu se$, $3.2~\mu sec$ and $12.8~\mu sec$ respectively. In the case of $N_u = 64$ and $N_u = 256$, the CP length is less than the channel delay spread thus, the system is exposed to frequency selective fading and ISI occurs. So, the OFDM system performance is degraded and the BER is increased. In the case of $N_u = 1024$, the CP length becomes greater than channel delay spread thus; the system is exposed to frequency flat fading. Thus, the BER performance of OFDM systems is decreased. Finally, the OFDM systems with large FFT size become more immune to large channel delay spread. But the large FFT size makes the OFDM systems more are susceptible to CFO.

## 5.4 Proposed schemes

In this section, the proposed dual bandwidth scheme and proposed dual bandwidth scheme using clustered pilot tones for CFO estimation and correction are simulated. A typical OFDM system based on WLAN [48] is considered. Specifically, each frame contains 280 OFDM symbols for 16QAM modulation technique. Training symbols are added at the beginning of the frame for coarse CFO estimation and channel estimation as specified in [48], and pilot tones used for fine estimation. The parameters used in the simulation OFDM symbol length 64 among them, 48 are information carrying subcarriers, 4 pilots. Furthermore, assume a multipath Rayleigh fading channel corrupted by AWGN and normalized CFO is 0.4; the channel is assumed to be unchanged during one frame and perfectly estimated at the receiver.

### 5.4.1 Proposed dual bandwidth scheme for CFO correction

The dual bandwidth scheme uses one pilot tone to estimate and correct the frequency offset assuming the pilot tone has a fixed value $Z_{m,l} = +1 \ \forall \ l$ [20].

### 5.4.1.1 The time response for dual bandwidth scheme

The figure 5.11 shows the phase output and the phase error using dual bandwidth scheme compared to [20] over multipath Rayleigh fading channel for QPSK modulation scheme.

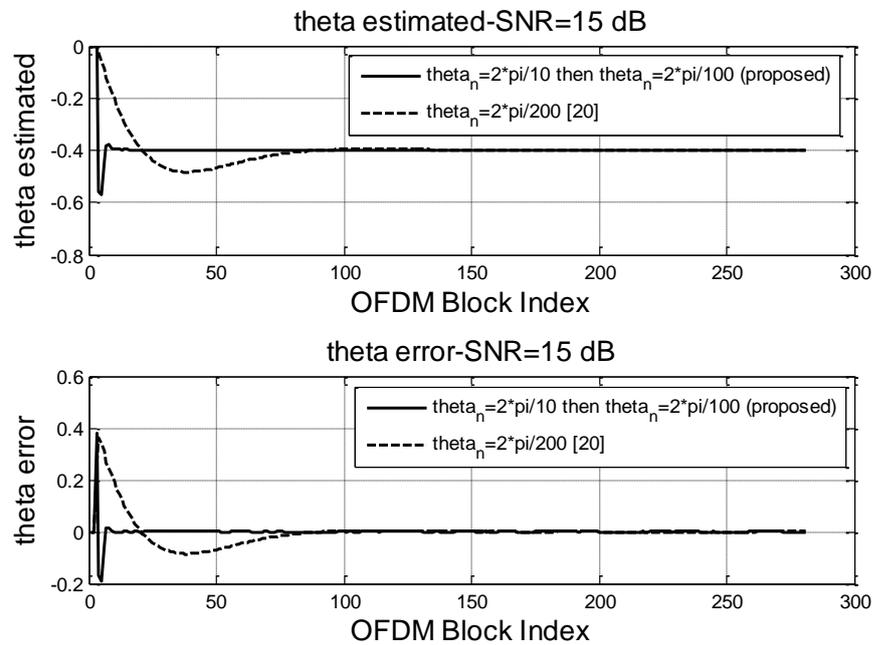

**Figure 5.11: The phase output and phase error using dual bandwidth scheme compared to [20] over multipath Rayleigh fading channel for QPSK modulation scheme.**

Table 5-8 shows OFDM symbol index versus the phase output $(\hat{\theta})$ using dual bandwidth scheme compared to [20] for QPSK modulation scheme.

**Table 5-8 : OFDM symbol index versus the phase output $(\hat{\theta})$ using dual bandwidth scheme compared to [20].**

| OFDM symbol index | phase output $(\hat{\theta})$ for proposed | phase output $(\hat{\theta})$ for [20] |
|---|---|---|
| 1 | 0 | 0 |
| 2 | 0 | 0 |
| 3 | 0 | 0 |
| 4 | -0.5563 | -0.02485 |
| 5 | -0.572 | -0.05788 |
| 6 | -0.438 | -0.09063 |
| 7 | -0.3813 | -0.1217 |
| 8 | -0.3794 | -0.15 |
| 9 | -0.3917 | -0.1774 |
| 10 | -0.3971 | -0.2036 |
| 11 | -0.396 | -0.2283 |
| 12 | -0.3967 | -0.2513 |

Figure 5.12 shows the phase output and phase error using dual bandwidth scheme compared to [20] over multipath Rayleigh fading channel for 16QAM modulation scheme.

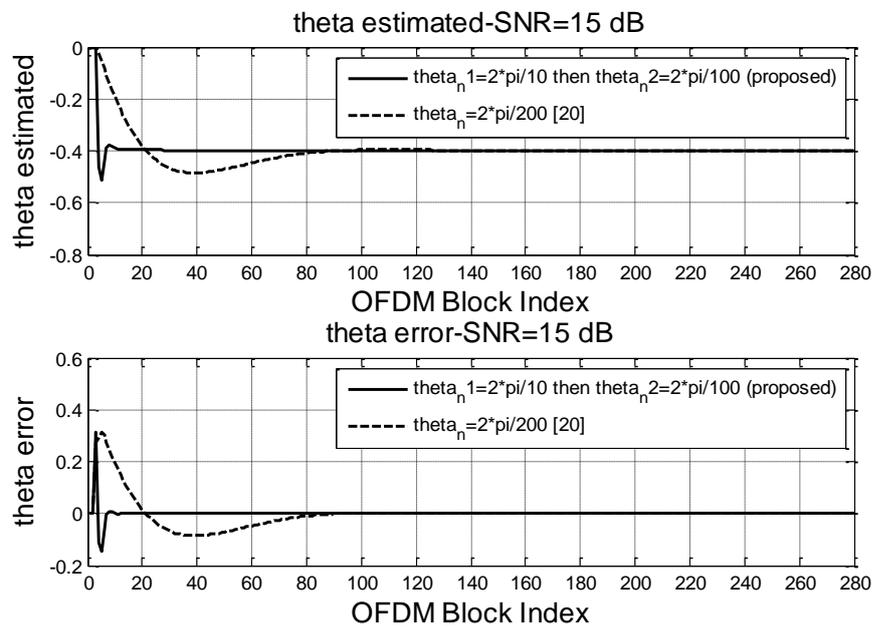

**Figure 5.12 : The phase output and the phase error using dual bandwidth scheme compared to [20] over multipath Rayleigh fading channel for 16QAM modulation scheme.**

Table 5-9 shows the OFDM symbol index versus phase output $(\hat{\theta})$ using dual bandwidth scheme compared to [20] for 16QAM modulation scheme.

**Table 5-9 : OFDM symbol index versus the phase output estimate $(\hat{\theta})$ using dual bandwidth scheme compared to [20].**

| OFDM symbol index | phase output $(\hat{\theta})$ for proposed | phase output $(\hat{\theta})$ for [20] |
|---|---|---|
| 1 | 0 | 0 |
| 2 | 0 | 0 |
| 3 | 0 | 0 |
| 4 | -0.4638 | -0.02422 |
| 5 | -0.5138 | -0.05123 |
| 6 | -0.4357 | -0.08097 |
| 7 | -0.3875 | -0.1109 |
| 8 | -0.377 | -0.1398 |
| 9 | -0.3816 | -0.1669 |
| 10 | -0.3885 | -0.1927 |
| 11 | -0.392 | -0.2174 |
| 12 | -0.3928 | -0.2408 |

The Figures 5.11 and 5.12 illustrate the dual bandwidth scheme phase accumulator output and phase error output compared to [20] for QPSK and 16QAM modulation respectively.

We conclude that, the acquisition time of dual bandwidth scheme may be achieved around twenty OFDM symbols (i.e. the acquisition time of dual bandwidth scheme is achieved around $64\ \mu sec$). But the acquisition time of [20] is achieved around ninety OFDM symbols (i.e. the acquisition time of [20] is achieved around $288\ \mu sec$). The dual bandwidth scheme is significantly faster than [20]. Thus; the enhancement of dual bandwidth scheme is 77.78 percent than [20]. Moreover, for larger $\theta_{n1}$ parameter, the acquisition time may be decreased again but large oscillations significantly occur resulting in performance degradation. The acquisition time of dual bandwidth scheme is robust to multipath Rayleigh fading disturbances compared to [20].

### 5.4.1.2 The constellation diagram for dual bandwidth scheme

Figure 5.13 shows the constellation diagram of QPSK modulation scheme: (a) offset free, (b) with CFO $\varepsilon = 0.4$.

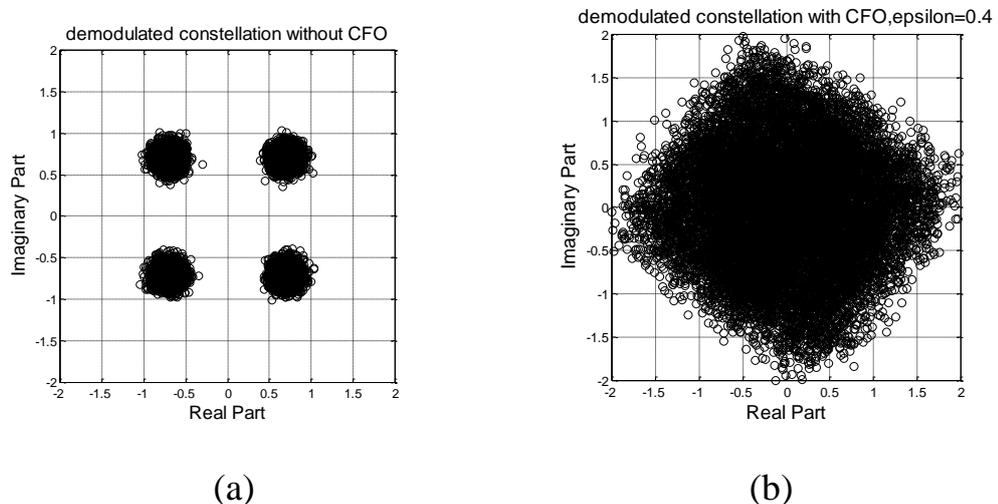

(a)                                                            (b)

**Figure 5.13 : Constellation diagram of QPSK modulation scheme:
(a) offset free, (b) with CFO $\varepsilon = 0.4$.**

Figure 5.14 shows the constellation diagram of QPSK modulation scheme after PLL: (a) using [20], (b) using dual bandwidth scheme.

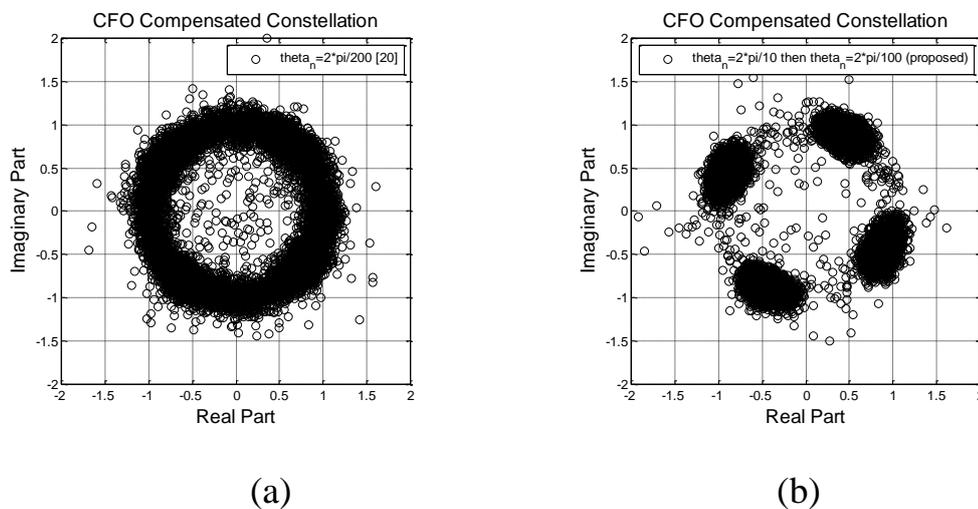

(a)                                                            (b)

**Figure 5.14 : The constellation diagram of QPSK modulation scheme after PLL:
(a) using [20], (b) using dual bandwidth scheme.**

Figure 5.15 shows constellation diagram of QPSK modulation scheme after phase offset correction: (a) using [20], (b) using dual bandwidth scheme.

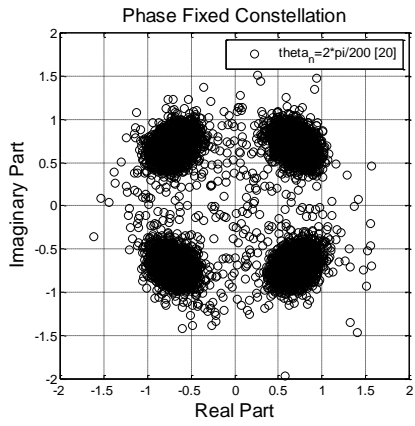 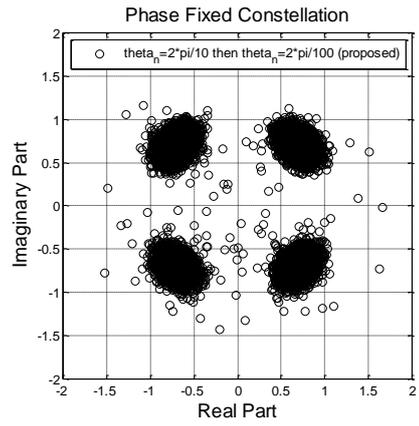

(a)                                      (b)

**Figure 5.15 : The constellation diagram of QPSK modulation scheme after phase offset correction: (a) using [20], (b) using dual bandwidth scheme.**

Figure 5.16 shows the constellation diagram of 16QAM modulation scheme: (a) offset free, (b) with CFO $\varepsilon = 0.4$.

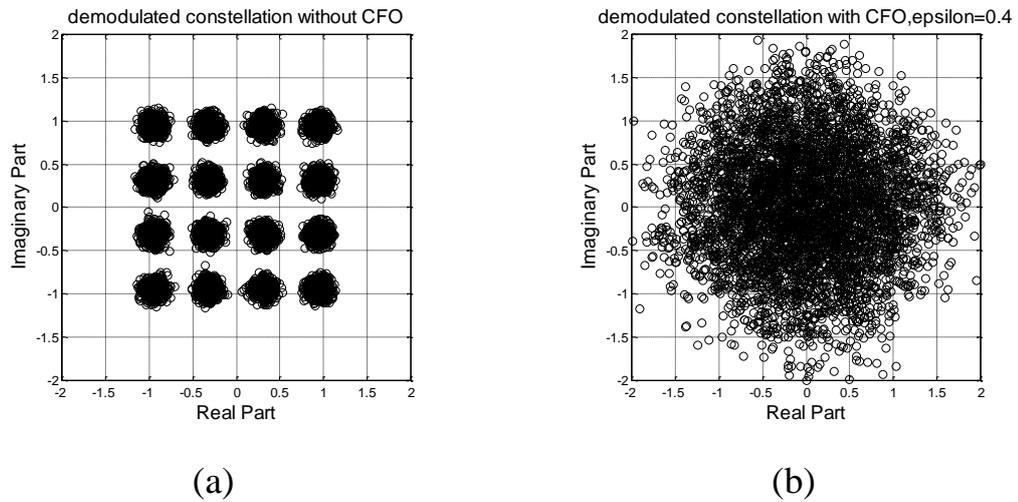

**Figure 5.16 : Constellation diagram of 16QAM modulation scheme:
(a) offset free, (b) with CFO $\varepsilon = 0.4$.**

Figure 5.17 shows the constellation diagram of 16QAM modulation scheme after PLL: (a) using [20], (b) using dual bandwidth scheme.

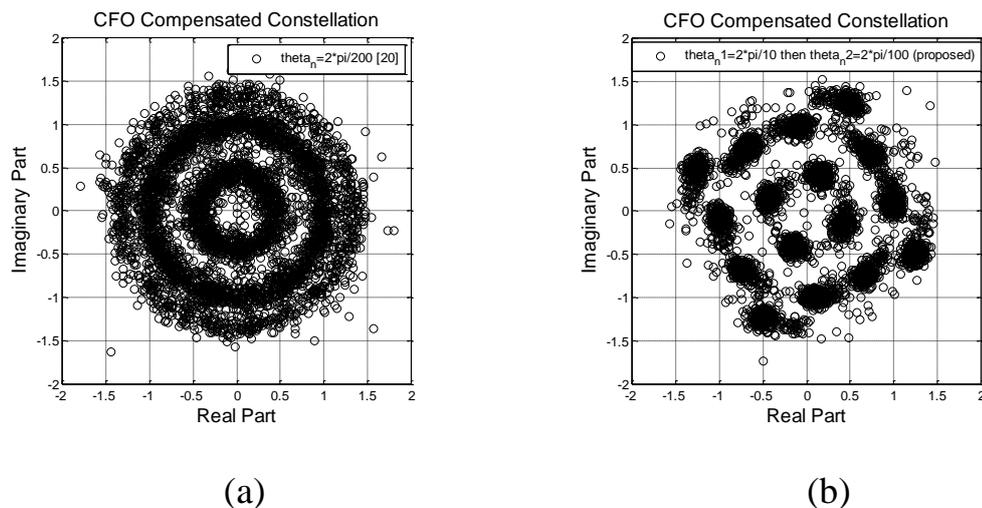

**Figure 5.17 : The constellation diagram of 16QAM modulation scheme after PLL:
(a) using [20], (b) using dual bandwidth scheme.**

Figure 5.18 shows the constellation diagram of 16QAM modulation after phase offset correction: (a) using [20], (b) using dual bandwidth scheme.

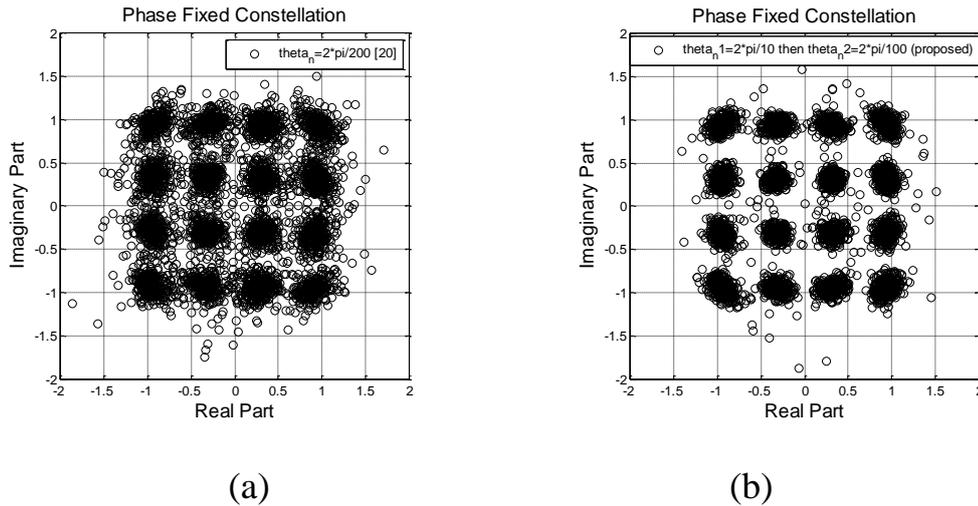

(a)                          (b)

**Figure 5.18 : The constellation diagram of 16QAM modulation scheme after phase offset correction: (a) using [20], (b) using dual bandwidth scheme.**

We conclude that the CFO causes the spin for all the subcarriers as shown in figures 5.13 (b) and 5.16 (b) where the CFO leads to the spin of all the subcarriers for QPSK and 16QAM modulation schemes respectively. The figures 5.14 and 5.17 depict the steady state constellation diagram after PLL using [20] and dual bandwidth scheme. The ICI has been significantly reduced using dual bandwidth scheme compared to [20] but all subcarriers from different OFDM symbols may have a phase offset. This residual phase offset can be corrected by computing the offset angle of the pilot tone and complex rotate each subcarrier with that offset angle using dual bandwidth scheme and [20] for QPSK and 16QAM modulation schemes respectively as shown in figures 5.15 and 5.18. Finally, we find that the noise decreasing significantly using the dual bandwidth scheme. We conclude that, the performance of the dual bandwidth scheme is better than [20].

### 5.4.1.3 The BER performance for dual bandwidth scheme

Figure 5.19 shows the BER versus $E_b/N_0$ for OFDM system using dual bandwidth scheme compared to [20] and an offset free system for QPSK modulation scheme over multipath Rayleigh fading channel.

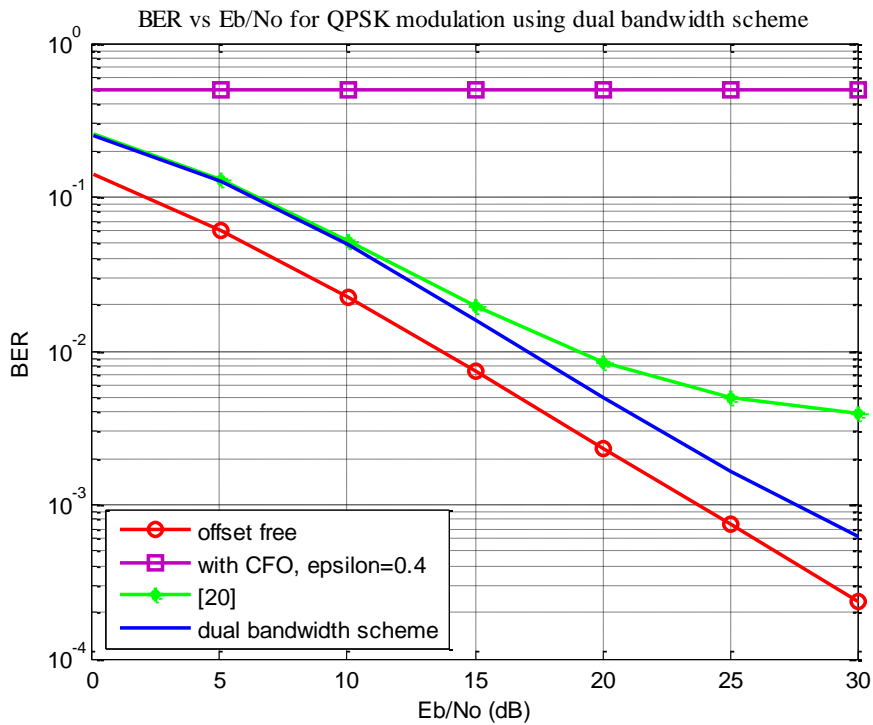

**Figure 5.19 : BER versus $E_b/N_0$ for OFDM system using dual bandwidth scheme compared to [20] and an offset free system for QPSK modulation scheme.**

Table 5-10 shows the BER versus $E_b/N_0$ for OFDM system using dual bandwidth scheme compared to [20] and an offset free system for QPSK modulation scheme over multipath Rayleigh fading channel.

**Table 5-10 : BER versus $E_b/N_0$ for OFDM system using dual bandwidth scheme compared to [20] and an offset free system for QPSK modulation scheme.**

| $E_b/N_0$ dB | BER | | |
|---|---|---|---|
| | Offset free | [20] | Dual bandwidth scheme |
| 0 | 0.1415 | 0.2559 | 0.2538 |
| 5 | 0.06124 | 0.1313 | 0.1277 |
| 10 | 0.02208 | 0.0524 | 0.04851 |
| 15 | 0.0073 | 0.01946 | 0.01588 |
| 20 | 0.002349 | 0.008418 | 0.005009 |
| 25 | 0.0007444 | 0.004976 | 0.001667 |
| 30 | 0.0002366 | 0.003897 | 0.0006282 |

Figure 5.20 shows the BER versus $E_b/N_0$ for OFDM system using dual bandwidth scheme compared to [20] and an offset free system for 16QAM modulation scheme over multipath Rayleigh fading channel.

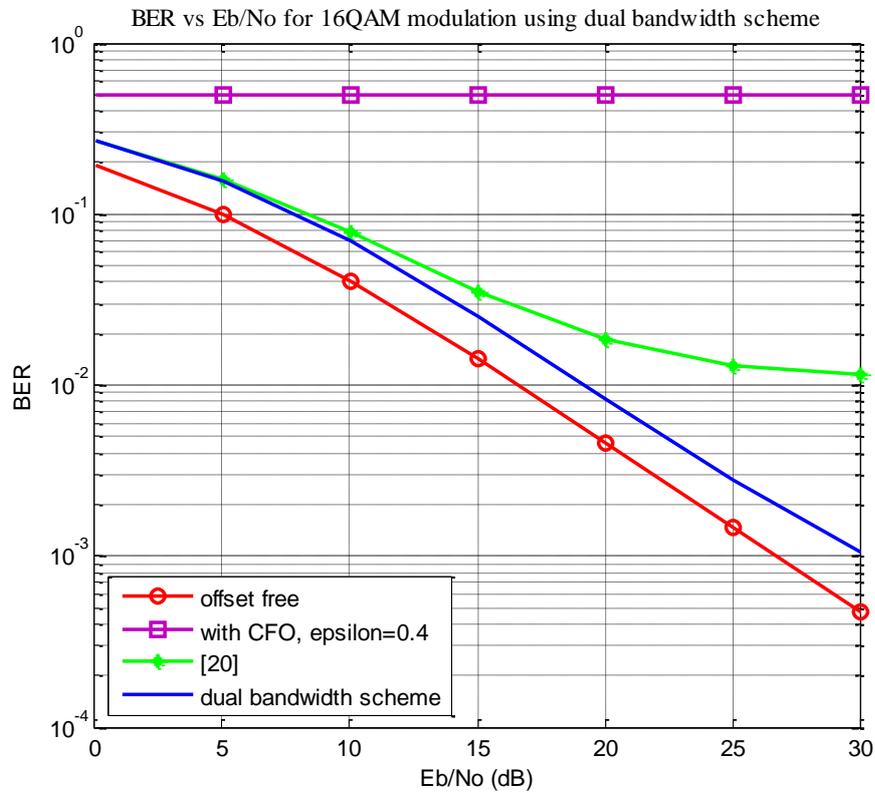

**Figure 5.20 : BER versus $E_b/N_0$ for OFDM system using dual bandwidth scheme compared to [20] and an offset free system for 16QAM modulation scheme.**

Table 5-11 shows the BER versus $E_b/N_0$ for OFDM system using dual bandwidth scheme compared to [20] and an offset free system for 16QAM modulation scheme over multipath Rayleigh fading channel.

**Table 5-11 : BER versus $E_b/N_0$ for OFDM system using dual bandwidth scheme compared to [20] and an offset free system for 16QAM modulation scheme.**

| $E_b/N_0$ dB | BER | | |
|---|---|---|---|
| | Offset free | [20] | Dual bandwidth scheme |
| 0 | 0.1925 | 0.2704 | 0.2675 |
| 5 | 0.09921 | 0.1607 | 0.1551 |
| 10 | 0.04033 | 0.07833 | 0.06972 |
| 15 | 0.0141 | 0.03514 | 0.02513 |
| 20 | 0.004616 | 0.01856 | 0.0082 |
| 25 | 0.001478 | 0.01311 | 0.002761 |
| 30 | 0.0004669 | 0.0114 | 0.001056 |

We conclude that, the dual bandwidth scheme and [20] have the same BER performance when $E_b/N_0$ is less than 10dB. For $E_b/N_0$ greater than 10dB, the BER performance of the dual bandwidth scheme is significantly improved compared to [20].

### 5.4.2 Improved dual bandwidth using clustered pilot tones scheme

In this scheme, further refinement to the dual bandwidth scheme using clustered pilot tones uses two pilot tones to estimate and correct the CFO assuming the clustered pilot tones has a fixed value $\{Z_{k,l}, Z_{k+1,l}\} = \{+1, -1\}$ [30].

Figure 5.21 shows the BER versus $E_b/N_0$ for OFDM system using improved dual bandwidth scheme compared to dual bandwidth scheme [47] and an offset free system for QPSK modulation scheme over multipath Rayleigh fading channel.

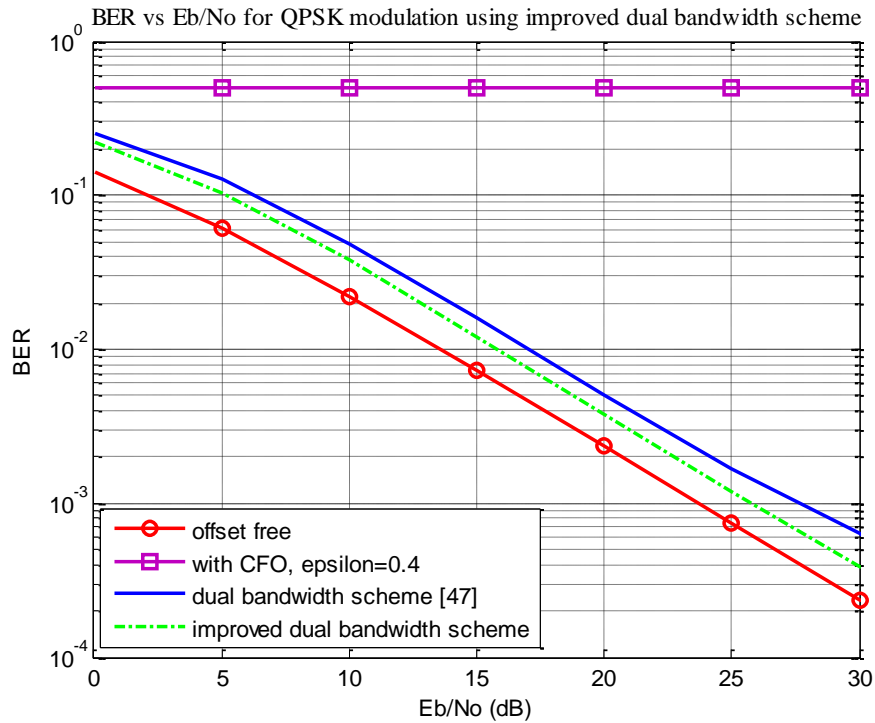

**Figure 5.21: BER versus $E_b/N_0$ for OFDM system using improved dual bandwidth scheme compared to dual bandwidth scheme [47] and an offset free system for QPSK modulation scheme.**

Table 5-12 shows the BER versus $E_b/N_0$ for OFDM system using improved dual bandwidth scheme compared to dual bandwidth scheme [47] and an offset free system for QPSK modulation scheme over multipath Rayleigh fading channel.

**Table 5-12: BER versus $E_b/N_0$ for OFDM system using improved dual bandwidth scheme compared to dual bandwidth scheme [47] and an offset free system for QPSK modulation scheme.**

| $E_b/N_0$ dB | BER | | |
|---|---|---|---|
| | Offset free | Dual bandwidth scheme | Improved dual bandwidth scheme |
| 0 | 0.1415 | 0.2538 | 0.2185 |
| 5 | 0.06124 | 0.1277 | 0.103 |
| 10 | 0.02208 | 0.04851 | 0.03774 |
| 15 | 0.0073 | 0.01588 | 0.01214 |
| 20 | 0.002349 | 0.005009 | 0.003744 |
| 25 | 0.0007444 | 0.001667 | 0.001179 |
| 30 | 0.0002366 | 0.0006282 | 0.0003898 |

Figure 5.22 shows the BER versus $E_b/N_0$ for OFDM system using improved dual bandwidth scheme compared to dual bandwidth scheme [47] and an offset free system for 16QAM modulation scheme over multipath Rayleigh fading channel.

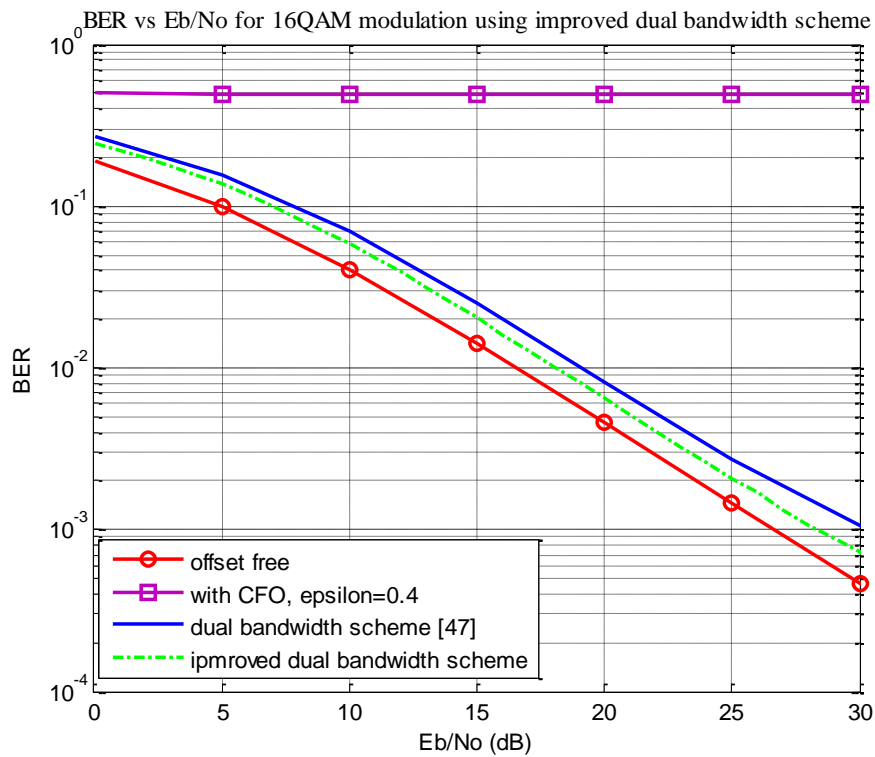

**Figure 5.22 : BER versus $E_b/N_0$ for OFDM system using improved dual bandwidth scheme compared to dual bandwidth scheme [47] and an offset free system for 16QAM modulation scheme.**

Table 5-13 shows the BER versus $E_b/N_0$ for OFDM system using improved dual bandwidth scheme compared to dual bandwidth scheme [47] and an offset free system for 16QAM modulation scheme over multipath Rayleigh fading channel.

**Table 5-13 : BER versus $E_b/N_0$ for OFDM system using improved dual bandwidth scheme compared to dual bandwidth scheme [47] and an offset free system for 16QAM modulation scheme.**

| $E_b/N_0$ dB | BER | | |
|---|---|---|---|
| | Offset free | Dual bandwidth scheme | Improved dual bandwidth scheme |
| 0 | 0.1925 | 0.2675 | 0.2443 |
| 5 | 0.09921 | 0.1551 | 0.1359 |
| 10 | 0.04033 | 0.06972 | 0.05832 |
| 15 | 0.0141 | 0.02513 | 0.02037 |
| 20 | 0.004616 | 0.0082 | 0.006525 |
| 25 | 0.001478 | 0.002761 | 0.002098 |
| 30 | 0.0004669 | 0.001056 | 0.0007246 |

We conclude that, the BER performance using dual bandwidth scheme and clustered pilot tones is simulated over AWGN and multipath Rayleigh fading channels for QPSK and 16QAM modulation schemes compared to dual bandwidth scheme [47] and an offset free system.

For QPSK modulation scheme, the BER = $4.851*10^{-2}$ is investigated at $E_b/N_0 = 10\ dB$ using dual bandwidth scheme [47] but it is investigated at $E_b/N_0 = 8.749\ dB$ using improved dual bandwidth scheme. The BER= $6.282*10^{-4}$ is investigated at $E_b/N_0 = 30\ dB$ using dual bandwidth scheme [47] but it is investigated at $E_b/N_0 = 27.84\ dB$ using improved dual bandwidth scheme. Thus, The BER performance using dual bandwidth scheme and clustered pilot tones is improved by about 1.25-2.16 dB than dual bandwidth scheme.

For 16QAM modulation scheme, the BER=$6.972*10^{-2}$ is investigated at $E_b/N_0 = 10\ dB$ using dual bandwidth scheme [47] but it is investigated at $E_b/N_0 = 9.067\ dB$ using improved dual bandwidth scheme. The BER=$1.05*10^{-3}$ is investigated at $E_b/N_0 = 30\ dB$ using dual bandwidth scheme [47] but it is investigated at $E_b/N_0 = 28.05\ dB$ using improved dual bandwidth scheme. Thus, The BER performance using dual bandwidth scheme and clustered pilot tones is improved by about 1-2 dB than dual bandwidth scheme.

# Chapter 6. Conclusions and Suggestions for Future Work

## 6.1 Conclusions

In this thesis, the wireless communication parameters and the basics of the OFDM systems have been studied. The OFDM systems has many features, such as robustness against multipath fading and narrow-band interference, high spectral efficiency and simple channel estimation and equalization, which are why it is an attractive method for wireless communication systems. In this thesis, the OFDM systems suffer from several problems such as the IQ imbalance, the PAPR, the phase noise, the STO and the CFO is discussed in more details. In this thesis, we evaluate the OFDM system for WLAN IEEE 820.11a.

The BPSK is more power efficient and need less bandwidth between all other modulation schemes used in an OFDM system. The 64QAM modulation scheme requires higher bandwidth and gives an excellent data rates with trade off increasing power as compared to others modulation schemes in an OFDM system. Thus, the BPSK has the lowest BER while the 64QAM has highest BER than others.

The CP eliminates ISI and ICI completely as long as the CP length is greater than or equal to channel delay spread. The cost due to CP is power loss and decreasing throughput for OFDM system.

Finally, the OFDM system for large FFT size is more immune for large channel delay spread. Thus, the BER performance is acceptable. But the larger FFT size requires greater number of calculations where the complexity of performing an FFT is dependent on the size of the FFT. Thus, the OFDM system becomes more complex.

In this thesis, we focus on the CFO and its effects on the performance of OFDM systems. CFO compensation schemes for OFDM systems have been studied. CFO compensation schemes are classified into NDA and DA schemes. The CFO correction based on type-2 control loop has large acquisition time. The large acquisition time makes the system is slow which degrade system performance. This problem motivates us to propose the dual bandwidth scheme to estimate and correct CFO for OFDM systems. The dual bandwidth scheme has small acquisition time which makes the system fast and enhanced system performance.

The BER performance is achieved for QPSK and 16QAM modulation schemes respectively for OFDM system. The proposed dual bandwidth scheme and [20] have the same BER performance when $E_b/N_0$ is less than 10dB. For $E_b/N_0$ greater than 10dB, the BER performance of the proposed dual bandwidth scheme is significantly improved compared to [20].

The constellation diagram is achieved for QPSK and 16QAM modulation schemes for offset free system with CFO $\varepsilon = 0.4$ at $E_b/N_0 = 15dB$. The constellation diagram is achieved for proposed dual bandwidth scheme and [20] is investigated at $E_b/N_0 = 15dB$. The noise is decreasing significantly for proposed dual bandwidth scheme. Thus, the performance of the proposed dual bandwidth scheme is better than [20].

The acquisition time of dual bandwidth scheme may be achieved around twenty OFDM symbols (i.e. the acquisition time of dual bandwidth scheme is achieved around 64 $\mu sec$). But the acquisition time of [20] is achieved around ninety OFDM symbols (i.e. the acquisition time of [20] is achieved around 288 $\mu sec$). The dual bandwidth scheme is significantly faster than [20]. Thus; the enhancement of dual bandwidth scheme is 77.78 percent than [20]. Moreover, for larger $\theta_{n1}$ parameter, the acquisition time may be decreased again but large oscillations significantly occur resulting in performance degradation. The acquisition time of dual bandwidth scheme is robust to multipath Rayleigh fading disturbances compared to [20].

The BER performance of the simulated system using dual bandwidth scheme and clustered pilot tones in both AWGN and multipath Rayleigh fading channels for QPSK and 16QAM modulation schemes compared to dual bandwidth scheme [47] and an offset-free system. For QPSK modulation scheme, the BER performance using improved dual bandwidth scheme is enhanced by about 1.25-2.16 dB compared to dual bandwidth scheme. For 16QAM modulation scheme, The BER performance using improved dual bandwidth scheme is enhanced by about 1-2 dB than dual bandwidth scheme.

## 6.2 Suggestions for future work

The proposed dual bandwidth scheme can be built to have a large bandwidth for rapid acquisition and a much narrower bandwidth for good tracking in the presence of noise. In this thesis, the proposed dual bandwidth scheme can be accomplished by switching in different capacitor values in a circuit of the loop filter. We suggest for future works, the proposed dual bandwidth scheme could be further accomplished by switching the loop resistance values of the loop filter or switching loop gain of the PLL. The proposed dual bandwidth scheme may be improved by optimizing the loop bandwidth. The proposed dual bandwidth and improved dual bandwidth schemes are implementing on DSP chip.

# References

[1] T. S. Rappaport, "Wireless Communications Principles and Practice", Second Edition, Prentice-Hall, 2002.

[2] J. G. Proakis, "Digital Communications", Fourth Edition, McGraw-Hill, 2001.

[3] A. R. S. Bahai, B. R. Saltzberg, and M. Ergen, "Digital Communications: Theory and Applications of OFDM", Second Edition, Springer, 2004.

[4] R. V. Nee and R. Prasad, "OFDM for Wireless Multimedia Communications", Artech House, Boston, London, 2000.

[5] Reto Ness, Jean-Paul Linnartz, Liesbet Van der Perre, Marc Engels, "Wireless OFDM Systems: How to Make Them Work", Mark Engels, Kluwer Academic Publishers, Dordrecht, The Netherlands, 2002.

[6] Fuqin Xiong, "Digital Modulation Techniques", Second Edition, Artech House, Boston, London, 2006.

[7] T. Jiang and Y. Wu, "An overview: peak-to-average power ratio reduction techniques for OFDM signals", IEEE Transactions on Broadcasting Technology society, Vol. 54, No. 2, pp. 257- 268, June 2008.

[8] Y. J. Kou, "Peak-to-Average Power-Ratio and Inter-carrier Interference Reduction Algorithms for Orthogonal Frequency Division Multiplexing Systems", Ph.D. Dissertation, University of Victoria, Dec. 2005.

[9] M. Morelli, C.-C.J. Kuo, and M.-O. Pun, "Synchronization techniques for orthogonal frequency division multiple access (OFDMA): A tutorial review" Proceedings of IEEE, Vol. 95, No. 7, pp. 1394-1427, July 2007.

[10] N. C. Beaulieu and P. Tan, "On the effects of receiver windowing on OFDM performance in the presence of carrier frequency offset", IEEE Transactions on Wireless Communications, Vol. 6, No.1, pp. 202- 209, January 2007.

[11] N. D. Alexandru and A. L. Onofrei, "ICI reduction in OFDM Systems using phase modified sinc pulse", Wireless Personal Communication, Vol. 53, No. 1, pp. 141-151, March 2010.

[12] J. G. Andrews, A. Ghosh, and R. Muhamed, "Fundamentals of WiMAX – Understanding Broadband Wireless Networking", Pearson Education, Inc., 2007.

[13] Pei-Yun Tsai and Tzi-Dar Chiueh, "OFDM Baseband Receiver Design for Wireless Communications", John Wiley and Sons (Asia), 2007.

[14] Y. Cho, J. Kim, W. Yang, and C. Kang, "MIMO-OFDM Wireless Communications with MATLAB", John Wiley and Sons, 2010.

[15] K. Fazel and S. Kaiser, "Multi-Carrier and Spread Spectrum Systems", John Wiley and Sons, 2008.

[16] Chi Chung Ko, Ronghong Mo, and Miao Shi ,"A New Data Rotation Based CP Synchronization Scheme for OFDM Systems", IEEE Transactions on Broadcasting Technology society, Vol. 51, No. 3, pp.315-321, September 2005.

[17] Chien-Chih Chen, Jung-Shan Lin, "Iterative ML Estimation for Frequency Offset and Time Synchronization in OFDM Systems", International Conference on Networking Sensing and Control,Vol.2, pp.1412-1417, 21-23 March 2004.

[18] A. Tarighat and A. Sayed, "Joint Compensation of Transmitter and Receiver Impairment in OFDM systems", IEEE Transactions on Wireless Communications, Vol.6, No.1, pp. 240- 247, January 2007.

[19] H. Shake and S. Fouludifrrd. "A New Technique for Estimation and Compensation of IQ Imbalance in OFDM Receivers", 8th International Conference on Communication Systems (ICCS), Vol.1, pp.224-228, 25-28 November 2002.

[20] Xiaofei Chen, Zhongren Cao, fredric harris, Bhaskar Rao, "OFDM carrier frequency offset correction based on type-2 control loop", IEEE International Conference on Acoustics, Speech and Signal Processing (ICASSP), pp. 3161- 3164, March 2012.

[21] Francois Horlin, Andre Bourdoux, "Digital Compensation for Analog Front-ends a New Approach to Wireless Transceiver Design", John Wiley and Sons, 2008.

[22] J. Y. Yu, M. F. Sun, T. Y. Hsu, and C. Y. Lee, "A Novel Technique for I/Q Imbalance and CFO Compensation in OFDM Systems", IEEE International Symposium on Circuits and Systems (ISCAS), pp. 6030–6033, 23-26 May 2005.

[23] Tim Schenk, "RF Imperfections in High-Rate Wireless Systems Impact and Digital Compensation", Springer, 2008.

[24] D. Tandur and M. Moonen, "Joint Adaptive Compensation of Transmitter and Receiver IQ Imbalance under Carrier Frequency Offset in OFDM Based Systems ", IEEE Transactions on Signal Processing, Vol.55, No.11, November 2007.

[25] Uma Shanker Jha and Ramjee Prasad, "OFDM towards Fixed and Mobile Broadband Wireless Access", Artech House, 2007.

[26] J. Armstrong, "Analysis of new and existing methods of reducing intercarrier interference due to carrier frequency offset in OFDM", IEEE Transactions on Communications, Vol. 47, No.3, pp. 365-369, March 1999.

[27] Y. Zhao and S.G. Haggman, "Inter-carrier interference self-cancellation scheme for OFDM mobile communication systems", IEEE Transactions on Communications, Vol. 49, No.7, pp. 1185–1191, July 2001.

[28] H. Schulze and C. Luders, "Theory and Applications of OFDM and CDMA: Wideband wireless Communications", John Wiley and Sons, 2005.

[29] G. Kalivas, "Digital Radio System Design ", John Wiley and Sons, 2009.

[30] Ye (Geoffrey) Li Gordon Stuber, "Orthogonal Frequency Division Multiplexing For Wireless Communications", Springer, 2006.

[31] W. Zhang, X. G. Xia and P. C. Ching, "Clustered Pilot Tones for Carrier Frequency Offset Estimation in OFDM Systems", IEEE Transactions on Wireless communication, Vol. 6, No. 1, pp.101-109, January 2007.

[32] Ming-Fu Sun, Jui-Yuan Yu, and Terng-Yin Hsu, "Estimation of Carrier Frequency Offset with I/Q Mismatch Using Pseudo-Offset Injection in OFDM Systems", IEEE Transactions on Circuits and Systems-I, Vol. 55, No. 3, pp.943-952, April 2008.

[33] J. van de Beek, M. Sandell, and P. O. B¨orjesson, "ML Estimation of Time and Frequency Offset in OFDM Systems", IEEE Transactions on Signal Processing, Vol. 45, NO. 7, pp.1800–1805, July 1997.

[34] Linling Kuang , Zuyao Ni, Jianhua Lu and Junli Zheng, " A Time Frequency Decision-Feedback Loop for Carrier Frequency Offset Tracking in OFDM Systems ", IEEE Transactions on Wireless Communications, Vol.4, No.2, pp. 367- 373, March 2005.

[35] S. Moridi and H. Sari, "Analysis of Four Decision-Feedback Carrier Recovery Loops in The Presence of Inter-symbol Interference", IEEE Transactions on communications, Vol. 33, No.6, pp. 543–550, June 1985.

[36] H. Liu and U. Tureli, "A High Efficiency Carrier Estimator for OFDM Communications", IEEE Communications Letters, Vol. 2, No. 4, pp. 104–106, April 1998.

[37] M. Li and W. Zhang, "A Novel Method of Carrier Frequency Offset Estimation for OFDM Systems", IEEE Transactions on Consumer Electronics, Vol. 49, No.4, pp. 965-972, November 2003.

[38] P. H. Moose, "A Technique for Orthogonal Frequency Division Multiplexing Frequency Offset Correction", IEEE Transactions on Communications, Vol. 42, No.10, pp. 2908–2914, October 1994.

[39] T. M. Schmidl and D. C. Cox, "Robust Frequency and Timing Synchronization for OFDM", IEEE Transactions on Communications, Vol. 45, No.12, pp. 1613–1621, December 1997.

[40] M. Morelli and U. Mengali, "An Improved Frequency Offset Estimator for OFDM Applications", IEEE Communications Theory Mini-Conference, Vol. 3, pp. 106–109, 6-10 June 1999.

[41] ETSI, "Digital Video Broadcasting: Framing Structure, Channel Coding and Modulation for Digital Terrestrial Television", ETSI EN 300 744 V1.4.1, January 2001.

[42] F. Classen and H. Meyr, "Frequency synchronization Algorithm for OFDM Systems Suitable for Communication over Frequency Selective Fading Channels", Proceedings of IEEE Vehicular Technology Conference (VTC), Vol.3, pp. 1655-1659, 8-10 June 1994.

[43] M. Speth, S. Fechtel, G. Fock, and H. Meyr, "Optimum Receiver Design for OFDM Based Broadband Transmission–part II: a case study", IEEE Transactions on Communications., Vol. 49, No.4, pp. 571-578, April 2001.

[44] Ronald. E. Best, "Phase-locked Loops Designs Simulation and Applications", McGraw-Hill, Fifth Edition, 1999.

[45] F. M. Gardner, "Phaselock Techniques", Hoboken, NJ: Wiley, Third Edition, 2005.

[46] Lioyd Temes, Mitchel E.Schultz, "Electronic Communication", McGraw-Hill, Second Edition, 1998.

[47] Mohammed. S. El-Bakry, Hamed. F. El-Shenawy, Abd El-Hady. A. Ammar "Improved Carrier Frequency Offset Correction for OFDM Systems Based on Type-2 Control Loop", 31st National Radio Science Conference (NRSC), pp.157-166, 28-30 April 2014.

[48] IEEE Standard 802.11a, "Wireless LAN Medium Access Control (MAC) and Physical Layer (PHY) Specifications", 1999.

[49] William C. Y. Lee "Mobile Cellular Telecommunications analog and digital systems", McGraw-Hill, Second Edition, 1995.

# الملخص العربي

نتيجة للطلب المتزايد علي تطبيقات اللاسلكية الوسائط المتعددة مما يتطلب تصميم أنظمة اتصالات لاسلكية ذات معدل تداول عالي ونظرا لمحدودية الطيف الترددي المتاح كمصدر ذو قيمة مما اوجب ضرورة الاستفادة منه بكفاءة في وجود الانظمة اللاسلكية الاخري.

أصبح الاكثار المتعامد بتقسيم التردد كتقنية تعديل متعددة الموجات الحاملة ذات أهمية كبري في أنظمة الاتصالات اللاسلكية للمميزات الاتية : كفاءة استخدام الطيف الترددي ومقاومة الخفوت الانتقائي للتردد نتيجة تعدد المسارات وقدرته علي التكامل مع التقنيات الاخري ورغم ذلك انه يعاني من النسبة العالية بالنسبة لمتوسط القدرة وكذا حساسيته لانحراف تردد الموجة الحاملة.

يؤدي انحراف تردد الموجة الحاملة الي التداخل بين الموجات الحاملة وكذا تدهور اداء انظمة الاكثار المتعامد بتقسيم التردد مما يتحتم استنباط واستخدام طرق دقيقة لتعويض انحراف تردد الموجة الحاملة لكي يكون اداء نظام الاتصالات مقبول.

ان انحراف تردد الموجة الحاملة يعني عدم توافق بين تردد الموجة الحاملة المستقبلة وتردد الموجة الحاملة الناتجة من المذبذب المحلي في الاستقبال ويحدث انحراف تردد الموجة الحاملة نتيجة سببين أولهما عدم التوافق بين المذبذب المحلي في الارسال والاستقبال وثانيا تأثير دوبلر وهو انحراف التردد نتيجة للحركة النسبية بين المرسل والمستقبل.

وينقسم انحراف تردد الموجة الحاملة الي انحراف رقم صحيح وانحراف كسري وتم عمل مسح للطرق المختلفة لتزامن التردد والتي تنقسم الي طرق باستخدام بيانات مساعدة وطرق لا تعتمد علي بيانات مساعدة وفي الطرق التي تستخدم بيانات مساعدة يتم ادخال بيانات تدريبية اضافية معروفة الي بيانات المعلومات المرسلة وهذه الطرق تعطي اداء أفضل ولكنها لا تستخدم الطيف الترددي بكفاءة بينما الطرق التي لاتعتمد علي بيانات مساعدة تستخدم بيانات المعلومات فقط وهي تعطي اداء اقل ولكنها تستخدم الطيف الترددي بكفاءة.

تتم عملية تزامن التردد علي مرحلتين : الاكتساب والتتبع ان الاكتساب يعني تقدير خشن لانحراف التردد وهي ذات مدي واسع ولكن بدقة منخفضة بينما التتبع يعني تقدير ناعم لانحراف التردد وهي ذات مدي اضيق ولكن بدقة عالية.

تبحث هذه الرسالة في طرق تعويض انحراف التردد باستخدام بيانات مساعدة وتركز علي عملية تتبع انحراف تردد الموجة الحاملة حيث تم عمل نموذج لنظام الاتصالات ذو الاكثار المتعامد بتقسيم التردد المثالي اولا اي بدون اي انحراف تردد الموجة الحاملة ثم نموذج في وجود انحراف تردد الموجة الحاملة وتم تقويم اداء نظام الاتصالات باستخدام المحاكاة وتم اسخدام معدل الخطأ النبضي ورسم التفلطح والخرج الطوري والخطأ الطوري كمعايير تقويم الاداء عند استخدام الانواع المختلفة للتعديل والحالات المختلفة لقناة الاتصال.

وفي هذه الرسالة تم اقتراح طريقة ذات عرضي نطاق لحلقة التحكم ذو النمط الثاني لتقليل زمن الاكتساب ولقد اثبتت النتائج ان الطريقة المقترحة اسرع وتؤدي الي تحسين اداء نظام الاتصالات ذو الاكثار المتعامد بتقسيم التردد. وكذلك تم تحسين طريقة ذات عرضي نطاق باستخدام نغمات اشارات الارشاد المجمعة واثبتت النتائج تحسن اكثر في اداء نظام الاتصالات ذو الاكثار المتعامد بتقسيم التردد.

تم تنظيم هذه الرسالة فى ستة فصول بيانها على النحو التالي :

الفصل الأول يتناول الدافع وراء هذه الرسالة وأهدافها وتنظيمها.

الفصل الثاني يتناول قنوات الاتصال اللاسلكية ومعاملات القناة اللاسلكية ويناقش المبادئ الأساسية لتقنية الإكثار المتعامد بتقسيم التردد كما يشرح مفهوم (CP) اللازمة لتجنب التداخل بين الرموز و التداخل بين الموجات الحاملة في القنوات الاتصال ويتناول أيضا مميزات وعيوب وتطبيقات تقنية الإكثار المتعامد بتقسيم التردد.

الفصل الثالث يقدم نموذج لنظام الإكثار المتعامد بتقسيم التردد بدون إعاقات كما يتم مناقشة التحديات التي تواجه تقنية الإكثار المتعامد بتقسيم التردد حيث تم تقديم تأثير(IQ imbalance) على أداء نظام الإكثار المتعامد بتقسيم التردد وأيضا تم توضيح أسباب النسبة العالية بالنسبة لمتوسط القدرة (PAPR) لأنظمة الإكثار المتعامد بتقسيم التردد وتأثيرها علي أداء مكبر القدرة. وتم تقديم تأثيرات انحراف زمن الرمز (STO) على أداء تقنية الإكثار المتعامد بتقسيم التردد وأخيرا تم تقديم انحراف تردد الموجة الحاملة لتقنية الإكثار المتعامد بتقسيم التردد وشرح أسباب انحراف تردد الموجة الحاملة و تقديم نموذج لنظام الإكثار المتعامد بتقسيم التردد فى وجود انحراف تردد الموجة الحاملة وشرح تأثير الانحراف الرقمي الصحيح والانحراف الكسري على أداء أنظمة الإكثار المتعامد بتقسيم التردد.

الفصل الرابع يشرح مراحل عملية تزامن الترددات وطرق تعويض انحراف تردد الموجة الحاملة حيث تم استعراض الطرق التي تستخدم بيانات مساعدة والطرق التي لا تعتمد علي بيانات مساعدة ومميزات وعيوب كل طريقة كما تم مقارنة الفرق بين تقدير انحراف تردد الموجة الحاملة باستخدام نغمات إشارات الإرشاد التقليدي و تقدير انحراف تردد الموجة الحاملة باستخدام نغمات إشارات الإرشاد المجمعة وعرض التعاريف والمعاملات الرئيسية لحلقة التحكم وتم اقتراح طريقة ذات عرضي نطاق لحلقة التحكم ذو النمط الثاني وتم تحسين طريقة ذات عرضي نطاق لحلقة التحكم ذو النمط الثاني باستخدام نغمات إشارات الإرشاد المجمعة .

الفصل الخامس يعرض النتائج العددية لنظام الإكثار المتعامد بتقسيم التردد في شبكة اتصالات محلية لاسلكية باستخدام معدل الخطأ النبضي في الحالات المختلفة لقناة الاتصال ولتقنيات التعديل المختلفة و مع وبدون CP وبأطوال CP مختلفة وأطوال مختلفة لتأخير انتشار القناة و أحجام مختلفة FFT .

ويعرض أيضا النتائج العددية للطريقة المقترحة ذات عرضي نطاق لحلقة تحكم ذو النمط الثانى وللتحسين لطريقة ذات عرضي نطاق لحلقة تحكم ذو النمط الثانى باستخدام نغمات إشارات الإرشاد المجمعة باستخدام رسم التفلطح والخرج الطوري و الخطأ الطوري معدل الخطأ النبضى لتقنيات التعديل المختلفة .

الفصل السادس يلخص النتائج التي تم الحصول عليها في هذه الرسالة وبعض اقتراحات البحوث المستقبلية.

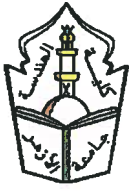

جامعة الأزهر
كلية الهندسة
قسم الهندسة الكهربية

# تقويم اداء أنظمة الاتصالات اللاسلكية متعددة الحوامل

رسالة مقدمه إلى قسم الهندسة الكهربية – كلية الهندسة – جامعة الأزهر
للحصول على درجة التخصص الماجستير

فى

هندسة الاتصالات الكهربية

مقدمة من

## محمد صبحي أحمد البكري

بكالوريوس هندسة الالكترونيات والاتصالات – معهد هندسة وتكنولوجيا الطيران

يعتمد من لجنة الممحتنين

**أ.د./السيد مصطفي سعد**
كلية الهندسة – جامعة حلوان

**أ.د./هشام محمد البدوي**
المعهد القومي للاتصالات

**أ.د./عبدالهادي عبدالعظم عمار**
كلية الهندسة – جامعة الأزهر

القاهرة – مصر
٢٠١٤

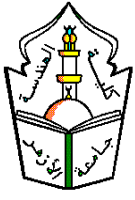

جامعة الأزهر
كلية الهندسة
قسم الهندسة الكهربية

# تقويم اداء أنظمة الاتصالات اللاسلكية متعددة الحوامل

رسالة مقدمه إلى قسم الهندسة الكهربية – كلية الهندسة – جامعة الأزهر
للحصول على درجة (التخصص) الماجستير

فى

هندسة الاتصالات الكهربية

مقدمة من

## محمد صبحي أحمد البكري

بكالوريوس هندسة الالكترونيات والاتصالات – معهد هندسة وتكنولوجيا الطيران

تحت إشراف

### أ.د./ عبد الهادي عبدالعظيم عمار
قسم الهندسة الكهربية
كلية الهندسة – جامعة الأزهر

### د./ حامد عبدالفتاح الشناوي
قسم الاتصالات الكهربية
المعهد العالي للهندسة – أكاديمية الشروق

القاهرة – مصر
٢٠١٤